\documentclass[12pt,ngerman,english]{article}
\usepackage[T1]{fontenc}
\usepackage[utf8]{inputenc}
\usepackage{geometry}
\usepackage{fancyhdr}
\usepackage{color}

\usepackage{afterpage}
\newcommand\blankpage{%
 \null
 \thispagestyle{empty}%
  \addtocounter{page}{-1}%
  \newpage}
  
\usepackage[nottoc,notlot,notlof]{tocbibind}
\usepackage{babel}
\usepackage{float}
\usepackage{units}
\usepackage{amsmath}
\usepackage{amssymb}
\usepackage{stmaryrd}
\usepackage{graphicx}
\usepackage{setspace}
\usepackage{subcaption}

\usepackage{cancel}

\usepackage{esint}
\usepackage[unicode=true,pdfusetitle,
 bookmarks=true,bookmarksnumbered=false,bookmarksopen=false,
 breaklinks=false,pdfborder={0 0 0},pdfborderstyle={},backref=false,colorlinks=true]
 {hyperref}
\hypersetup{
 linkcolor=blue,urlcolor=blue,citecolor=blue,filecolor=blue}
 \usepackage{pdfpages}
 \usepackage[export]{adjustbox}
 
 \usepackage{listings} 
\usepackage{soul} 

\makeatletter

\usepackage{upgreek}
\usepackage[usenames,dvipsnames]{xcolor}
\def\barp{p}
\def\barq{q}
\def\i{{\rm i}}
\def\e{{\rm e}}
\def\r{h_1}
\def\s{h_0}
\def\ep{\epsilon}

\makeatother

\begin{document}
\thispagestyle{empty}
\begin{center}
\textbf{\Large{Model-based and empirical analyses of stochastic fluctuations in economy and finance}}

\par\end{center}{\Large \par}

\bigskip{}

\begin{center}
\textbf{\Large{DISSERTATION}}
\par\end{center}

\bigskip{}

\begin{center}
\textbf{\Large{zur Erlangung des akademischen Grades}}
\par\end{center}

\begin{center}
\textbf{\Large{Doctor rerum naturalium}}

\textbf{\Large ({Dr. rer. nat.})}

\bigskip{}

\bigskip{}

\bigskip{}

\bigskip{}

\textbf{\Large{vorgelegt}}

\par\end{center}

\bigskip{}

\begin{center}
\textbf{\Large{dem Bereich Mathematik und Naturwissenschaften}} 

\textbf{\Large{der Technischen Universität Dresden}}

\bigskip{}

\bigskip{}

\textbf{\Large{von}}

\bigskip{}

\bigskip{}

\begin{center}
\textbf{\Large{}{\em RUBINA ZADOURIAN}}

\bigskip{}

\bigskip{}

%\textbf{\Large{}{geboren am 10.10.1982  in Iran, Shiraz}}

\bigskip{}

\bigskip{}

\bigskip{}

%\textbf{\Large{}{Eingereicht am 21.02.2018}}

\par\end{center}

\par\end{center}

\bigskip{}

\begin{center}
\textbf{\Large{}{Die Promotion wurde am Max-Planck-Institut für Physik Komplexer Systeme in Dresden und am Mathematischen Institut der University of Oxford durchgeführt.}}
\par\end{center}

\afterpage{\blankpage}

\newpage
\thispagestyle{empty}

\textbf{\Large{List of publications related to this PhD work:}}
\vspace{1 cm}

\textbf{1. \underline{ZADOURIAN}, Rubina; GRASSBERGER, Peter. Asymmetry of cross-correlations between intra-day and overnight volatilities. EPL (Europhysics Letters), 2017, 118. Jg., Nr. 1, S. 18004.}

\textbf{\url{https://doi.org/10.1209/0295-5075/118/18004}}

\vspace{1.2 cm} 

\textbf{2. \underline{ZADOURIAN}, Rubina; SAAKIAN, David B.; KLÜMPER, Andreas. Exact probability distribution functions for Parrondo's games. Physical Review E, 2016, 94. Jg., Nr. 6, S. 060102.}

\textbf{\url{https://doi.org/10.1103/PhysRevE.94.060102}}

\vspace{1.2 cm} 

\textbf{3. \underline{ZADOURIAN}, Rubina; KLÜMPER, Andreas. Exact probability distribution function for the volatility of cumulative production. Physica A: Statistical Mechanics and its Applications, 2017.}

\textbf{\url{https://doi.org/10.1016/j.physa.2017.12.003}}

 \vspace{1.2 cm}

\textbf{4. LAFOND, Francois; GOTWAY BAILEY, Aimee; BEKKER Jan David; REBOIS, Dylan; \underline{ZADOURIAN}, Rubina; MCSHARRY, Patrick; FARMER, J. Doyne. How well do experience curves predict technological progress? A method for making distributional forecasts.Technological Forecasting $\&$ Social Change (2017)}

\textbf{\url{https://doi.org/10.1016/j.techfore.2017.11.001}}
\vspace{1.2 cm}

\textbf{5. CHEONG, Kang Hao; SAAKIAN, David B.; \underline{ZADOURIAN}, Rubina. Allison mixture and the two-envelope problem. Physical Review E, 2017, 96. Jg., Nr. 6, S. 062303.}

\textbf{\url{https://doi.org/10.1103/PhysRevE.96.062303}}
\vspace{1.2 cm}

\afterpage{\blankpage}
\newpage

\thispagestyle{empty}
\tableofcontents{}
\thispagestyle{empty}

\newpage

\thispagestyle{empty}
\topskip0pt
\vspace*{\fill}
\begin{center}
\section*{\textbf{\LARGE{}{Part I}}}
\section*{\textbf{\huge{}{Introduction}}}
\end{center}
\vspace*{\fill}
\vspace*{10cm}
\newpage
\setcounter{page}{1}

\section{Introduction}

The scientific, financial and technological progresses
 in the past decades led to an emergent understanding
of complex systems \cite{Cilliers,Waldrop}. In many
domains scientists realized that the development
of new ideas requires to grasp concepts originating from research on
complex behaviors. Complexity has shown to be a link between
 many disciplines in science and
thus become a ubiquitous interdisciplinary field of 
study. As a result, discovering and realizing applications
of complex systems shed light on a wide variety of
fields and endeavors \cite{Kaneko,Miller}.

\bigskip

Complex systems are those in which many
different parts are interacting simultaneously
and non-linearly with each other, which is responsible
for new collective phenomena, usually exhibiting chaotic and emergent behaviors. The basic mechanisms of climate, earthquake,
 human brain and financial markets exhibit features of complex and
 collective behaviors occurring in real life \cite{Com,Mann}.
One of the main goals of the
 analyses of such systems is the manifestation of the
        spontaneous change of properties in temporal and
         spatial structures. Moreover, in a complex system
          the interaction among components occurs in such
           a way, that the system as a whole cannot be fully
            described by analyzing each single
             constituent of the system. 
             Interactions lead to situations, in which 
              the system can show new behavior
               and possesses properties, which are dissimilar from the
                behavior and properties of its components.
               Such systems often act under
                conditions far from equilibrium and may be
                 influenced by their memory. Another 
                 characteristics of complex systems is that
                  they are usually open, i.e. the interactions
                   not only exist between the components
                    of the system but also with its environment \cite{Yam}.
                    
                   \bigskip

As a perfect example of man-made complex systems,
 in the present thesis, financial and economic systems
  are under consideration. Such systems exhibit features
   typical of complex behavior discussed above \cite{Farmer1}: 
   In financial markets there exists a large number of
    participants with various activities. 
   A large number of agents, banks, companies 
   and their mutual evolving networks provide an
    appropriate ground for the investigation of
     complexity and randomness. The financial 
     markets are certainly an open system and continuously
      interact with political systems, science, 
      agriculture, technology, etc. Moreover, the
       financial and economic systems are characterized
        by the dynamics of supply and demand, that
         means the process can not be in equilibrium.
          In addition, the financial markets exhibit a long 
          term memory effect and are significantly 
          influenced by their history \cite{Ding}. 
          
 \bigskip
 
Furthermore, the discipline of statistical mechanics, 
based on probability theory, in theoretical physics 
opened a new avenue for better understanding of 
complexity. More precisely, probability theory is 
applied to explore the behavior of mechanical 
systems by analyzing the uncertainties of the 
states of the system. In particular, the first
 victorious effort to understand the complex
  behavior was the development of equilibrium 
  thermodynamics, which allowed to extract 
  the properties of macroscopic systems and 
  describe their behavior in terms of only a few 
  emergent macroscopic properties entering state
   functions and laws \cite{Gibbs1}. 
     The discovery of the laws of thermodynamics
      played a seminal role for describing the
       macroscopic behavior of various systems 
       beyond ordinary thermodynamics \cite{Tolman}-\cite{Fisher}.

        \bigskip

One of the main objectives in statistical mechanics
 is to deal with the properties of large systems,
  represented by ensembles, including many degrees
   of freedom, by providing exact methods to connect
  the macroscopic and microscopic properties.  
  The main reason for our need of statistical
   methods, stems from the lack of our knowledge
    of the gigantic number of microscopic degrees
     of freedom. As a result, the purpose of statistical
      mechanics is to invent methods to deal with this
       complexity and incompletely known systems.

          \bigskip

One of the inspirations that statistical mechanics
 provides for financial markets stems from
  various similarities, e.g. a large number
   of elementary constituents interact in a rather simple manner, 
   often by few body interactions. In fact, in physics
    there are established models 
    (Ising spin systems, percolation models,
     interaction networks, etc.) 
     that are built on this property and are used to model
      economic systems \cite{Saberi,Hosseiny}.
      Such interdisciplinary approaches provide a setting
 that enables researchers for better understanding
  various real-life problems, which are mainly based
   on turn to a new view of complexity.

     \bigskip

Another branch, which has an analogy and parallelism
 with statistical mechanics is information theory. 
 This is another approach, that neatly describes
  the notion of the information flow, which constructs
   the structural ground for setting up probability
    distributions on the basis of fundamental knowledge of uncertainty and information. 
Shannon in his stunning work introduced entropy 
as a measure of the uncertainty of the information
 contents and in his papers \cite{Shannon,Shannon1}
  he could completely describe the mathematical
   background of communication theory. 
   The concept of ``energy flow'' from statistical
    mechanics was somehow replaced by
     the notion of ``information flow''. Shannon's
      communication theory was essential for science,
       in particular for engineering and technological sectors.
        In the present thesis, the elements of information
         theory are considered to analyze the 
         temporal correlations
         in the financial time series \cite{Rubina}.
         
           \bigskip
           
  Nonlinear dynamics has been also devoted to develop
 methods and tools \cite{Kantz} in different areas
  (ergodicity, chaos, stochastic processes, etc.) 
  for studying the complex phenomenon, that appears
   in various natural or man-made systems \cite{Fuchs}.
    Moreover, it is known that the presence of perturbations, 
    artifacts and irregularities in the data introduce nonlinear
     effects which often occur in complex networks. 
     Therefore in the present thesis also nonlinear effects
      and dependencies such as mutual information,
       Spearman's correlation coefficient, $K$-nearest neighbor 
        statistics are intensively considered.
        
         \bigskip

Econophysics history dates back to Daniel Bernoulli for his pioneering
 work, in which he proposed a theory for risk aversion and utility
  \cite{Bernoulli}. Notable scientists with a physics background were often awarded  prizes in economy, such as F. Black, R. F. Engle and J. Tinbergen. The latter two scientists were honored by the Nobel Prize. J. Tinbergen won the prize for 
   the development of econometric modeling, R. F. Engle for his time
    varying volatility (ARCH model) and F. Black is known for his well-known
     Black-Scholes model.  Physicists
    have progressively offered a novel approach to
     the economic sciences by devising new methods
      to explain a variety of phenomena, leading to a 
      number of models and stylized facts \cite{Anderson}-\cite{Luciano},
 just to name a few. Especially, in the past three decades, they
  intensively studied fluctuations and collective behavior in 
  human-driven dynamics in economy and finance, inspired
   by the idea, that such systems are usually multi-agent 
   systems similar to many particle systems, with  well 
   defined interaction rules. Some notable physicists who 
   have a great impact in the progress of the field are
   J. P. Bouchaud, R. Cont,  J. D. Farmer, L. Pietronero and H. E. Stanley.
  For a more comprehensive exposition
      of the field I recommend the book written 
      by J. P. Bouchaud and M. Potters: ``Theory
of Financial Risk and Derivative Pricing: From Statistical Physics
to Risk Management''.

\begin{center}
\text{* * *}
\end{center}
   
     \bigskip

The objective of this thesis is the investigation
 of {\em complexity}, {\em asymmetry},
  {\em stochasticity} and {\em non-linearity} 
  of the financial and economic systems by 
  using the tools of statistical mechanics and
   information theory. More precisely, this thesis
    concerns {\em statistical-based modeling} and {\em empirical 
    analyses} with applications in finance, forecasting,
     production processes and game theory. In these
      areas the time dependence of probability 
      distributions is of prime interest and can be 
      “measured” or exactly calculated for model
       systems.
       
         \bigskip
       
        The correlation coefficients and 
       moments are among the useful quantities 
       to describe the dynamics and the correlations
        between random variables. However, the 
        full investigation can only be achieved if the
         probability distribution function of the
          variable is known; its derivation is one of the
           main focuses of the present thesis.
            Asymmetry and forecasting of financial
             time series will be addressed here as well.
              Moreover, the derivation of stochastic
               properties of production processes is also
                intensively under consideration. Furthermore, 
                being motivated by the usefulness and 
                importance of the notion of {\em volatility},
                 the empirical (in financial sector) 
                 and theoretical analyses (in production process)
                 of it, in the above mentioned areas are of major consideration.
                 Finally, the exact probability distribution functions
                     for two of the well-known
                      models in game theory, which appear at the intersection of 
                      different disciplines,
                       are also derived.

 \bigskip

For the more precise discussion, I go through each 
chapter by presenting the details. The thesis contains
 five chapters as follows: The first chapter is the introduction part. 
  In the second chapter
  I review some mathematical and numerical tools
   that are relevant to the analyses done in the thesis.
    The third, fourth and fifth chapters I comment on below in detail.

    \bigskip

    \subsection*{Part I}

In the third chapter I investigate the correlated structure
 of financial time series and attempt to unveil the fine structure
  of volatility, based on empirical analysis.
  
   The recent financial
       crises and risk portfolios require a better 
       understanding of financial market 
       dynamics. One of the
   important challenges in finance is to forecast the
    financial market volatility, which has an important
     meaning for portfolio management and risk 
     measurement (e.g. if the volatilities are 
     predicted to be high for a given asset, 
     then the investors may reduce their 
     commitments to the asset, in order to
      avoid risk). It is well known that fluctuations
        in equity prices traded on any stock exchange
         are weakly correlated, and hardly allow any
          non-trivial forecasting. This is different 
          for the amplitudes of these fluctuations,
           namely for volatilities.  With some modern
            options it is possible to make profits 
            with forecasts for volatilities
(although forecasts of signed fluctuations
 would be more easy to turn into
money, if they were possible), whence
 volatilities have been studied
extensively in the econometric literature
 \cite{Engle,Bouchaud}. 
 
   \bigskip

As first shown in \cite{french}, the statistics of volatilities
is not uniform. Rather, there is a marked daily structure, with high
volatility during the opening hour of the market and a more calm period
around noon. Also, equity prices at the opening of the trading session are in general
different from the closing prices on the previous trading day, showing
that there is a non-trivial overnight dynamics. 

  \bigskip

The general consensus seems
that overnight volatility is useful for predicting subsequent intraday
volatility \cite{Blanc}-\cite{Chicheportiche}. This is an important
result. But prediction involves a model (GARCH \cite{Gallo},
SEMIFAR \cite{Chen}, or different versions of the stochastic volatility
model (SVM) \cite{Tsiakas}-\cite{Lee}), and none of the papers cited
above reported \textit{model independent} analyses of the raw data themselves.
This is so in spite of the fact that data analyses that are not involving any
model and using only elementary methods and minimal assumptions would
be most useful for understanding the \textit{data} and \textit{basic mechanism(s)} underlying
the phenomena, which is the purpose of the analyses in the
 third chapter. Such analyses based on raw
  data will allow us to understand the relationships between
   different observed quantities in order to perform data based predictions.
   
     \bigskip
     
     One of the main concerns in the third chapter of the thesis
 is with cross-correlations between intraday and overnight
  volatilities. Due to strong non-stationarity in the data,
   the analyses are based on different approaches, 
   namely linear (Pearson) and non-linear methods
    (Spearman, mutual and KNN statistics). For example, 
    the Spearman's correlation coefficients, being based on rank statistics, are known to be 
    much more robust. 
    
     \bigskip

I will discuss the finding of a quite  stunning time asymmetry
 in the short time cross correlations between intraday and
  overnight volatilities (absolute values of log-returns of
   stock prices). While overnight volatility is significantly 
   (and positively) correlated with the intraday volatility
    during the following day (allowing thus non-trivial
     predictions), it is much less correlated with the
      intraday volatility during the preceding day.
       While the effect is not unexpected in view
        of previous observations, its robustness and
         extreme simplicity are remarkable. The work has been published in \cite{Rubina}. (For the
          empirical part, the time series of stock prices
           in different countries are used).
           
           \bigskip

 At first sight this strong asymmetry looks very 
strange, in particular
since time asymmetry is usually considered 
to be very weak in financial
data. Many popular models (most noticeably
 all models of the ARCH family)
are time symmetric by construction,
 and where time asymmetry is seen
\cite{Zumbach} it is only seen in very special observable.
But the findings are indeed compatible with previous analyses 
\cite{Chen}-\cite{Edmonds}.

The asymmetry found in this thesis 
 is also interesting and arguable for the risk
  management by considering larger time scales
   and by comparing day/night to night/day results
    between more distant nights and days. It is remarkable that an extensive
statistical study of intraday and overnight returns and volatilities
was recently made in \cite{Wang}, but since that
 analysis was not
guided by any theoretical considerations, 
the finding described above
was missed. 

 \bigskip

Furthermore, the analyses have been done by taking
 into account foreign markets in different time zones
  by considering Asian, American and European markets.
   For the applied content, I used mainly free database
     downloaded from yahoo finance website \cite{yahoo}.
Moreover, in this thesis I analyze the sentiment index
 presented on Boerse Stuttgart website \cite{Stuttgart}
  and also VIX index reported by Chicago Board Options
   Exchange (CBOE) \cite{CBOE}, which are based on
    the {\em Sentiment} and {\em Fear} index of traders
     and investors. The discussion will be followed by analyzing the predictive power of VIX which tracks the Standard $\&$ Poor's 500 U.S. stock market index. Finally, it is worthwhile to mention
      that the motivation of such analyses comes from
       the concept of \textit{self-fulfilling prophecy}, which states that
        humans with their expectations can influence and
         control some dynamics.

  \bigskip

\subsection*{Part II}

The study of fluctuations and collective
 phenomena in human-driven dynamics is one of the
  main goals of the fourth chapter, which describes
   a stochastic model, where the action of many
    interacting components is presented by random noise.
I study here a model-based analyses
 and stochastic properties of the volatility of the
  cumulative production based on the experience
   curve hypothesis. 
   
   \bigskip
   
    The concept of experience
    curve and its empirical evidence were presented
     in Wright’s \cite{Wright} seminal paper, in which
      he first discovered the relationship of experience and quantity of the products.
      Wright’s curve is known in the literature as a ``learning curve'',
 since it is based on ``the more learning by more producing
  hypothesis'', for describing the price-experience relationship. Otherwise said,
   the higher the experience in producing a specific
    product is, the lower its production costs are, when the inflation
     is factored out. More precisely, the learning curve states that the
      empirically reduction of cost follows a constant
       proportion rate, as the production duplicates.

\bigskip

The motivation of the analyses of this part comes from the fact that the experience curve
effect can be
 observed in any business, any industry, and 
 any cost element \cite{Handerson}. The learning curve is a way for measuring production efficiency, to forecast production costs and to predict the future prices. As a result, the study of the effect is important, for instance for reducing costs in the production process, which also can lead to lower prices of products in the market. Some researchers discuss its usefulness for forecasting and planning the deployment of industrial
 and technological activities \cite{Lafond}-\cite{Martino}. Especially in \cite{Lafond} by using historical data a method for making distributional forecasts is progressively developed.

\bigskip

Despite the wide variety of empirical evidence of the
 experience curves, there is a lack of theoretical
   framework of the concept.
 Motivated by this fact I present here the
  mathematical framework for describing a relationship
   between the volatility of cumulative production and
    the volatility of production (experience) itself. In analogy to the
concept of learning curves, which is a relation between the input and the
output of a learning process, one of the main findings shows the recursion relation
between previous and next probability distribution
 functions of the system, which characterize their volatility.
Knowing distribution functions of the cumulative
 production and its volatility can describe the
  marketing and movement of products. Also it will
   allow us to calculate various
     quantities, such as mean, variance,
      higher order moments and price volatility correlation.

      \bigskip

In the first part of investigations, by using the
 steepest descent method, for a model system
  the volatility of cumulative production for
   the narrow distribution of noise is calculated. 
   To explain this finding the stochastic properties of cumulative
     production are studied, by assuming that 
     production is a geometric random walk
   and empirically cumulative production 
growth follows a smooth exponential behavior in the presence
 of noise. The result is stunning and potentially powerful.
       Also it is tested for 51 products and technologies
        and the agreement is pretty good \cite{Lafond}.
         In the long time limit, the results show that
          cumulative production grows at the same rate
           as production, and the
            volatility of this growth rate is lower than
             the volatility of production. This is due to
              the fact that production tends to grow
               exponentially, so that cumulative production
                tends to grow exponentially with lower fluctuations. 
                The investigation of cumulative production is a key parameter
   here and by considering the experience curve as a time series
    model, a mathematical framework for the volatility of the
     time series is derived. The finding can lead to the prediction of 
     the volatility of cumulative production based on the volatility of experience
      or production. The work has been published in \cite{Lafond}.
                 \bigskip

In the further course of this chapter, the generalization
 of the above analysis is carried through: The first part,
  described above, addressed the effects of narrow,
   normally distributed noise in the production process.
    Due to its wide applicability in industrial and 
    technological activities and motivated by the
     fact that there exists a large number of cases
      where the distribution describing a underlying
       phenomenon is not Gaussian, (e.g.~the price
        fluctuations of most financial assets \cite{Bouchaud}),
         the mathematical foundation for an {\em arbitrary}
          distribution function of the process is also
           presented, which is expected to pave the
            future research on forecasting of the production process.
          As a consequence of the generalization
             of the method, the results yield a systematic
              control over the validity of the finding
               obtained for the narrow distributions discussed above.
                 More precisely, in the present work, 
                 a recursion relation of integral type that
                  replaces simulations by highly accurate
                   numerical integration is derived. 
                   The results show how different types
                    of noise affect the cumulative production
                     within the model, based on the learning
                      by doing hypothesis. The work has been published in \cite{Rubin}. Finally, higher order moments
                       such as skewness and kurtosis of the process will
                        be addressed as well.

  \bigskip

      The
        method of the work used for the production process, can be applied 
        also to different domains \cite{Qu}-\cite{Matsumoto},
         especially it can characterize the descriptor
          of a risk measure \cite{Vanduffel}, 
          which is one of the main challenges in finance.
          In fact, with the same approach and analogy, in the 
  present thesis, the computation of the capital from
   investment by depreciation, based on the
    perpetual inventory method in economy is presented. (Note that
    the perpetual inventory method is concerned with
     the estimation of the net capital from the cumulative
      capital stock, the analysis of annual depreciation, etc \cite{OECD}).

\subsection*{Part III}

The study of probability is considered in many
 domains and dates back to the seventeenth century. 
 One of the fields which is very closely related to it, 
 is the gambling and game theory \cite{Gibbson}: 
 from dice to risk benefit analysis, i.e. ``games of
  chance'' is composed of random events and 
  variables. Motivated by this fact, in the fifth
   chapter of this thesis I study Parrondo's games
    \cite{pa96}-\cite{ab10}, 
   which exhibit interesting phenomena at the intersection
of game theory, statistical finance and physics, see \cite{ab10} for
interdisciplinary applications.

\bigskip

Parrondo's paradox states that playing two losing games in a random or periodic order can result in a winning outcome  \cite{pa96}-\cite{so15}.
 Of course, the opposite situation is also
possible; a random combination of two winning games can give a losing game.
Parrondo invented a game-theoretic model of a Brownian
  flashing ratchet \cite{pr92}-\cite{ch14}, thus producing a discrete-time model of the ratchet effect \cite{pa96}. 
In the case of Brownian ratchets, a particle
moves in a potential, which randomly changes between two versions. For each
there is a detailed balance condition. However, for random switches between
the two potentials, there is on average a directed motion. This phenomenon is
fundamentally related to portfolio optimization \cite{so15}, and
corresponds to the ``volatility pumping'' strategy in portfolio
optimization. For a two-asset portfolio one half of the capital is kept in the
first asset, the other half in the second asset with high volatility
\cite{vo}.

\bigskip

In the present thesis, by using the Fourier transform, the exact
probability distribution functions for both the capital dependent and history
dependent Parrondo's games are calculated. In certain cases there are found strong
oscillations near the maximum of the probability distribution with two
limiting distributions for odd and even number of rounds of the game.

Parrondo's paradox is
 also used in financial risk management by revealing
  the statement, that losing strategies combined 
  and can turn into a winning strategy. So the
    investigation of Parrondo strategies is useful
     for instance for cases in which declining birth
      and harmful processes combine in a beneficial way.

  Especially intriguing is an anti-Parrondo effect or {\it
    Verschlimmbesserung} where reduced confidence in a measurement results
  from an increase in the number of observations that are in agreement
  \cite{gu16}. The Parrondo's games have been also applied
   in \cite{Boman,Almberg} as a toy model for studying the
    dynamics in the stock market and in \cite{Maslov}, it has been
     illustrated a Parrondo-like model for switching between
      poor performing investments. The stochastic and
       universal portfolio models \cite{Fernholz,cover} are
        also closely related to Parrondo's model. Later other modifications of the games were invented
such as the two-envelope problem and the Allison mixture
  \cite{ab14} where random mixing of two random sequences creates
  autocorrelation \cite{ad09a}. In the further course of this PhD work, the exact probability distribution functions of the two-envelope problem are also derived. Here, we will give an
  integral representation for the exact probability distribution of the
   model, then calculate both the mean capital growth rate and
    variance of the capital distribution after a large number of games. 
The work has been published in \cite{Rub}.

\bigskip

  For applications it is most important to find the capital
 growth rate and the variance of the distribution. This
  is the reason that the present thesis not only deals
   with a few characteristics of the model, but with
    the exact distribution function and its asymptotic 
    behavior. The existence of degenerate largest
     eigenvalues of the time evolution creates
      oscillations in the probability distribution 
      of the capital and results in the existence
       of two limiting distributions. This is a typical
        situation with real data of stock fluctuations
         in financial markets, and surprisingly it is explained 
         by a simple model in the present thesis.  The method of the
work can be applied to Brownian ratchets, molecular motors and portfolio
optimization. The work has been published in \cite{Zadourian}.
         
         \bigskip

I begin below with technical explanations of the above
 statements by exploring the realm of this 
 field by examples. In the further discussion I switch to ``we'', since I include
  here much material from the published work, which resulted
   from collaborations by being in different places such
    as Max-Planck Institute for the Physics of Complex
     Systems, Dresden, Germany  (Prof.~Holger Kantz); 
     Jülich Research Center
      Germany  (Prof.~Peter Grassberger); Mathematical Institute of University of 
      Oxford and Institute for New Economic Thinking at the
       Oxford Martin School, UK (Prof.~J.~Doyne Farmer); 
       Sapienza University of Rome, Italy (Prof.~Luciano Pietronero) and 
       Physics department of University of Wuppertal, 
       Germany (Prof.~Andreas Klümper).
        
\newpage
 
 \topskip0pt
\vspace*{\fill}
\begin{center}
\thispagestyle{empty}
\section*{\textbf{\LARGE{}{Part II}}}
\section*{\textbf{\huge{}{Review of some mathematical and numerical tools}}}
\end{center}
\vspace*{\fill}
\vspace*{10cm}
\newpage

\section{Review of some mathematical and numerical tools}

In the present thesis quantities based on probability
 distributions are of prime interest, which can be measured
  either numerically or calculated analytically. For 
  completeness we first review in detail several tools 
  of statistical dependencies between random variables
   used in this thesis. 
   
   The correlation tools were
    invented for measuring the degree of relationship
     between two random variables. In data analysis, 
     one of the important goals for using the correlation is to perform data based predictions. In case of correlation 
      between two random variables, the knowledge
       of one of them, by using adequate tools, 
       allows to predict the value of the other random
        variable. Here we discuss Pearson's and Spearman's
        correlation coefficients. 
        
         The powerful tools such as Fourier 
         transformation, convolution and the steepest descent 
         method, are described in detail as well. In this chapter
          we review also the 
        elements of information theory, such as mutual
         information, entropy, conditional mutual 
         information, etc.  For 
         the numerical part we thoroughly analyze two
          useful methods, namely the histogram and 
          the $K$-nearest neighbor statistics. In particular,
           in this thesis these statistical tools are utilized
            for estimating the mutual information, which in turn
             serves for analyzing the correlated structure of 
             financial time series. In the Appendix,
             one can find the programs for the analysis
              of mutual information, written in Matlab,
              considering the autoregressive time series. 
 
\subsection{Pearson's correlation coefficient}

The Pearson's correlation coefficient is a measure
 of linear dependency between two random 
 variables and for the random variables $X$ and $Y$ is given by the following expression: 

\begin{equation}
\rho_{X,Y}=\frac{E[XY]-E[X]E[Y]}{\sqrt{E[X^2]-[E[X]]^2}\sqrt{E[Y^2]-[E[Y]]^2}},
\end{equation}

where $E$ denotes the expectation value.

In the literature, the formula can be found often written as follows: 

 \begin{equation}
 \rho_{X,Y}=\frac{cov(X,Y)}{\sigma_{X}\sigma_{Y}},
 \end{equation}
 
 where $cov(X,Y)=E[XY]-E[X]E[Y]$ is the covariance between random variables and $\sigma_{X}$, $\sigma_{Y}$ are the standard deviations of $X$ and $Y$, respectively. So,
  the Pearson's correlation coefficient implies a normalized covariance 
 of the two random variables.

Pearson's correlation coefficient is mostly used for sample data statistics.
  One of the nice properties of Pearson's correlation
    coefficient is its invariant feature under a change
     of scale of the two random variables \cite{pear}.

Due to our interest in time series analysis of 
stock markets, it is useful to introduce also the autocorrelation, which
 is a function of time delay. This is one of the typical measures for 
 finding for instance repeating phenomena between observations with respect to time lag.

One can measure autocorrelation by considering the values of 
the process at different time domains and by 
using the above defined Pearson's correlation coefficient. Otherwise
 said, in case when we are dealing with the Pearson's correlation
  between values of the random process at different times, then 
  one speaks about autocorrelation of the random process and 
  between times $s$ and $t$ is defined as

\begin{equation}
\rho_{t,s}=\frac{E[(X_t-\mu_{t})(X_{s}-\mu_{s})]}{\sigma_{t}\sigma_{s}},
\end{equation}

where we supposed the process has mean $\mu_{t}$ and variance $\sigma_{t}^2$ at time $t$ and the mean $\mu_{s}$ and the variance $\sigma_{s}^2$ at the time $s$ respectively.

If we consider a stationary process, in which the mean $\mu$ and
 the variance ${\sigma}^2$  are time independent, then the autocorrelation
  between two different points
in the $X_{t}$ stationary process is defined as follows: 

\begin{equation}
\rho_{\tau}=\frac{E[(X_t-\mu)(X_{t+\tau}-\mu)]}{\sigma^2}\nonumber.
\end{equation}

Thus, for the stationary processes the 
autocorrelation is a function of the time 
difference between two points. 
 Trend stationary and autoregressive 
processes are typical examples of models with
 autocorrelation. The latter process will be discussed later in detail.

\subsection{Spearman's correlation coefficient}

Pearson's correlation coefficient can also be used
 by considering rank statistics of data. In this case
  the obtained correlation coefficient is called Spearman.
    Phrased in simple words, if we want to deal with the 
    Spearman's correlation coefficient we need to
     convert the scores to the ranked data and apply 
     the Pearson's correlation coefficient formula 
     to the set of ranks. 
     
     The Pearson's correlation
      shows the degree of linear relationship between
       two variables, whereas the Spearman's correlation
        is a good choice for measuring the consistency
         of the relationship, including nonlinear effects; also it is an
          appropriate measurement tool, for studying  
           highly skewed or kurtotic distributions.  
         
    Especially the Spearman's correlation gives a better result, 
         if the data has outliers, which is typical in financial time series, whereas
          the Pearson's correlation coefficient is not sensitive to the effect of outliers. 
          This statement comes from the fact that Spearman's coefficient limits
           the values of outliers to the value of their ranks.

 Spearman's coefficient is a measure of the monotonic relationship between two variables.  We like to note, that in
a monotonic case we have strictly increasing or decreasing, but not mixed relation. Every time when $X$ increases by the same increment of the amount of $Y$, then
   the correlation is perfectly consistent and the data points fit perfectly on a
    straight line. In a case of an increasingly negative monotonic relationship, 
             $X$ increases in rank when $Y$ decreases. (Note that the amount
              of increase (decrease) of the variables might not be the same).
       A positive (negative) Spearman coefficient indicates an increasing
      (decreasing) monotonic relationship between $X$ and $Y$.

      When two random variables are consistently 
          related to each other, their ranks will be linearly
           related to each other. Let us interpret this statement through an example. In a perfectly 
           consistent positive relationship, when the variable $X$ increases,
            the $Y$ variable is also increasing. That means the smallest value
             of $X$ corresponds to the smallest value of $Y$. We 
             observe the same for other values. This means that the rank of $X$ and $Y$ increase
              consistently. As a consequence, the ranks are linearly related to
              each other, i.e. we observe a perfect fit on a straight line.
               In other words, each increase in $X$ is followed by an increase in $Y$, 
               although the relationship can be non-linear. This example 
               illustrates that a consistent relationship between variables creates
                a linear relationship, when the values of the variables are converted to their ranks.
                 Based on this, the Pearson's correlation coefficient formula is used for 
                 measuring a linear relationship considering the ranked data.

                  Spearman's correlation coefficient is nonparametric, that means, it
does not make any assumption on the distribution of the sample data.
On the other hand when both variables are normally distributed the use of the Spearman's
correlation coefficient does not give any additional information.

              Based on rank statistics Spearman's correlation
             coefficient is known to
              be more robust and also
              insensitive to the difference between raw data. 
              These statements will be discussed more precisely 
               in the next chapter
              by revealing them through financial time series analysis.

For obtaining the Spearman's correlation coefficient, one can also transform the data $X$ and $Y$ to a uniform distribution and then calculate the Pearson's correlation coefficient between them.

There are two methods to calculate Spearman's 
correlation coefficient. As discussed above, it can
 be obtained with the same formula as Pearson's, 
 by converting the observations of stochastic variables $X$ 
 and $Y$ to their ranks $r=$rg${X}$ and $s=$rg${Y}$ in the sample as follows:

\begin{equation}
\rho_{r,s}=\frac{E[rs]-E[r]E[s]}{\sqrt{E[r^2]-E[r]^2}\sqrt{E[s^2]-E[s]^2}}.\nonumber
\end{equation}

Let us organize the data of $n$ observations of $X$ and $Y$ random variables. If the data does not have tied ranks (all ranks are unique)
  it is convenient to use the following expression:

\begin{equation}
\rho_{r,s}=1-\frac{6\sum_id_{i}^2}{n(n^2-1)},
\end{equation}
 
where $d_{i}=r_{i}-s_{i}$ is the difference between paired
 ranks. In this formula the Spearman's correlation
   coefficient is calculated by the ratio of the
    sum of the squared differences in the ranks of the paired
     data values, to the number of variables pairs. There are
      different ways to prove this formula, for example, by
       considering the Pearson's correlation coefficient formula 
       applied to a sample and by keeping in mind that $\rho$ is
        invariant under changes in location and scale of the
         variables. Note that in the case of tied ranks, formula (4) is not valid.

\subsection{Fourier transform}

As discussed above, one of the core and important investigations in this
 thesis is the study of probability distribution functions
  for different problems in economy. Knowing the
   distribution function of the underlying quantity
    and phenomenon, tells us a lot about the system under 
    consideration. One of the tools that frequently is used in probability theory and its application is the so called Fourier transformation.
    
 The Fourier transform of functions is known to be very useful for
  instance for solving linear problems with some kind of
   translational invariance and may serve as 
    characteristic function for stochastic distributions.

   The Fourier representation of a function $f:R \longrightarrow \mathcal{C}$ and 
   its Fourier transform $F(k)$ are given by:

\begin{equation}
f(x)=\frac{1}{2\pi}\int_{-\infty}^{\infty}F(k)e^{ikx}dk,
\end{equation}

\begin{equation}
F(k)=\int_{-\infty}^{\infty}f(x)e^{-ikx}dx.
\end{equation}

The integral in Eq.~(5) is also called the inverse
 Fourier transform. The approach has many applications
  in mathematical science,  physics and engineering: 
  for example the Fourier transform is widely used in
   the study of quantum mechanics, signal processing and
    wave propagation \cite{Ablowitz}. In this work, this powerful tool
    will be used for finding the exact probability distribution functions
     of important models in game theory and
       financial risk, the so called Parrondo's games,
        which we defer to chapter five.
 
In analogy to Eqs. (5) and (6), for the
 case of functions in $(-L,L)$ intervals with
  periodic boundary conditions  there exist the following expressions:

\begin{equation}
f(x)=\sum_{n=-\infty}^{\infty}F_{n}e^{\frac{in{\pi}x}{L}},
\end{equation}

% e^{\frac{{i}{n}{\pi}{x}}{L}

\begin{equation}
F_{n}=\frac{1}{2L}\int_{-L}^{L}f(x)e^{-\frac{in{\pi}x}{L}}dx,
\end{equation}

where ${F_{n}}$ are called the Fourier coefficients. 

The Fourier transform is also useful for the calculation of moments.
It is known that the moments of the distribution functions are given by

\begin{equation}
m_n=\int_{-\infty}^{\infty}x^{n}f(x)dx,\nonumber
\end{equation}

where the moments of $f(x)$ are derived through successive 
derivatives as follows:

\begin{equation}
m_n=(-i)^n\frac{d^n}{dk^n}F(k)\mid_{k=0}.
\end{equation}

Due to this
 equation the Fourier transform $F(k)$ is also known as the characteristic function. Since $f(x)$ is normalized, one always has  $F(0)=1$ 
and therefore we have the $m_0=1$ normalization condition.
From the theoretical point of view the knowledge
 of all moments is equivalent to the knowledge of
  the distribution function. 

Another important quantity is the cumulant,
 which is defined from the characteristic function as follows: 
 
\begin{equation}
c_n=(-i)^n\frac{d^n}{dk^n}\log{F(k)}\mid_{k=0}.
\end{equation}

Similarly to mean and variance, one of the nice
 features of cumulants is their additivity, i.e. for
  independent random variables the cumulant of the sum is
   given by the sum of the individual cumulants. 
   This stems from the fact that by multiplying the 
   characteristic functions, their logarithm add. The additivity
    of cumulants simply follows from the linearity of the derivative.
    
     Since the Gaussian distribution is ubiquitous it is  also worthwhile to
      mention that for this distribution all cumulants of order larger than two are zero.

\subsection{Convolution}

The Convolution is a mathematical tool, which
 maps two functions to a new function. The 
 convolution is the integral of the product of 
 the two functions after one is reversed and 
 shifted and is given by the following expression:

\begin{equation}
(f*g)(x)=\int_{-\infty}^{\infty}{f(\tau)g(x-\tau)d\tau}=\int_{-\infty}^{\infty}{f(x-\tau)g(\tau)d\tau}.
\end{equation}

The applications of convolution can be seen
 in different domains, such as probability,
  statistics, finance and for solving differential
   equations. Hence we are interested in its
    application to the calculation of the distribution
     function for the sum of independent random
      variables, let us discuss this through an example:
       If we denote the variation of the price of an asset
        between this month and the next month by
         $\Delta_{X}$ and similarly between the next
          subsequent months by $\Delta_{Y}$, then
           one may ask what the distribution of 
           the total variation of the asset price
           within the whole period could be. Let us
            suppose that these two variables are 
            distributed according to some independent
             distributions $P_{1}(\Delta_{X})$ and
              $P_{2}(\Delta_{Y})$. As a matter of fact,
               the distribution of the sum of these two
                random variables is given by the following
                 equation, which is simply the convolution 
                 of the respective distributions:

\begin{equation}
P(x)=P_{1}*P_{2}=\int{P_{1}(\tau)P_{2}(x-\tau)d\tau}.
\end{equation}

One can generalize the concept to $n$ 
 independent random variables and get the following expression:

\begin{equation}
P(x)=\int_{-\infty}^{\infty}{P_{1}(\tau_{1})P_{2}(\tau_{2})...P_{n}(\tau_{n})\delta \left(x-\sum_{j=1}^{n}\tau_{j}\right)d\tau_{1}d\tau_{2}...d\tau_{n}},
\end{equation}

which simply owes to the characteristics
 of independently distributed random variables.
  The special case, Eq. (12) arises for $n=2$ and by integrating out $\tau_{2}$.
  
 It is worthwhile to mention the 
convolution theorem, which states that the Fourier 
transform of a convolution is the product of Fourier
 transforms. The statement is true for various Fourier-related transforms and can be expressed as: 
 
\begin{equation}
{F}{(f*g)}={F}(f)\cdot{F}(g),\nonumber
\end{equation}

where ${F}(f)$ and ${F}(g)$ are the Fourier transforms of 
 $f$ and $g$ functions respectively.

The transition of the convolution to the pointwise product is
 very useful, especially for performing numerical analysis; while 
 for the convolution one should tackle with $\mathcal{O}(n^2)$ 
 orders, this formula supports to reduce the quadratic order to
  $\mathcal{O}(n\log{n})$, which helps  to implement an algorithm faster.

Finally, in the further course of the thesis, this valuable tool
will be used for finding the relationship between the cumulative
production and the production itself (in a presence of noise), by analyzing their
probability distribution functions and volatilities. 

\newpage
   
\subsection{Steepest descent or saddle point method}

The saddle point method described, e.g. in \cite{Ablowitz, Elvezio}
 presents an approximation formula for any
  {\em narrow} probability distribution function.
The essence of the saddle point method is 
to approximate the integral by taking into 
account only that portion of the range of the 
integration where the integrand takes large values. 

The method is a powerful approach for considering
the large $k$ asymptotic of integrals of the form:

\begin{equation}
I(k)=\int_{C}f(z)e^{k\phi(z)}dz,
\end{equation}

where $f(z)$ and $\phi(z) \in \mathcal{C}$
 are analytical functions of $z$.

The idea of the approach is based on the application
 of the analyticity of the integrand by changing 
 the contour C to a new contour $C'$ where
  $\phi(z)$ has an imaginary part, which is a constant. 
  That means the Eq. (14) becomes: 

\begin{equation}
I(k)=e^{ikv}\int_{C'}f(z)e^{ku(z)}dz,
\end{equation}

where $\phi(z)=u+iv$.

Considering the fact that $\phi(z)$ is analytic 
and $v$ is constant on $C'$, the derivative
 of $u$ perpendicular to $C'$ is also zero.
  Therefore along this path $C'$, the increase 
  of $u$ is maximal (path of steepest ascent) or the 
  decrease of $u$ is maximal (path of steepest descent). For the
  calculation of the integral we utilize the latter one,
   therefore the method is called steepest descent. The
    path of the steepest descent includes a point $z^*$ for 
    which $\phi'(z^*)=0$. Such a point is called
     saddle point and therefore the method is often 
     called the saddle point method.

The idea is to deform the contour into the steepest
 descent curve, which includes the saddle point and the contribution
  near the saddle point is the dominant one. If one deforms the integration contour, the
value of integral will remain unchanged and the contour
levels of the real part of $\phi(z)$ are everywhere orthogonal
to the contour line of the imaginary part of the $\phi(z)$ function. 
This statement ensures that the imaginary part of $\phi(z)$ is constant.
 More precisely (by using the above notations), 
we should deform the contour of integration $C$ into
 a new integration path  $C'$, so
 that the new integration path passes through zero(s) of   
  $\phi'(z)$ and that the imaginary part of the $\phi(z)$ is 
  constant. 
    
But this is not always the case and sometimes the contour cannot be  deformed into
    a curve including a saddle point, e.g. if one encounters
    singularities of the integrand. In practice, the steepest
     descent method requires the identification of critical points $\phi'(z^*)=0$
      close to the original curve $C$ and to control the occurrence of singularities
       between $C$ and the steepest descent curve $C'$.  
       
       For large $k$ we have the following result:
       
       \begin{equation}
       I(k)\simeq\sqrt{\frac{2\pi}{-k\phi''(z^*)}}f(z^*)e^{k\phi(z^*)}.\nonumber
       \end{equation}

In general, the saddle point of a multivariate function
 $S=S(z_{1}, z_{2}..., z_{t})$ is defined by the system of equations $\partial_i
S(z^*)=0, \; \partial_i =\partial/\partial_{z_i}, \;$ $i \in \{1,...t\}$ for
which one can write
\begin{equation}
\label{2}
k\phi(z)\equiv{S(z)}=S(z^*)+ \sum_{ij}(z_i-z^*_i)(z_j-z^*_j) G_{ij}
+{\cal O}(\{(z-z^*)^3\}),
\end{equation}
where $z^*$ is the solution of the saddle point equations and
$G_{ij}=\frac{1}{2}\partial_i\partial_j S(z)\vert_{z=z^*}$.

For the further analysis of the method we 
need to analyze the function around the saddle
 point, which is a dominant point. 

The point $z^*$ is called saddle point of 
 order $N$ if the first $N$ derivatives disappear:

\begin{equation}
\frac{d^m{\phi}}{dz^m}\vert_{z=z^*}=0,\quad     m=1,..N, \quad  \frac{d^{N+1}{\phi}}{dz^{N+1}}\vert_{z=z^*}\ne0.
\end{equation}

  The above statements will be more clear for the
   reader by following the calculations in the fourth chapter
    for two analytical expressions.
    This powerful method will be used here for the analysis of
 stochastic properties of cumulative production in a presence of 
  narrow distribution of noise. The investigation leads to 
   an important
 result in the production process, which we defer to chapter 4.
  In the fifth chapter the method is also employed for
   obtaining elementary and nice expressions in the
    large time limit, by considering model-based analysis.

\newpage

\subsection{Elements and quantities of information theory}

Information theory sheds light on many problems
that have been troubling communication engineers
and scientists for years. It provides a universal
tool for measuring the amount of information 
based on uncertainty and knowledge of the 
underlying system. The words communication
and information are important for all of us and
in many domains the tools of information theory
can be used. Indeed, it has a broad scope in
areas, such as engineering, statistical physics, mathematics and economy. 

The background of information theory dates
 back to the early 1940, when the researchers
  thought it is impossible to send information 
  with negligible probability of error. Shannon 
  showed indeed how it is possible to make the 
  probability of error to come to zero for all rates of information
  below the channel capacity. 
  
  The channel 
  capacity is one of the fundamental concepts 
  of information theory. It is exactly
   the maximum amount at which one can share 
   the information in the channel and have the 
   corresponding information at the output with
    a low probability of error.
   If we define the input random variable with $X$
  and the output $Y,$ then the capacity $C$ is determined:
  
  \begin{equation}
C=\max{I{(X,Y)}},
\end{equation}

where $I(X,Y)$ is the mutual information between these
 random variables, which we will comment on in detail later.
  
According to the above discussion, that means for any information rate $R<C$, it is always
 possible to transmit information with small error. Contrariwise,
  for any information rate larger than the channel capacity it is
   impossible to transmit the information with small error \cite{mut}.
  Mobile phones could be a simple example of a noisy channel for describing
 the transmission process and while there is noise, it is possible to maximize the rate of information through the channel.
 
  In information theory, the channel coding
    is exactly aimed to find such codes that can be utilized to transmit
     data over a noisy channel with a small coding error. 
      In order to describe the
transmission and data comparison, the quantities
such as entropy, relative entropy, mutual information and
conditional mutual information were introduced.

The basic and fundamental quantities of information
 theory are constructed in terms of probability 
 distributions that underlie the process. Thus, 
 from the mathematical point of view, information
  theory is closely tied with statistical mechanics and
   probability theory \cite{Cover}.

Furthermore, since the elements of information theory measure a relationship and dependency between
  variables, they are appropriate tool for describing
   correlations between different quantities in different
    domains such as statistics, automation, physics, 
    biology, economy and finance \cite{mut}. In this 
thesis the application of this elegant theory appears
  in the latter one, namely
  for the study of the
  dependent structure of financial
 data. Let us conclude this initial discussion about
  the information theory by recalling Shannon's statement from
   his ``The Bandwagon'' letter \cite{band}, in which he wrote the importance of information theory
   as follows:  ``In short, information theory is currently
    partaking of a somewhat of a heady draught of general popularity''.

In the further course, we review some important quantities and elements of the information theory.

\subsubsection{Entropy}

 Entropy and relative entropy are devised to describe
 the uncertainty or disorder in the system. Entropy
  has properties that satisfy the intuitive notion of 
  measuring the information rate, thus it is sometimes
   called the self-information of a random variable.

Let us introduce the mathematical framework of the
  entropy, by focusing on discrete random variables,
 since in the thesis the application will be to real valued time series. According to Shannon the
  entropy of a discrete random variable $X$ with a 
  probability
   function $p(x)$ is defined by: 

\begin{equation}
H(X)=-\sum_{x}{p(x)\,{\log}\,p(x)}.
\end{equation}

From the above definition, it is clear that the 
entropy is a non-linear functional of the probability distribution
 of the random variable. The entropy is non-negative and is measured in bits or nats (the base of the logarithm
defines the unit of measurement). As it is clear from its notion, if there is no uncertainty
    of the outcome of the event, then the entropy is
     simply zero. 
     
     If we consider that the discrete random variable $X$ has $\{x_{1}, x_{2},...,x_{n}\}$ values, then the upper bound of the entropy is $H(X)\le\log(n)$.

 Furthermore, the entropy is characterized by the following properties:
 
 \begin{itemize}
 
\item Continuity: The amount of change in probabilities
is proportional to the amount of change in the entropy.

\item  Symmetry: The outcome is invariant under a permutation of the $x_i$'s.

\item Maximum: The entropy is maximum, if all events occur with the same probability.

\item  Additivity: For independent subsystems (or if the interactions between them are known) the entropy of a whole system is characterized
  by the sums of the entropies of all subsystems \cite{ent}.
  
\end{itemize}

Let us now extend the definition of entropy of a
 single variable to a pair of random variables. The
  joint entropy $H(X,Y)$ of a pair of discrete random
   variables $X$ and $Y$ is defined as follows:

\begin{equation}
H(X,Y)=-\sum_{x}\sum_{y}p(x,y)\,{\log}\,p(x,y),
\end{equation}
where $p(x,y)$ is the joint probability distribution
 function of the pair of random variables. Note that for simplicity we use everywhere the following sloppy notation for the marginal probabilities: $p_{x}(x)\to p(x)$.
 
The relationship between joint and marginal probability 
 distributions is defined by: $p(x)=\sum_{y}p(x,y)$
  and the conditional probability $p(x|y)$ is expressed by the joint 
  and marginal probabilities as follows: $p(x,y)=p(x|y)p(y)$. 

Another important quantity is the conditional entropy
 $H (X{\mid}Y)$, which is the entropy of a random
  variable conditioned on the knowledge of another
   variable and is given by the following relation: 

\begin{equation}
H(X{\mid}Y)=-\sum_{x}\sum_{y}p(x,y)\,{\log}\,p(x{\mid}y).
\end{equation}

Finally the relationship between joint and conditional
 entropy is given by the following expression:

\begin{equation}
H (X, Y ) = H (Y) + H (X|Y).
\end{equation}

Proof:

\begin{align}
H(X,Y)&=-\sum_{x}\sum_{y}p(x,y)\,{\log}\,p(x,y)\nonumber\\
&=-\sum_{x}\sum_{y}p(x,y)\,{\log}\,\left[p(x|y)\,p(y)\right]\nonumber\\
&=-\sum_{x}\sum_{y}p(x,y)\,{\log}\,p(y)
-\sum_{x}\sum_{y}p(x,y)\,{\log}\,p(x|y).\nonumber
\end{align}

Using the relationship between joint and marginal 
distribution discussed above and also from Eq.~(21) we obtain

\begin{equation}
H(X,Y)=-\sum_{y}p(y)\,{\log}\,p(y)-\sum_{x}\sum_{y}p(x,y)\,{\log}\,p(x|y)=  H (Y) + H (X |Y).\nonumber
\end{equation}

\subsubsection{Mutual Information, relative entropy and conditional mutual information}

One of the reasons for the intensive investigations
 of mutual information (MI) is its theoretical background
  \cite{Cover}. Furthermore, as mentioned above, 
  mutual information in contrast to the linear correlation
   coefficient is efficient for studying dependencies 
   that include non-linearity and do not exhibit 
   themselves in the covariance.

The mutual information shows the amount of 
information that one random variable contains 
about another. More precisely, mutual information 
is the reduction of the uncertainty of one random
 variable due to the information content of another random variable.

Let us now define the MI's mathematical framework,
 which is based on the notion of entropy and is 
 given by the following expression:

\begin{equation}
I(X,Y)=\sum_{x}\sum_{y}p(x,y)\log{\frac{p(x,y)}{p(x)p(y)}}.
\end{equation}

From the above definition, it is obvious that the 
mutual information is symmetric in $X$ and $Y$, i.e. $I(X,Y)=I(Y,X)$ and due to Jensen's inequality it
 is always non-negative. From the mathematical and 
 intuitive notion of the concept, it is clear that the mutual
  information is only zero if the two random variables
   are independent. The latter statement is also true 
   for quantities based on Renyi entropies \cite{Ren}. 
   However mutual information is particular in its
    notion and mathematical background. 
    
For obtaining a better understanding
  of the mutual information
   let us express it by entropies. We can rewrite equation (23) as
   
\begin{equation}
I(X,Y)=\sum_{x}\sum_{y}p(x,y)\log{\frac{p(x|y)}{p(x)}}\\
=\sum_{x}\sum_{y}p(x,y)\log{p(x|y)}-\sum_{x}\sum_{y}p(x,y)\log{p(x)}.\\\nonumber
\end{equation}

By use of Eqs.~(19), (21) and the relationship between marginal and joint distributions we obtain  

\begin{equation}
I(X,Y)=H(X)-H(X|Y).\nonumber
\end{equation} 

This relation between mutual information
 and the entropies of the variables explicitly
 shows that the mutual information 
 describes the reduction of uncertainties due to the
 knowledge of other variables, in this case $Y$.

We discussed above that in case of 
two independent random variables the mutual information
 is zero. Indeed, Eq. (23) shows this statement,
  since in this case we can rewrite the joint entropy 
  as $p(x,y)=p(x)\cdotp(y)$ and we get a $\log{1}=0$ identity. This statement also follows 
  directly from the intuitive notion and definition of mutual
   information, which quantifies the amount of information 
   that two random variables share among each other. Given this, it is clear that
    if the variables are independent, then information 
    about one of them does not give any knowledge about
     the other. On the other hand if $X$ and $Y$ are 
     related to each other, as a deterministic function,
      there is a share of information between them.
       In the extreme case, where $X$=$Y$ we have
 
 \begin{equation}
 I(X,X)=H(X)-H(X|X)=H(X).\nonumber
 \end{equation}
 
 This equation tells us that the mutual information of a random 
 variable with itself is simply equal to the entropy of the variable.
 
Mutual information appears to be a special case of
a more general quantity called relative entropy or
Kullback–Leibler entropy, which is a measure of the
difference or ``distance'' between two probability
distributions. If we denote the probabilities by $q$
and $p$, the Kullback–Leibler entropy is defined as follows: 

\begin{equation}
D(p{\mid}{\mid}q)=\sum_{x}p(x)\log\frac{p(x)}{q(x)}.
\end{equation}

Like the mutual information it is always non-negative 
and diverges if and only if $p=q$.

It is worthwhile also to define the conditional mutual information,
which describes a reduction of uncertainty in $X$ based on
the knowledge of $Y$ when $Z$ is given and is defined as follows:

\begin{equation}
I(X;Y{\mid}Z)=\sum_{z}p(z)\sum_{x}\sum_{y}p(x,y{\mid}z)\log\frac{p(x,y{\mid}z)}{p(x{\mid}z)p(y{\mid}z)}.
\end{equation}

As a matter of fact, the conditional mutual information
 is the expectation value of mutual information of two 
 random variables conditioned on the third and for discrete random variables $X, Y$ and $Z$ it is always non-negative.

Similar to the mutual information we can express the conditional mutual information through entropies:

\begin{equation}
I(X;Y{\mid}Z)=H(X{\mid}Z)-H(X{\mid}Y, Z).\nonumber
\end{equation}

As a result the relationship between mutual and conditional mutual information is given by 

\begin{equation}
I(X;Y{\mid}Z)=I(X;Y,Z)-I(X,Z).\nonumber
\end{equation}

\newpage

\subsection{Histogram method}

The histogram or binning method is a statistical technique,
 which proposes a multivariate histogram for the estimation
  of the probabilities and was first introduced by Pearson \cite{Pea}.
  The approach has a long history and 
  it is the most common and widespread method for the 
  estimation of the density of probability function.

For the construction of a histogram the sample space is divided
 into a number of bins. The idea of the method is based on 
 counting the points in different bins. Let us denote by
  $n_{x}(i)$ the number of points of variable $X$ falling 
  into the $i$th bin and consequently $n_{y}(j)$ shows the
   number of points of variable $Y$ falling into the $j$th bin. 
   Furthermore, $n_{x,y}(i,j)$ indicates the number of points 
   that fall into the bin $i$th as well as into the bin $j$th; these points result 
   from the intersection and joint realization of the data points. The idea 
   is to approximate the density of the probability function
    by the fraction of points, that fall into the corresponding bin.

Let us consider $X$ and $Y$ random variables, with the marginal
  $p_{x}=\int{dy{p(x,y)}}$ and $p_{y}=\int{dx{p(x,y)}}$ density
   functions and use the histogram approach for the estimation of mutual information. For the finite sum equation (23) approximately can be written:

\begin{equation}
I(X,Y)\approx I_{binned}(X,Y)\equiv \sum_{ij}{p(i,j)\log{\frac{p_{x,y}(i,j)}{p_{x}(i)p_{y}(j)}}},
\end{equation}

where $p_{x}(i)=\int_{i}dx\,p(x)$, $p_{y}(j)=\int_{j}dy\,p(y)$ and $p(i,j)=\int_{i}\int_{j}dxdy\,p(x,y)$. (We closely follow the notation used in \cite{Grassberger}).

We consider $N$ bivariate measurements, which are assumed to be independent identically distributed (iid) realizations of a random variable. After normalization, for the marginal and joint probabilities
we have: $p_{x}(i)\approx n_{x}(i)/N$, $p_{y}(j)\approx n_{y}(j)/N$ and $p(i,j)\approx n_{x,y}(i,j)/N$. In the limit $N\longrightarrow \infty$ indeed from the right hand side of Eq. (26) we get the MI for $X$ and $Y$ variables. 

$I(X,Y)$ in (26) is weakly dependent on the binning if the distribution of $(X,Y)$ is rather two dimensional and continuous, i.e. if it is well approximated by a continuous $p(x,y)$. In the other extreme case, where $X=Y$, then the mutual information is binning dependent. In 2.6.2 we discussed that for this extreme case the mutual information is equal to the entropy of the variable: $I_{binned}=-\sum_{i}{p_{i}\log{p_{i}}}$. If we consider that $X$ has a uniform distribution in $[0,1]$, M bins and $p_{i}=\frac{1}{M}$, then the mutual information $I(X,X)=\ln{M}$ is a function of the bin size, e.g. a coarse graining from $M$ to $M/2$ bins does not leave the mutual information invariant, it reduces its value by $\log{2}$.

There is no need to take a similar size for all bins, rather
 some estimators \cite{Fraser, Vajda} are flexible with
  regard to the bin size. Although such estimators are 
  better than estimators with fixed bin size, there are still 
  systematic errors by approximating the 
  probabilities with the frequency ratios and also by 
  discretizing mutual information. Nevertheless,
   this error could be minimized \cite{Grass} for the
    finite size corrections.

While the histogram method is easy to grasp, it has
several disadvantages: It is discrete and changes with
the choice of the initial parameters and bin width.
Even when using the same bin size, different initial 
conditions may change the histogram completely, e.g.
the discontinuities of the estimation are an 
artifact of the chosen bin locations. Otherwise said, we often observe bins with large filling 
and on the other hand bins that remain empty. Another problem
is the dimensionality, because the number of bins 
grows exponentially with the number of dimensions.
Due to these artifacts the histogram method is 
      not appropriate for the advanced level of 
      analyses and therefore we study
       another method called $K$-nearest neighbor statistics, 
       which we will comment on in the next section.

There are also other estimators, such as kernel density
 \cite{Moon, Steuer}, which are related to histograms,
  but they have properties such as continuity by a suitable
   choice of the kernel. Furthermore, the kernel density estimators
    are particularly advantageous when the data set is small.

\subsection{$K$-nearest neighbor statistics}
 
Here we discuss the $K$-nearest neighbor  statistics (KNN)
 for the estimation of mutual information presented 
 in \cite{Grassberger}, which in contrast to the above 
 described histogram method is data efficient, adaptive 
 and has minimal bias. The approach is non-parametric, 
 that means it does not make any assumption on the
  distribution of the underlying data. 

Let us consider metrics on the spaces underlying 
 $X$, $Y$ and $Z= (X,Y)$. The distance between the
  fixed point and its neighbors  is denoted by $d_{i,j}=\parallel{z_{i}-z_{j}\parallel}$ 
  and the maximum norm for the space $Z$ is

\begin{equation}
\parallel{z_{i}-z_{j}}\parallel=\max{\{\parallel{x_{i}-x_{j}}\parallel,\parallel{y_{i}-y_{j}}\parallel\}},
\end{equation}

where $\vert\vert.\vert\vert$ denotes the norms in the spaces underlying $X$ and $Y$ and they can be chosen differently. 

We denote the distance from $z_{i}$ to its $K$th neighbor
 by $\frac{\epsilon(i)}{2}$, where $K$
  is a positive integer. Similarly 
     $\frac{\epsilon_{x}(i)}{2}$ and
      $\frac{\epsilon_{y}(i)}{2}$ are
       the distances between the same points 
projected into $X$ and $Y$ sub-spaces. Obviously $\epsilon(i)=\max\{\epsilon_{x}(i), \epsilon_{y}(i) \}$. 

The value of the $K$ parameter is significantly important for the performance of the statistics. The notion of {\em near} is general 
  and the best choice depends
   on the data and the underlying problem. Typically the value of $K$ is
   chosen empirically and the parameter with the best
    result and accuracy is picked up. 
     If $K = 1$, then one speaks about a
     single nearest neighbor. 
     
For the estimation we follow two algorithms: In the
 first algorithm we consider the number $n_{x}(i)$
  of points $x_{j}$ that have a distance to  $x_{i}$ strictly
   less than $\frac{\epsilon(i)}{2}$ and consequently
    we follow the same procedure for $y$.

The formula, that does not require the investigation
 of probability distribution for the estimation of 
 mutual information is

\begin{equation}
I(X,Y)=\psi(K)-\langle\psi(n_{x}+1)+\psi(n_{y}+1)\rangle+\psi(N),
\end{equation}
where $\psi$ is the digamma
 function and $N$ is the number 
 of points in the entire sample.

In the second algorithm (which we use more
 frequently) we replace $n_{x}$ and $n_{y}$
  by the number of points in the respective subspaces:

\begin{equation}
\parallel{x_{i}-x_{j}}\parallel\leqslant\epsilon_{x}(i)/2
\end{equation}
and 

\begin{equation}
\parallel{y_{i}-y_{j}}\parallel\leqslant\epsilon_{y}(i)/2
\end{equation}

Finally, the second estimator of MI is

\begin{equation}
I(X,Y)=\psi(K)-1/K-\langle\psi(n_{x})+\psi(n_{y})\rangle+\psi(N).
\end{equation}

Both algorithms work pretty well and give very similar results.

\subsection{Analytical example of mutual information}

Mutual information can be obtained analytically if we
 consider normally distributed random
  variables $X$ and $Y$ with $0$ mean, $\sigma_{x}^2$ and $\sigma_{y}^2\equiv a^2\sigma_{x}^2$
    variances and covariance equal to $\sigma_{x}^2a^2$, with joint probability $p(x,y)$ and
    correspondingly marginal probabilities $p(x)$ and $p(y)$. Conditional probability of $x$ conditioned on $y$ is a Gaussian distribution with the mean value $y$ and the variance $\sigma_{\eta}^2$.
     In this case there exists a fully analytical expression \cite{Vajda}, 
     which is simple and potentially powerful and is equal to

\begin{equation}
I(X,Y)=-\frac{1}{2}\ln(1-a^2),
\end{equation}

where in a limit $a\to1$ we have $I(X,Y)\to\infty$. Let us derive it in detail by recalling the mutual information formula:

\begin{equation}
I(x,y)=\int_{R^2}{p(x|y)p(y)\log{\frac{p(x|y){p(y)}}{p(y)p(x)}dxdy}},
\end{equation}

where the integration is over the entire plane.

For the distribution functions we have the following expressions

\begin{equation}
p(x)=\frac{1}{\sqrt{2{\pi}}{\sigma_{x}}}e^{-\frac{x^2}{2{\sigma_{x}}^2}},
\end{equation}

\begin{equation}
 p(y)=\frac{1}{\sqrt{2{\pi}}{{a}\sigma_{x}}}e^{-\frac{y^2}{2{\sigma_{x}}^2{a^2}}},
 \end{equation}
 
and 
 
 \begin{equation}
 p(x|y)=\frac{1}{\sqrt{2{\pi}}{\sigma_{\eta}}}e^{-\frac{(x-y)^2}{2{\sigma_{\eta}}^2}}.
 \end{equation}
 
For consistency, we need
 
 \begin{equation}
\sigma_{x}=\frac{\sigma_{\eta}}{\sqrt{1-a^2}}.
\end{equation}

For simplicity we replace $\sigma_{\eta}=1$ \footnote
{We like to note that the final result will be literally valid
 for any $\sigma_{\eta}$; in fact it is independent 
 of $\sigma_{\eta}$.}  in Eqs. (34-36) and obtain the following: 

\begin{multline}
I(x,y)=\frac{-\sqrt{1-a^2}{\log(\sqrt{1-a^2})}}{2{\pi}{a}}\int{e^{-\frac{x^2}{2}+xy-\frac{y^2}{2a^2}}}dxdy\\+\frac{\sqrt{1-a^2}}{2{\pi}{a}}
\int{e^{-\frac{x^2}{2}+xy-\frac{y^2}{2a^2}}}\left(\frac{-x^2a^2}{2}+xy-\frac{y^2}{2}\right)dxdy.
\end{multline}

Let us consider the components separately: Thus we have 

\begin{equation}
I_1=\frac{-\sqrt{1-a^2}{\log(\sqrt{1-a^2})}}{2{\pi}{a}}\int{e^{-\frac{x^2}{2}+xy-\frac{y^2}{2a^2}}}dxdy
\end{equation}

and 
 
\begin{equation}
I_2=\frac{\sqrt{1-a^2}}{2{\pi}{a}}\int{e^{-\frac{x^2}{2}+xy-\frac{y^2}{2a^2}}}\left(\frac{-x^2a^2}{2}+xy-\frac{y^2}{2}\right)dxdy
\end{equation}

In the further calculations we use the following three well known formulas: 

\begin{equation}
\int_{-\infty}^{\infty}{e^{-x^2}dx}=\sqrt{\pi},
\end{equation}
   
\begin{equation}  
\int_{-\infty}^{\infty}{{x}e^{-x^2}dx}=0, 
\end{equation}

\begin{equation}  
\int_{-\infty}^{\infty}{dx{x^2}e^{-x^2}}=\frac{1}{2}\int_{-\infty}^{\infty}{dxe^{-x^2}}=\frac{\sqrt{\pi}}{2}.
\end{equation}

By making substitutions ($\zeta=y\sqrt{\frac{1-a^2}{2a^2}}$ and $\xi=\frac{x-y}{\sqrt{2}}$), which diagonalize the exponents and by using the complete
 square technique, for the $I_1$ we get the following expression:

\begin{equation}
I_{1}=-\frac{1}{2}\ln(1-a^2).
\end{equation}

Following the same procedure for the second part 
of the integral we get $I_2=0$. Thus the final expression
 for the mutual information is

\begin{equation}
I(X,Y)=-\frac{1}{2}\ln(1-a^2).
\end{equation}

\subsection{Numerical example for mutual information analysis}

In this subsection we demonstrate an example for the 
numerical analysis of mutual information, using
 the above discussed statistical tools, by considering
  the autoregressive and linear time series.

In statistics, autoregressive models AR(p) describe
time series, in which the output variable depends
linearly on its own previous values (back up to $p$ time steps).
Together with the moving average (MA) model it describes 
more general types of time series, called ARMA and ARIMA.
The Autoregressive model of order $p$ is given by the following formula:

\begin{equation}
X_{t}=c+\sum_{i=1}^{p}a_{i}{X}_{t-i}+\epsilon_{t},
\end{equation}

where $a_{1}, a_{2},..., a_{p}$ are
the parameters of the model and $c$ is a constant. 
$\epsilon_{t}$ is white noise with zero mean and a 
constant variance ${\sigma}^2$.
 
Let us consider the simplest model, namely AR(1):

\begin{equation}
X_{t}=c+a{X}_{t-1}+\epsilon_{t}.
\end{equation}

If we consider $|a|<1$ condition, then the process is stationary. 
That means the mean of the process remains the same $\mu=E(X_{t})=E(X_{t-1})$ hence we get: 

\begin{equation}
\mu=\frac{c}{1-a}.\nonumber
\end{equation}

\begin{figure}[h!]
\vspace*{-2.5cm}
\begin{centering}
\includegraphics[scale=0.5]{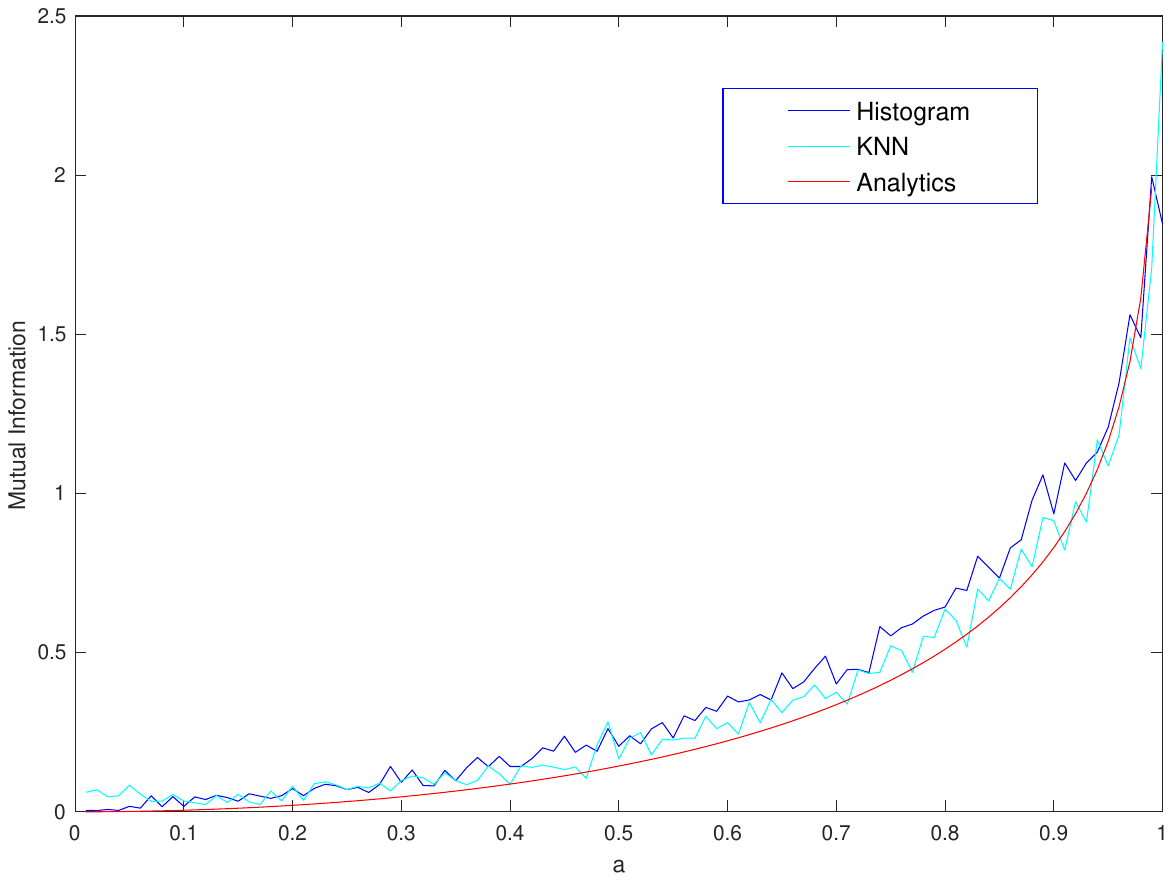} 
\vspace*{-3cm}
\par
\end{centering}
\protect\protect\protect\caption{The plot shows the 
 mutual information analysis for the AR(1) and the linear time
 series based on the past values of the first time 
 series, by including a Gaussian noise. The variable
  $a$ is taken from the interval $[0,1]$, for $a=1$ the
   mutual information diverges. The
  parameters chosen for the algorithms are: $N=1000$
        (length of data set), $K=5$ and the bin width is
       equal to 4.93. The figures show that the KNN
       algorithm coincides better with the analytical 
         curves for the larger $a$'s, which shows  more efficiency
    of the KNN algorithm.}
    \end{figure}

For the numerical analysis we consider the case 
where $c=0$ and according to the above equation for this case the mean of the process
 is zero and the variance is equal to

\begin{equation}
\hbox{Var}(X_{t})=\frac{\sigma_{\epsilon}^2}{1-a^2},
\end{equation}

where $\sigma_{\epsilon}$ is the standard deviation of $\epsilon_{t}$.

As second time series we consider a simple linear
 model, related to the first time series as follows: $Y_{t}=aX_{t-1}$.
 
 Finally we estimate the mutual information
 between them by using both histogram and 
 KNN methods, described in detail above. 
Since rescaling by any factor $a$ is irrelevant
 for the mutual information, the computed quantity is the time delayed mutual
information of the AR(1) process for time lag $1$.

Figure 1 illustrates the result and relationship between 
  mutual information and the $a$ parameter, 
  where $a$ is taken from the interval $[0,1]$. 
  For $a=1$ mutual information diverges, where the 
  result qualitatively coincides with the analytical result for
   the mutual information for the Gaussian case.
   
The red line in Figure 1 shows the
     result for the analytical expression of mutual information,
      namely formula (45), which we derived for the Gaussian
       distribution. The cyan line illustrates the result by
        using the KNN algorithm and the blue line shows the 
        mutual information analysis by considering the histogram
         method. We see there are discrepancies with the analytically
          obtained curve, which 
         may be caused
           by approximations of the numerical algorithms.
                
Finally, two programs, based on histogram and
     KNN statistical methods for the estimation of 
     mutual information can be found in the Appendix.

\newpage
 
 \topskip0pt
\vspace*{\fill}
\thispagestyle{empty}
\begin{center}
\section*{\textbf{\Large{}{Part III}}}
\section*{\textbf{\huge{}{Empirical analysis of financial time series}}}
\end{center}
\vspace*{\fill}
\vspace*{10cm}
\newpage

\section{Empirical analysis of financial time series}

Understanding the volatile feature of financial markets
 and their complex behavior intrigues many researchers. 
 As a consequence, different stylized facts and models 
 for investigation of financial time series are devised.
Financial time series analysis is a broad field, which includes
 the theoretical and empirical studies of assets valuation 
 over time. 
 
 The focus of this part is on financial volatility,
  which plays an important role for the risk management 
  and portfolio optimization. More precisely, the risks are measured by the variance
  of the asset returns and the square root of the variance
   is volatility. Volatility is simply the degree of variations 
   and fluctuations of a financial time series, i.e. it shows 
   the amount of uncertainty in the change of the price of 
   the asset's value. In particular, volatility is interpreted 
   as the statistical measure of the distribution of returns
    for a given financial asset per market index.

 It is clear that the price range can increase (decrease)
 impressively over a short time period in both directions. 
 The lower volatility simply indicates the smaller fluctuations
  in financial assets and in analogy the volatility of an asset
   is termed to be high, if the prices fluctuate rapidly \cite{F}. 
 
 The most popular techniques for modeling the volatility
  are the generalized autoregressive conditional
   heteroskedasticity (GARCH) models invented in
    \cite{Rob, Bollerslev}. The financial return volatilities
     can be explained well with aforementioned models and their family. 

If the market is hectic, volatilities are large, and usually
 it will take some time until the market has become calm
  again. This fact is related to the volatility clustering 
  phenomenon. According to Mandelbrot \cite{Mandelbrot},
   large changes tend to be followed by large changes, of 
   either sign, and small changes tend to be followed by 
   small changes. The fact of volatility clustering as an 
   aggregation has inspired researchers and has a major
    impact on the development of stochastic processes 
    in finance. Therefore works are devoted to explain 
    the origin of the volatility clustering in terms of 
    reflection and the behavior of traders, see e.g. \cite{Cont}.
     The autoregressive conditional heteroskedasticity (ARCH) 
     and the GARCH models are the most common models
      for describing the clustering phenomenon. These are autoregressive (AR) models, where the variance of the process is a random variable which itself follows an AR process.
      The observation of volatility clustering gives
       evidence of predictability of volatility. In other
        words, there is a memory effect in the size of 
        price change. Moreover, the stronger influence of negative
         returns than of positive returns on the future
          volatility and possible predictability have been
           studied in \cite{Sheppard}. 
           
           \bigskip

In the first part of this chapter, we discuss 
some asymmetry in financial time series. More 
precisely, we analyze the correlation between
 intraday and overnight volatilities by considering
  different measuring tools. In the second
   part of the chapter, we discuss the concept of 
   self-fulfilling prophecy and its influence in financial 
   market, by considering the sentiment and fear index 
   of investors. Let us unveil the above statements by 
   revealing them through technical analysis.

\subsection{Asymmetry of cross-correlations between intraday and overnight volatilities}

The volatility of equity returns is higher during
 exchange trading hours than during non-trading 
 hours. It has been shown that the variance of returns
  from the open-to-close of the trading day is over 
  six time larger than the variance of close-to-open
   returns over a weekend, e.g. see \cite{Fama, Oldfield}.
    The difference between the prices at the opening and 
    closing hours has various reasons, e.g. more public 
    information is clustered during business hours. In particular,
     it has been shown that the volatility decreases from 
     the opening hour until the early afternoon and rises 
     thereafter. It was also found that the
      variance of the return is higher for open-to-open 
      than for close-to-close periods \cite{lockwood}.

Finally, it has been repeatedly shown that the 
overnight dynamics
is qualitatively different from that during the day
 \cite {Gallo}-\cite{Edmonds}.
  Various reasons have been proposed for this: 

\begin{itemize}
\item A foreign equity which is mainly traded on some foreign market (that
is open during the night hours of the market studied) reflects mostly
its activity in their overnight volatility, and this activity might
be very different \cite{Chan} from the market under consideration.

\item The majority of news relevant for fundamental stock price evaluation
(company profits, employment rates, general econometric forecasts,
wars and natural disasters, ...) are released overnight \cite{Corral},
and there exists a correlation between frequency of news releases
and volatilities \cite{Edmonds}. 

\item While the market can react during the day to any outside perturbation,
it cannot do so during the night, which might also explain the higher
volatility immediately after the market opening \cite{Gallo}. 
\end{itemize}

Let us use the index $k$ to count trading days (i.e. skipping weekend
and other non-trading days), and denote by $o_{k}$ and $c_{k}$ the
opening and closing prices of one particular equity. Intraday log-returns
of this equity are defined as
\begin{equation}
d_{k}=\ln\frac{c_{k}}{o_{k}}\;,\label{eq:1}
\end{equation}
while overnight log-returns are
\begin{equation}
n_{k}=\ln\frac{o_{k}}{c_{k-1}}\;.\label{eq:2}
\end{equation}
Thus overnight returns are indexed by the index of the following day.
In case of weekends and holidays the over-``night'' returns include
all changes during the entire non-trading period. Volatilities are
in principle defined through the variances of log-returns as observed
over an extended time span. But when discussing them on a fine grained
temporal scale, they are usually replaced by the absolute values of
the log-returns (see e.g. footnote 11 in \cite{Tsiakas}). We will
follow this usage.

\begin{figure}[H]
\begin{centering}
\includegraphics[scale=0.35]{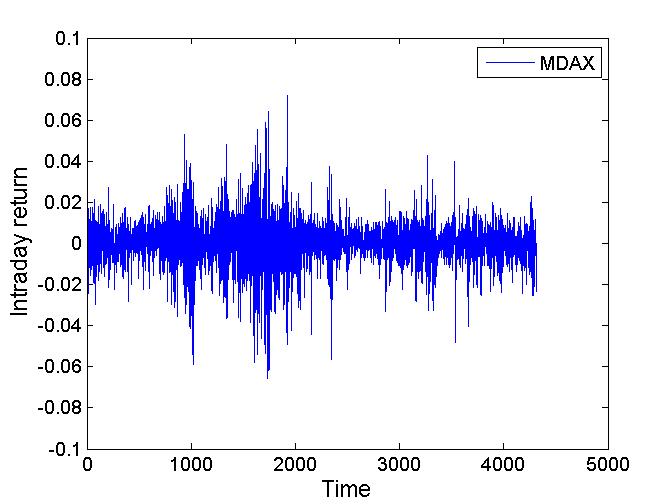} 
\includegraphics[scale=0.35]{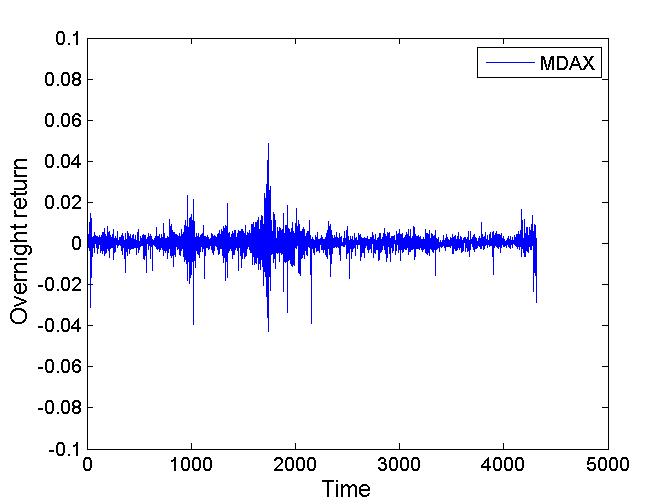}
%\par
\end{centering}
%\protect\protect\caption
\caption{Historical time series of stock intraday (panel a) and
overnight (panel b) returns for the MDAX index. Data shown refers to all trading days
between date and date.}
\label{corr-1.fig}
\end{figure}

The data we studied consists of 21 individual stocks traded at various
stock exchanges (Exxon, Shell, General Electric, Ford, Goldmann-Sachs, Bank of America,
Citigroup, IBM, Microsoft, Cisco, AIG, BP, Caterpillar and Ford all traded at NYSE;
Siemens, Deutsche Bank, Lufthansa, VW and Bayer traded in Frankfurt; and Sony $\&$ Mitsubishi 
traded in Tokyo) and 10 market indices and exchange-traded
funds (TecDax, MDax, DAX, Dow Jones, S$\&$P 100, Nasdaq, EuroSTOXX 50, SIM, S$\&$P/ASX and PowerShares QQQ). They were
mostly downloaded from Yahoo (https://finance.yahoo.com/), the rest
from finanzen.net (http://www.finanzen.net). The time sequences cover
between 10.4 and 45 years, with between 2612 and 13478 data points.
Before using them, we cleaned them from some of their artifacts (missing
data, wrong data, ...), but not of all. There are many
 websites that report daily financial data, but
for the validation of the data it is useful
to check the same dataset from different 
sources. Despite the availability of cleaning data tools provided
  by different software producers, we did not use them, first 
  because of their costs and second mostly data cleaning software
  takes long time. Instead we performed some statistical
   methods, such as calculation of mean, variance, etc., which can help to check
     whether there are some anomalous values among them (for example
     too low minimum or too high maximum values). But for the better 
     understanding of data characteristics we visualized them via graphics and
     scatter plots which
       could show us for example outliers. Detecting 
       outliers are an important issue, however one should
        not immediately remove them, because of their 
        significant influence on statistical parameters such
         as mean and variance. Instead it is essential to
          figure out why they occur and to understand 
          whether on that date something has happened
           in the financial world. But in cases where it was
            obvious that it is simply wrong data we cleaned them.

For instance, we did \textit{not}
remove jumps due to stock splitting. After cleaning, they show the
typical features well-known from previous analyses, such as fat tails,
short-time correlations in the returns, and long-time correlations
in the volatilities. For typical examples, see Figure 2. Notice that
these data still have outliers (mostly negative, due to crashes, bad
annual reports,...). The negative outliers occur mostly for the overnight
returns, consistent with the previous observation that negative news
are disseminated mostly when the markets are closed. The long autocorrelations
of the volatilities are seen both for daytime and overnight.\\

\begin{figure}[!]
\vspace*{-2cm}
\begin{centering}
\includegraphics[scale=0.5]{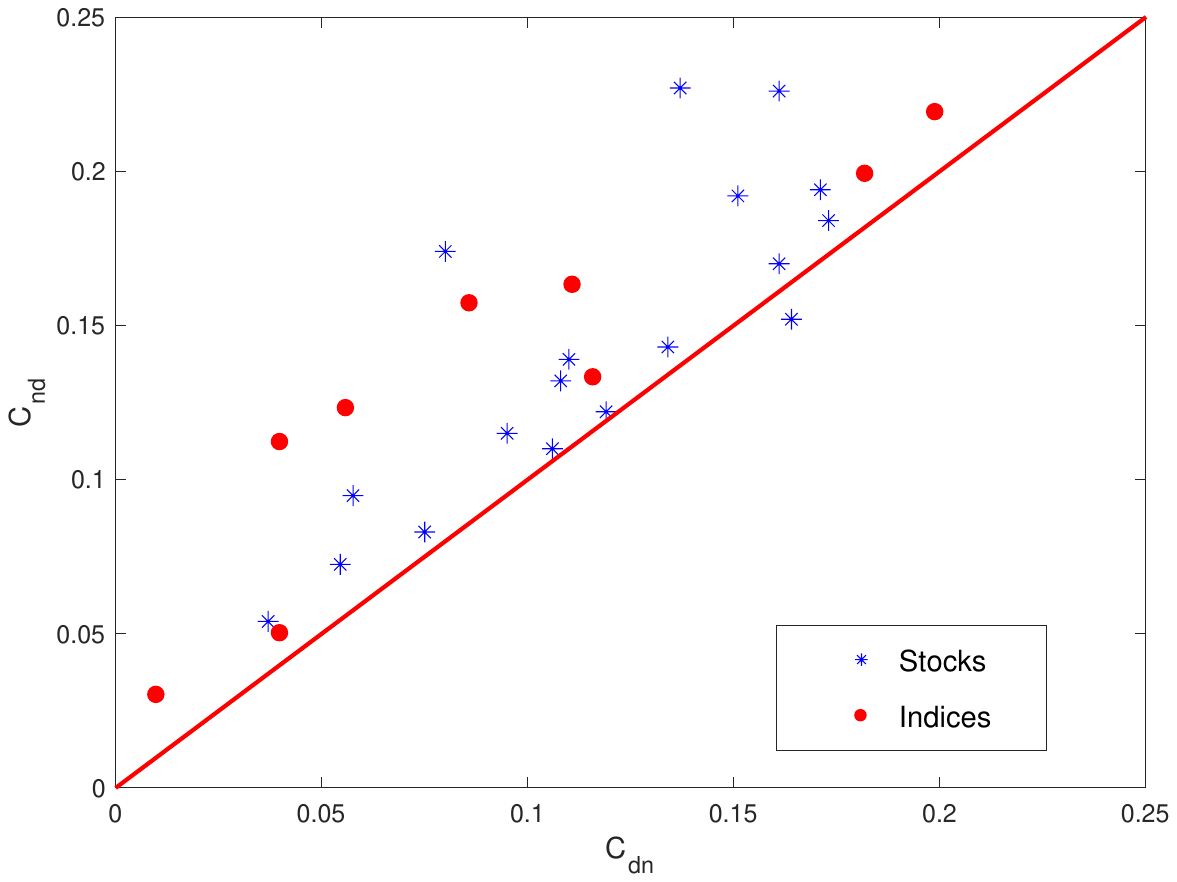} 
\par
\end{centering}
\vspace*{-3cm}
%\protect\protect
\caption{Data for 31 equities. Each dot corresponds to one equity. 
The Spearman correlation between
intraday volatilities and overnight volatilities during the \textit{subsequent}
night are plotted on the x-axis, while the correlations with the \textit{preceding}
night are on the y-axis.}
\label{corr-2.fig}
\end{figure}

\subsubsection{The observation of asymmetry using Spearman's correlation coefficient  }

Our main concern is with cross-correlations between intraday and overnight
volatilities. Due to the artifacts, irregularities, and strong
non-stationarity in the data, we did not use only simple Pearson coefficients.
Instead we mostly used Spearman coefficients
\cite{Press}. Spearman correlation coefficients, as discussed above, show a
dependency between two variables, which are based on rank statistics. These
are known to be much more robust. It implies monotonic relationships
including non-linearity. To a certain extent the Spearman correlation
coefficient is insensitive to the difference between raw data. For these
reasons it shows more robustness for our finding.

\begin{figure}[!]
\vspace*{-0.8cm}
\begin{centering}
\includegraphics[scale=0.4]{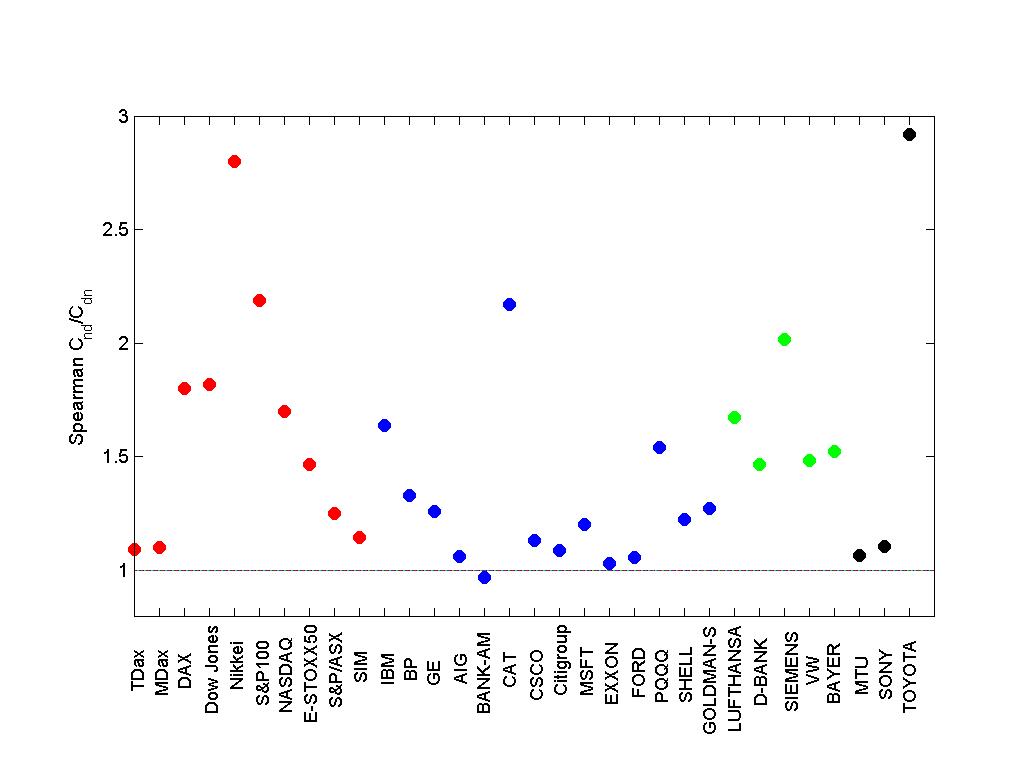} 
\par
\end{centering}
\protect\caption{Ratios $C_{nd}$/ $C_{dn}$ for the 32 equities and
indices shown also in Figure~3. For stocks, the colors indicate the stock 
exchanges where they are traded.}
\label{corr-3.fig}
\end{figure}

Our main results are shown in Figure~3 and 4. In Figure~3
we show for each equity two cross-correlations between the ranks $r_{d_k}$
and $r_{n_k}$ of the two volatilities $|d_k|$ and $|n_k|$:
\begin{equation}
C_{nd}=\frac{\langle r_{d_k}r_{n_k} \rangle-\langle r_{d_k} \rangle\cdot\langle r_{n_k} \rangle}{\sigma_d\;\sigma_n}\;,\label{eq:3}
\end{equation}
\noindent is the rank correlation between the intraday volatility and the volatility
during the \textit{preceding} night ($\sigma_{d}$ and $\sigma_{n}$
are the square roots of the rank variances), while
\begin{equation}
C_{dn}=\frac{\langle r_{d_k} r_{n_{k+1}}\rangle-\langle r_{d_k} \rangle\cdot\langle r_{n_k} \rangle}{\sigma_{d}\;\sigma_{n}}\;,\label{eq:4}
\end{equation}
gives the analogous correlation with the \textit{following} night. (Note that the averages are taken across all trading days, along with the respective nights). We see that
in almost all cases
\begin{equation}
C_{nd}>C_{dn}.
\end{equation}

\begin{figure}[!]
\begin{centering}
\vspace*{-5cm}
\includegraphics[scale=0.7]{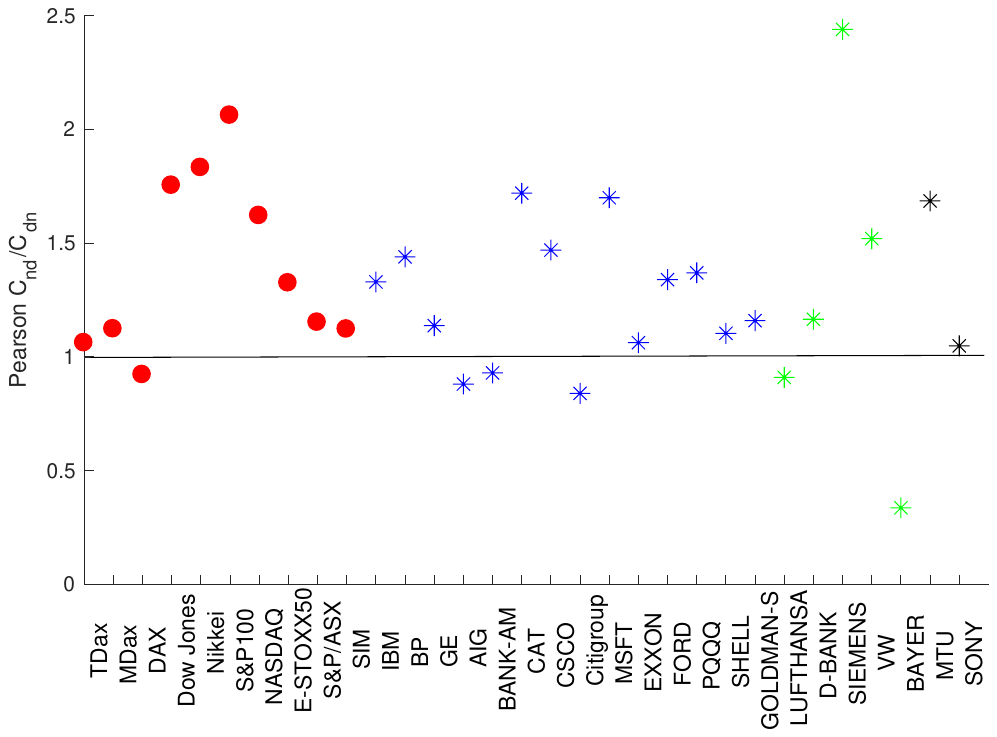} 
\par
\end{centering}
\vspace*{-4.5cm}
\caption{Ratios $C_{nd}$/ $C_{dn}$ for the 31 equities and
indices are shown by using Pearson correlation coefficient.}
\label{corr-4.fig}
\end{figure}

For some equities the difference is small, but for others it can be
more than a factor of two. In only one case the inequality was violated.
Thus the overnight volatility is much stronger correlated with the
volatility during the following day than during the preceding day.
Otherwise said, overnight volatilities seem to influence strongly
what goes on during the following trading day, but do not seem to
be strongly influenced by what was going on during the day before.

The ratios $C_{nd}/C_{dn}$ for the equities used in Figure~ 3
are plotted also in Figure~4, where we have also specified
the equities. The first 10 entries in this figure are market indices,
while the others correspond to individual stocks. We see no big differences,
except that aggregated indices show a somewhat stronger effect. There
are also no noticeable differences related to the place where the equity
is traded, to the length of the time series, and -- in case of individual
stocks -- to the type of company.\\

\subsubsection{The validity of analyses by using Pearson and mutual information}

Next we like to demonstrate that the found asymmetry is not
an artifact of the chosen correlation coefficient. It is most clearly seen
by use of the Spearman correlation coefficient, but still visible by use of
Pearson coefficients, see Figure~\ref{corr-4.fig}. In comparison to
Figure~4, however, we observe in Figure~\ref{corr-4.fig} a
sizable fraction of data points lying below 1.

Alternatives to the Spearman's correlation coefficient are
Kendall's $\tau$ \cite{Kendall} or mutual information \cite{Cover}, both of
which are known to be similarly robust.  Let us finally consider here mutual
information for proving our results. 

\begin{figure}[!]
\vspace*{-5cm}
\begin{centering}
\includegraphics[scale=0.7]{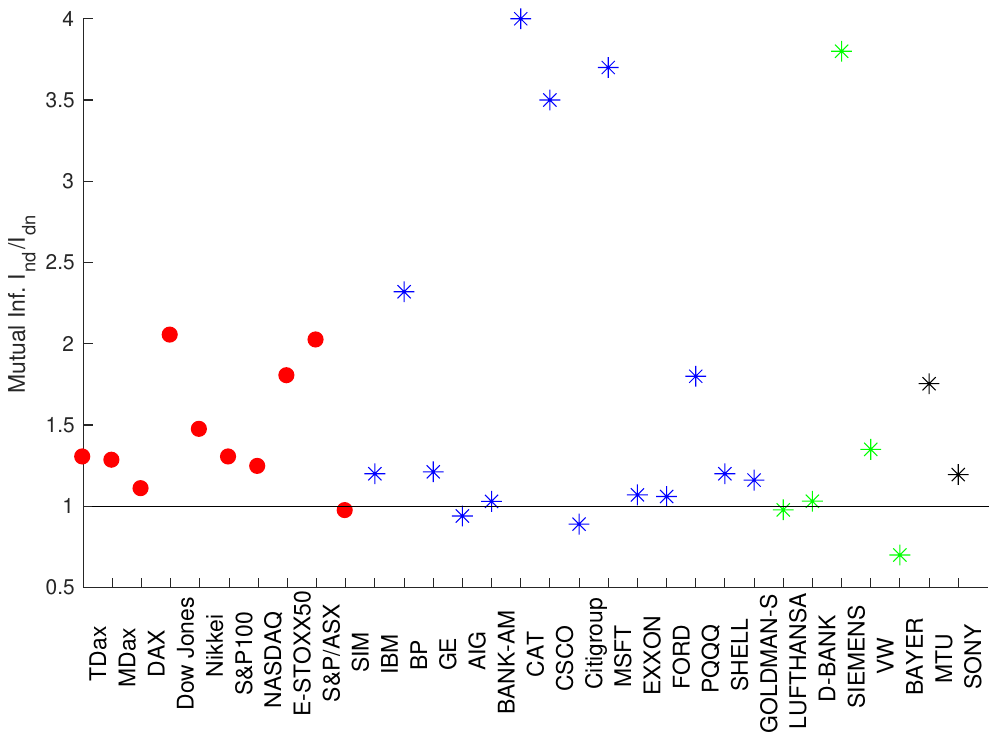}
\par
\end{centering}
\vspace*{-4.5cm}
\caption{Ratios $I_{nd}$/ $I_{dn}$ for the 31 equities and
indices are shown for mutual information.}
\label{corr-5.fig}
\end{figure}

%
%The Spearman correlation index describes a relationship between two variables
%without making any assumptions about the distribution of the variables.
%

As described in the previous section, mutual information
 is a reduction of uncertainty knowing
another random variable and it tells us how much two
 random variables share
information among each other.

Let us recall the equation by considering two random
 variables $X$ and $Y$ with a joint
  probability function $p(x,y)$ and marginal probability
   functions $p(x)$ and
  $p(y)$:
  
\begin{equation}
I(X,Y)=\sum_{x}\sum_{y}p(x,y)\log{\frac{p(x,y)}{p(x)p(y)}}
\end{equation}

where in our case $X$ and $Y$ are the intraday and overnight
volatilities. The results are shown in Figure~\ref{corr-5.fig}, which
shows the robustness of our finding. $I_{nd}$ is the mutual information for
$(x_{k},y_{k})=(d_{k},n_{k})$ and $I_{dn}$ is the mutual information for
$(x_{k},y_{k})=(d_{k},n_{k+1})$. Note 
that in this analysis we use again the absolute values of returns.

For mutual information estimation we use two approaches,
histogram and $K$-nearest neighbor statistics. As discussed in the 
previous section, while the histogram is easy to
comprehend, it has several disadvantages: it depends on the bin width and
different initial choices can change the outcome. In contrast to this
conventional method based on binning, the $K$-nearest neighbor statistics
are based on entropy estimation of $K$-nearest neighbor distances. This
means that they are data efficient, adaptive and have minimal bias
\cite{Grassberger}.

Moreover, in the present thesis investigations by taking into
account foreign markets in different time zones are also considered. Our analyses show that
European markets at their opening time are largely driven by Asian and
American markets. On the other side we see that American markets are largely
driven by both European and Asian market on the same day, whereas the analyses show stronger
correlation between American and Asian markets for the closing prices. There
is also evidence that the Asian market is affected by American markets on
the previous day. 

Another interesting question is about the relation between
 stocks, conditioned on a third one.
  Motivated by such analyses, the Pearson partial correlation
    coefficient has been also performed in \cite{Dror} and also 
    conditional mutual information can answer such questions.
   
While the intraday price dynamics is largely influenced by ``chartist''
behavior, the overnight dynamics is mostly influenced by facts exogenous
to the stock market (or at least not directly related to the day-to-day
price evolution of the considered equity) and thus of ``fundamentalist''
nature. What our results suggest is that ``fundamentalist'' information 
is more useful in prediction than ``chartist'' information.

The present analysis cannot of course specify which of the possible
external influences (foreign stock markets, company performance reports,
news about general economic indicators such as employment rates and
forecasted economic growth, wars, economic crises, natural disasters,
...) is of greatest importance for the overnight dynamics, but such
information could possibly be obtained by performing a larger study
similar to the present one in which equities are grouped according
to business sectors, stock exchanges, trading volume, bull \textit{versus}
bear markets, etc. Another improvement suggested by our analysis could
consist in replacing the simple cross correlations by partial correlations
or by transfer entropies \cite{Schreiber}, testing in this way for
linear or non-linear Granger causality \cite{Barnett}. It would be
of interest to see whether the asymmetry found in the present thesis
is also present at larger time scales, by comparing day/night to night/day
results between more distant nights and days. The very fact that
 different regions in the phase space
of a recurrent system can have different powers of predictability
has been known for long time \cite{DFarmer}.

Finally, with the hindsight gained from this analysis, we might also
turn to signed returns (in contrast to volatilities) and test whether
some parts of a full 24 hours day have more influence on periods
than others.

  \newpage

\subsection{Self-fulfilling prophecy in finance}

The motivation of this part stems from the desire, whether
 it is possible to make some forecast of an asset price change
  by using the sentiment and fear index of investors and traders.
   In general, the inspiration of such analyses has come from a
    concept called self-fulfilling prophecy, whence because of 
    its nice and important notion, let us comment on in detail.

 Self-fulfilling prophecy is a well known concept, which claims 
that social beliefs and expectations often influence social reality
 \cite{Jump}-\cite{Popp}, i.e. the concept describes a
  connection between beliefs and their outcomes in the 
  situation under consideration.  In the literature, the term 
  self-fulfilling prophecy is associated with two sociologists 
  W. I. Thomas and Robert K. Merton. Merton described the
   idea as a cultural belief that becomes true because people 
   believe that this is true already, i.e. it can take the form 
   of a genuine prediction. The basic idea behind this concept
    was posited by Thomas, in the literature well known as the 
    Thomas theorem \cite{Jump1}. According to the Thomas
     theorem: ``If men define situations as real, they are 
     real in their consequences''. After all, Merton popularized
      the concept of self-fulfilling prophecy in his work 
      and defined as follows: ``The self-fulfilling prophecy is,
       in the beginning, a false definition of the situation 
       evoking a new behavior, which makes the originally false
        conception come true''. In conclusion, the idea points
         out a general and in some way a crucial key about the
          social construction of reality, based on prediction.

The famous example of such a phenomenon is a run on a bank,
 which was postulated by Robert Merton. The hypothesis
  describes how false rumors could lead to the financial
   bankruptcy and collapse of the bank, despite its financial
    solvency, i.e. in spite of the comparative liquidity of 
    the bank's assets, a false rumor of insolvency  affected
     the actual outcome and brought the insolvency of the bank.

The concept has been used and tested in many fields such
 as sociology, psychology, education \cite{Bra, Wil}, 
 economy \cite{Roger} and medicine. For example in the latter 
 one, the self-fulfilling prophecy occurs as a placebo effect
 \cite{Har}-\cite{Ox}, which in turn, is medically ineffectual treatment.

In order to gain a better understanding of the attitude and
 the goal involved in self-fulfilling prophecy, psychologists
  have tried to improve the dynamics of the process, 
  considering the constraints and the conditions, under 
which the expectation leads to its own fulfillment. In
general, one should mention that self-fulfilling prophecy
is not just wishing or expecting something and it will lead to its fulfillment.
Otherwise said, wishing alone does not make the expectation
to become real:  It is a dynamical process, which 
consists of a series of steps and these steps are crucial to the triggering of a whole idea.
 The steps simply depend on the underlying phenomenon.
Obviously the concept begins with
the step of having a wish or an expectation 
and it ends with the step of the fulfillment of the 
expectation.

The self-defeating prophecy is the logical converse and 
complementary the opposite of the self-fulfilling prophecy
concept: it is a prediction that prevents what it predicts from
happening \cite{James}. As the opposite of the self-fulfilling 
prophecy, the idea can be taken by referring to any situation
that provokes an opposite behavior to the initial expectations.
    It is again a dynamical process composed of series of steps, 
    which act and begin to operate in such a way, that the
     expectations will not become true.
   
 There is another interesting concept called self-fulfilling
 crisis, which refers to a situation that argues the financial
  crisis is not directly related to the bad economic
  conditions, but it is a consequence of pessimistic
   expectations of traders \cite{Obstfeld, Paul}. 
   
Let us now tie the concept to specific financial data 
in order to explore its strength and to make it more intelligible.

 \subsubsection{Sentiment Index}
 
 In the past decades, sentiment, fear or in general opinion
 mining analyses in the financial sector have become a burning
issue \cite{Pang, feldman}. The most important purposes
of such analyses is the predictability of possible trends in
the stock market prices. These predictions are driven by
analyzing the opinions, feelings and emotions of investors and therefore
are called sentiment and fear analyses. The latter one is known as a
 VIX index and we will discuss it in detail in the next section.

 Many researchers have been attracted to experiment
 with such tests, because often financial markets react 
 and move due to the human emotions. In particular, we investigated the analysis of Euwax 
   Sentiment, reported on Boerse Stuttgart website 
   \cite{Stuttgart}. 
   Euwax Sentiment is a private investor index based on the 
   products of DAX, XDAX and DAX/XDAX combinations. Positive
    values indicate expectation that the market prices are 
    rising and  in analogy negative values indicate expectations
     of the falling of the prices.

The Euwax Sentiment index is based on the orders which are 
submitted and executed within 60 seconds.
The data consists of real trading transactions, providing traders
    with high accuracy of frequency, which presents an up to date picture of the market change and evaluation.
 
  It is well known that one of the major facts for forecasting the
   daily volatility of stock market prices is the observation of high
    frequency data analyses \cite{Hair}-\cite{Martens}. Thus the sentiment
  and fear index could be one of the useful signals for prediction
   purposes.
   
The sentiment score is given according to the following formula:

\begin{equation}
\frac{{\rm Number\, of \,calls - Number\, of\, puts\,}}{{\rm Number\, of\, calls\,+ Number\, of\, puts\,}}\cdot100,
\end{equation}

\bigskip

where the scale of values is ranged between  
[-100, 100]. Long term product investments and purchases
and short term product sales create a positive sentiment
 index, whereas short product purchases and long products sales
  indicate a negative sentiment. Due to the division by the overall 
  number of orders, the index value does not depend on the 
  absolute number of the executed orders per day.

 Let us briefly review (especially for physicists) what in general option pricing and the above ``call'' and ``put'' words in finance mean.
 An option is a contract, that gives the option holders the right (but not the obligation) to execute a particular transaction (buy or sell) with the option (contract) writer, by following some terms. Calls and puts are two main types of options. More precisely, an option that transfers the right to the owner to buy at a specific price is referred to as a call; Vice versa, an option that gives the right to the owners to sell at a specific price is referred to as a put.

Investors buy puts, in case where they think the price of an asset will fall and sell if they think it will rise. If the market experiences a downward turn this is a worst case scenario for put sellers and vise versa, the maximum profit is gained when the price of the asset is above or at the option's strike price at the expiration time. In contrast to put option, investors buy calls when they think the price of the particular asset will rise and sell a call, if they think the price will fall.
If the stock will not be purchased at the specified price before the expiration date, then the stock loses its value.

The value of an option is determined by various quantities such as the current stock price, the intrinsic value (current stock price – strike price (call option), or 
strike price – current stock price (put option)), time to expiration and the volatility. 
The well-known Black-Scholes model is the most widespread option pricing model, which 
as an input uses these parameters, by giving as an output the market value of the option.

While the price of a stock is rising, more likely the price of a call option will rise and the price of a put option will fall. If the stock price has a downward scenario, then one most likely observes the reverse situation of the call and put options.

We analyzed the correlated structure of DAX and the
 sentiment index related to it by using different measurement 
 tools such as mutual information. For the analysis we used 
 the intraday chart presented on the Boerse Stuttgart website,
  which reports every minute from 9:05 to 20:00 based on the
   executed orders of the last thirty minutes.\\

\subsubsection{CBOE's volatility Index (VIX)}
   
Let us in analogy focus on the fear index analysis, by illustrating 
some figures, which show the usefulness of such analyses and 
make the above statements visible. Fear or VIX index, is the 
symbol of the  ``CBOEs Volatility Index'', which is a measure
of the implied volatility of  S$\&$P 500, calculated and 
reported by  Chicago Board Options Exchange (CBOE) \cite{CBOE}.
 In a short time the index achieved popularity between traders
  and became an important benchmark for the U.S. stock market. 
  VIX continues to be used widespread by traders, in risk
   management, forecasting and different financial sectors.

    \begin{figure}[!]
 \vspace*{-0.5cm}
\begin{centering}
%\textbf{Comparison of $S\&P 500$ volatility and VIX}
%\textbf{Comparison of $S\&P 500$ volatility and VIX}\par\medskip
\vspace*{-3cm}
\includegraphics[scale=0.6]{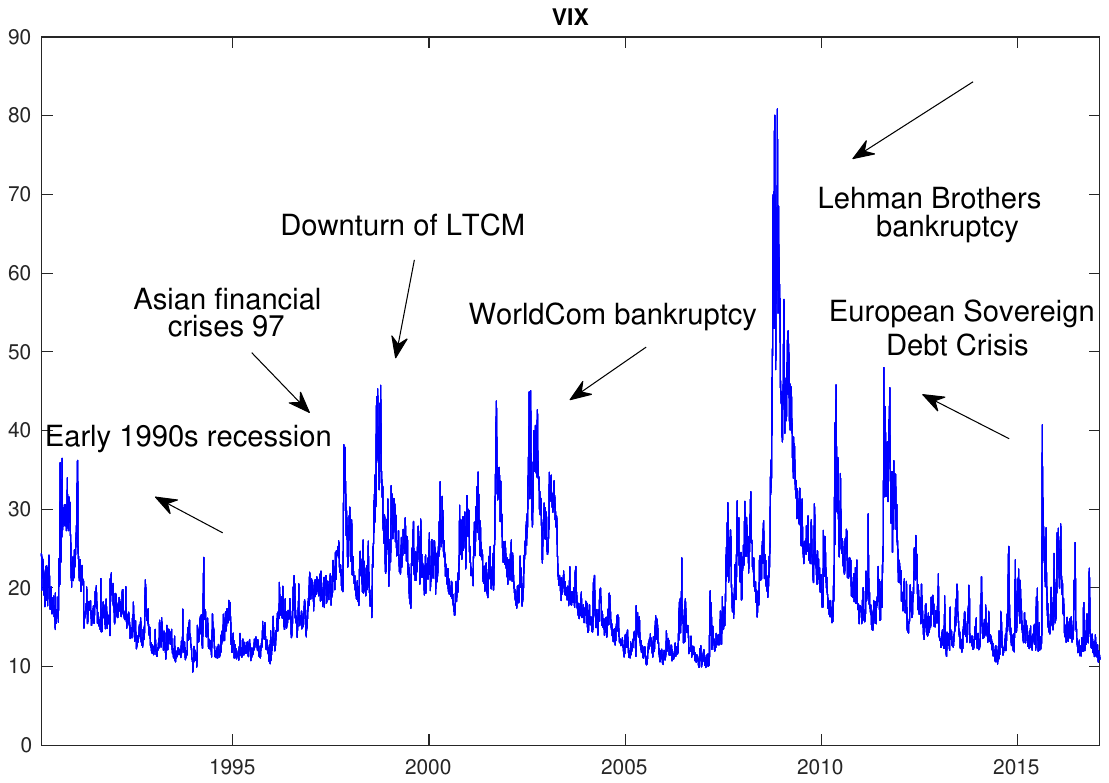}
\vspace*{-4cm}
\par
\end{centering}
\caption{The plot shows the time series of VIX and the different crises, 
with its minimum value 9.3, maximum value 80.8 and mean 19.6.}
\end{figure}

The current VIX index demonstrates an expectation of stock
  market volatility in the near future. More precisely, it indicates
 the expectation of the
 implied stock market volatility for S$\&$P 500 index
  over the next 30 days.
 
 VIX is based on stock options rather than stock prices, so
  it is built on the call and put options.
   In general the idea of creating a volatility index from
option prices dates back to 1973, after the idea of 
exchange-traded options emerged. In the further 
course of such analysis, CBOE was 
inspired by these ideas and extended the concept of 
the early efforts. In the literature the VIX term is associated
 with CBOE's consultant  Robert Whaley, who in 1992
  developed the volatility instrument based on option prices, which is quoted in
  percentage points \cite{white}.

The VIX was the first effective attempt at tracking
 the volatility index. As it was launched, the index was an indicator of
  the implied volatility of  S$\&$P 100. Later in 2004 it was 
  developed based on put and call options of a broader index
   S$\&$P 500. There exists more than 20 
years VIX historical prices, so 
 scientific papers related to VIX,  
 dates back since the end of the 80-s.

Figure 7 shows the VIX index 
    fluctuations since 1988 until present. Here we see for example that the VIX peaked
  at the end of August 1998, at 42-44 percent, 
  which coincides with the LTCM downturn. The chart illustrates
  that before the global crash
      of 2008, the VIX peaked above 30 percent in August
       2007 and after two months we see the stock market
        poked. It followed then with the 2008 global financial
         crisis and Lehman Brothers bankruptcy. In the figure one can see how the different crises such as the 
           Asian financial crises 97, Early 1990s recession, WorldCom bankruptcy,
            European Sovereign Debt Crisis gave rise to the anxiety of investors.

Values of VIX above 30 are considered to be large and result from investors'  
     anxiety and uncertainty,      
             whereas values below 20 are usually
       associated with less hectic and calm market expectations. The higher VIX demonstrates that investors anticipate the huge moves 
 in either direction.\\

\subsubsection{Calculation of the historical volatility and VIX}

\begin{figure}[!]
\vspace*{-2cm}
\hspace*{-0.5cm}
\begin{centering}
\includegraphics[scale=0.4]{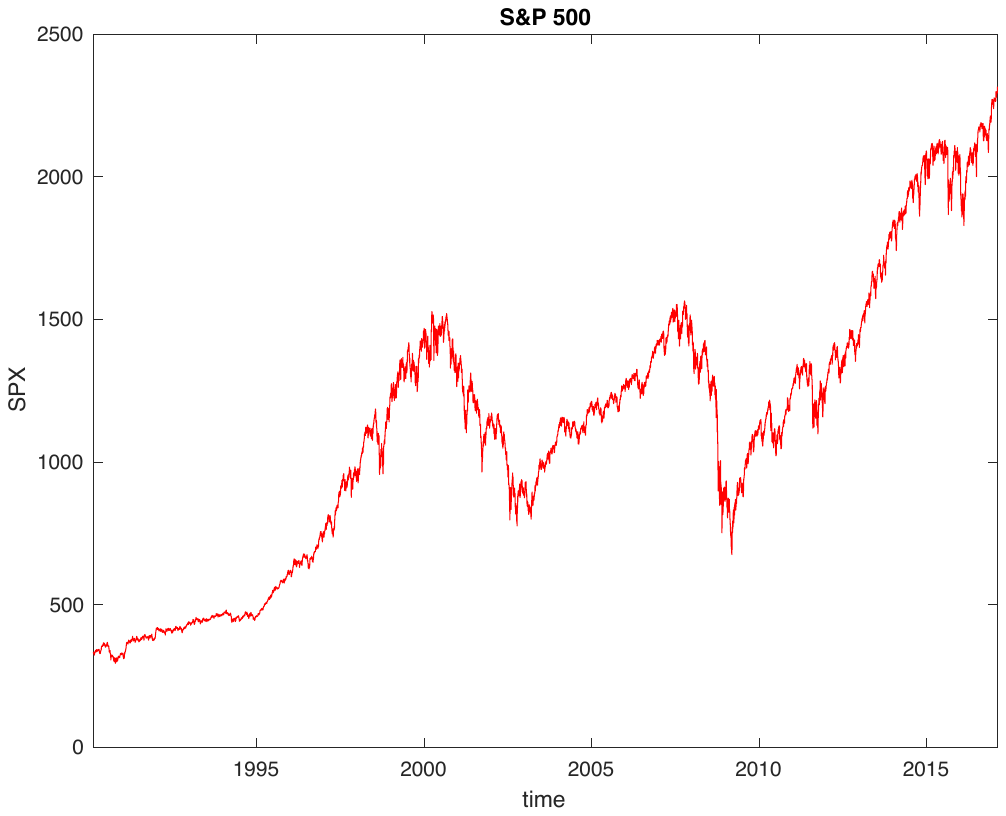} 
\includegraphics[scale=0.4]{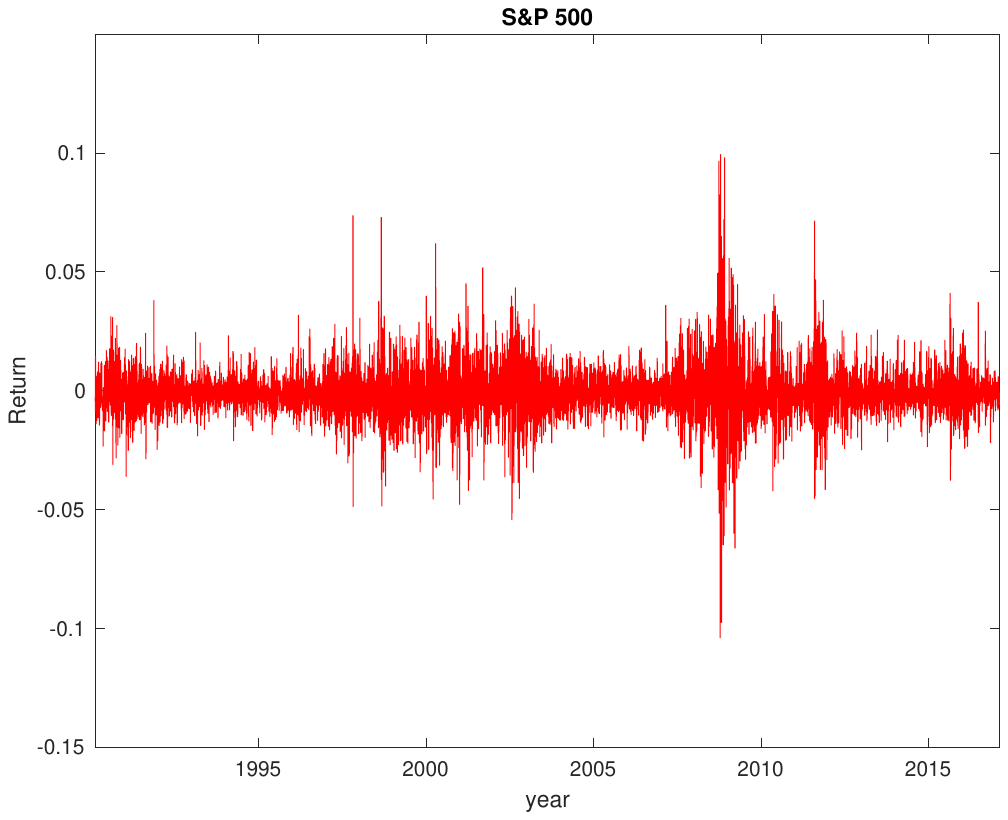} 
\vspace*{-3cm}
%\par
\end{centering}
%\protect

\caption{Time series of S$\&$P 500 index (panel a) and returns of S$\&$P 500 (panel b), which are obtained by $r_{t}=\frac{p_{t+1}-p_{t}}{p_{t}}$, where $p_{t+1}-p_{t}$ indicates a price change.}
\end{figure}

In this subsection we discuss how we obtained the historical
or realized volatility (we call it simply volatility) of S$\&$P 500
and the values of VIX index. 

In general volatility describes how much the asset prices are far from their mean values and is measured as the standard deviation of logaritmic returns.

In Figure 8 (panel a) one can see the evaluation of the time series of S$\&$P 500, where
time sequences cover 30 years, involving about
8000 data points. The data was mostly downloaded 
from Yahoo's website. Fortunately, unlike the case of the last
chapter on equity data, here we have not
 seen wrong data and errors, which made the analysis easier.
 
 Most financial analyses study returns
instead of prices of stocks, because first they have more
interesting statistical properties and second the return of
an asset has a scale-free summary of the 
investment opportunity \cite{Camp, Tsay}. Finally, as
discussed above, the returns are the descriptor of the risk
      measurement and volatility. Panel b in Figure 8 illustrates the returns of S$\&$P 500,
       which are obtained by the following formula: 
       $r_{t}=\frac{p_{t+1}-p_{t}}{p_{t}}$, where $p_{t+1}-p_{t}$
        indicates a price change.   
        
Not being confused with different names for volatilities, 
      first of all we like to mention the difference between implied
       and historical volatility: The implied volatility of an option contract is the expected
 value of the volatility, which will return the value equal to
  the current market price of the option, i.e its input in 
  option pricing models such as Black-Scholes 
  demonstrates a theoretical value identical to the present
   market price of the option. As its definition and the 
   name says, implied volatility is a forward looking measure,
    which differs from historical volatility, which in turn
     measures historic and past returns of an asset.

For the calculation of the volatility one uses the historical standard deviation, and for the annualized one we have
 
\begin{equation}
\hbox{Volatility}=\sqrt{\frac{\sum_{i=1}^{n}(r_{i}-\bar{r})^2}{n-1}}\cdot\sqrt{252}\cdot100,
\end{equation}

where the index $i$ is a counter representing each trading day and $n$
the number of trading days in the underlying time 
frame. The value of $r$ represents compounded 
daily returns and $\bar{r}$ is
the mean of returns over the time window of these $n$ days. Obviously for
 obtaining a historical volatility chart one 
can take different values of $n$. 
The constant value of $252$ represents the annualization 
factor, i.e. the  number of trading days in the U.S. The
exact number of trading days on each year might be
slightly differs, however for simplicity in the literature
this number is taken, for having a constant (approximate) value in the formula.
Finally by multiplying with $100$ we turn to a percentage scale,
by quoting the values as an percentage.

 As mentioned above, VIX is constructed for a constant 30-calender
day horizon and thus represents the
expectation of stock market volatility during the following month. Therefore for the study of the relationship between S$\&$P 500 volatility and VIX in Eq. (56) we consider $n=21$ trading days.

Let us now discuss briefly about the VIX index calculation.    
    The construction of VIX basically follows from the Black-Scholes
   theoretical framework. More specifically, The estimation of the future volatility for the asset underlying the option contract is derived from the Black-Scholes formula, where the model assumes the price of the underlying assets follows a geometric Brownian motion with constant drift and volatility.

Although VIX tracks the S$\&$P 500 volatility,
 the VIX index is being made of options rather than stocks and like the sentiment index, the VIX index calculation is based on 
  calls and puts. It includes the combination of
    multiple options and derives a cumulative value of volatility. 
 Each option reflects market's expectation of future 
 volatility and VIX indicates  
    how volatile are the options
   until their expiration date. The formula for VIX index calculation includes 
 different parameters such as risk free interest rate, strike
  prices and the intervals between them and it is constructed by
   the variance of a near and a next-term options.
   
Let us briefly recall the VIX formula from the CBOE website:

\begin{equation}
\sigma^2=\frac{2}{T}\sum_{i}\frac{\Delta K_{i}}{K_{i}^2}e^{RT}Q(K_{i})-\frac{1}{T}\left(\frac{F}{K_{0}}-1\right)^2, \qquad \hbox{where} \qquad \hbox{VIX}=\sigma\cdot100. \nonumber
\end{equation}

In the above formula we have the following parameters as input: $T$: time to expiration, $F$: forward index level desired from index option prices, $K_{0}$: first strike below the forward index level $F$, $K_{i}$: strike price of the $i$th out-of-the-money option, $\Delta K_{i}$: interval between strike options, $R$: risk-free interest rate to expiration and $Q(K_{i})$  is the midpoint of the bid-ask spread for each option with strike $K_{i}$.
(For more information about the parameters 
we refer to the ``White Paper'' \cite{white}
presented by the CBOE website.)

We downloaded VIX historical data from Yahoo's website since 1988 to present
and for consistency compared with the data downloaded 
from CBOE website.

\begin{figure}[!]
\vspace*{-3.5cm}
\begin{centering}
\includegraphics[scale=0.6]{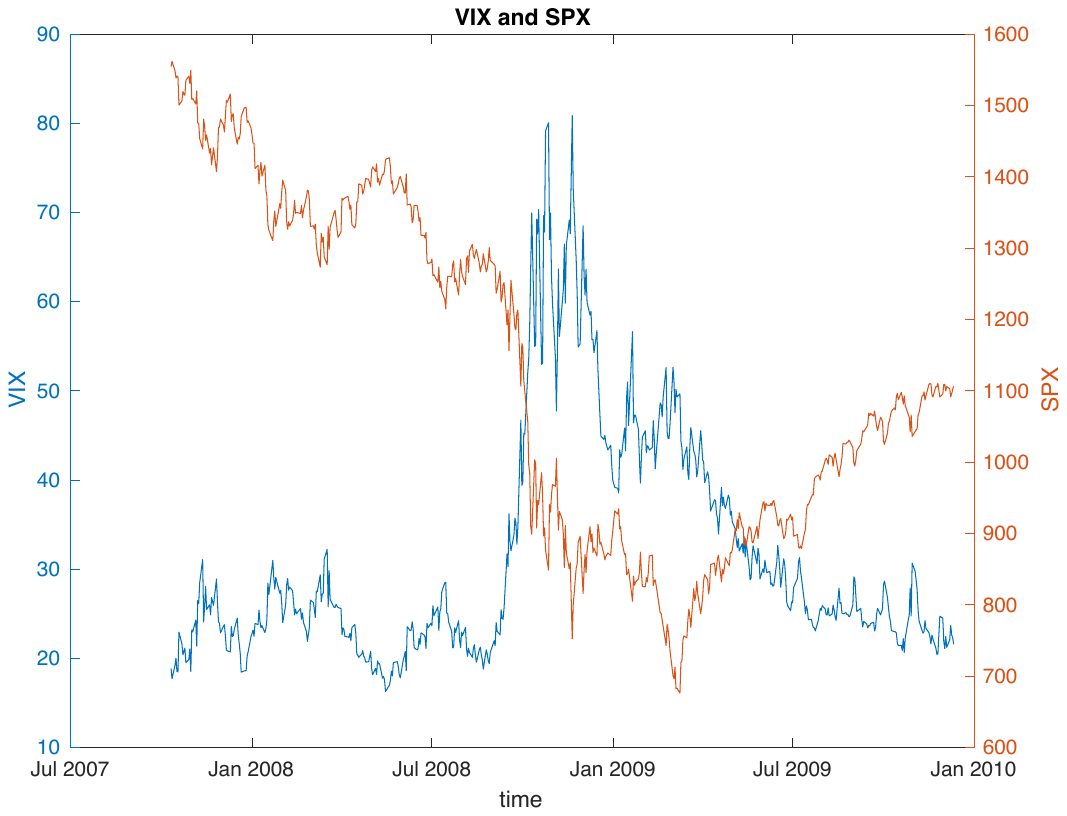} 
\vspace*{-4cm}
\par
\end{centering}
\caption{The comparison of VIX (blue) and S$\&$P 500 (red). 
The plot illustrates there are periods, with the VIX's price tending
 to increase despite the market is going down.\\}
 \end{figure}

Furthermore, CBOE has established other ``fear gauge'' indices
for different stocks as well, for example  VXN tracks the NASDAQ 100 and
  VXD tracks the Dow Jones Industrial average.
Similar to its better-known 
counterpart  VIX, 
the higher the VXN index, the greater
   expectation for Nasdaq-100 volatility. The function of 
   the index is the same as VIX, this time being constrained as  
    the indicator of market nervousness about the technology
     sector. Its highest level was observed in October 2008, in
      the global financial crises, by reaching 86.52$\% $ (like
       VIX, it is quoted in percentage terms). The methodology of 
       VXN calculation is identical to that used for the VIX.\\

 \begin{figure}[!]
 \vspace*{0.5cm}
\begin{centering}
%\textbf{Comparison of $S\&P 500$ volatility and VIX}
\textbf{Comparison of $S\&P 500$ volatility and VIX}\par\medskip
\vspace*{-3cm}
\includegraphics[scale=0.5]{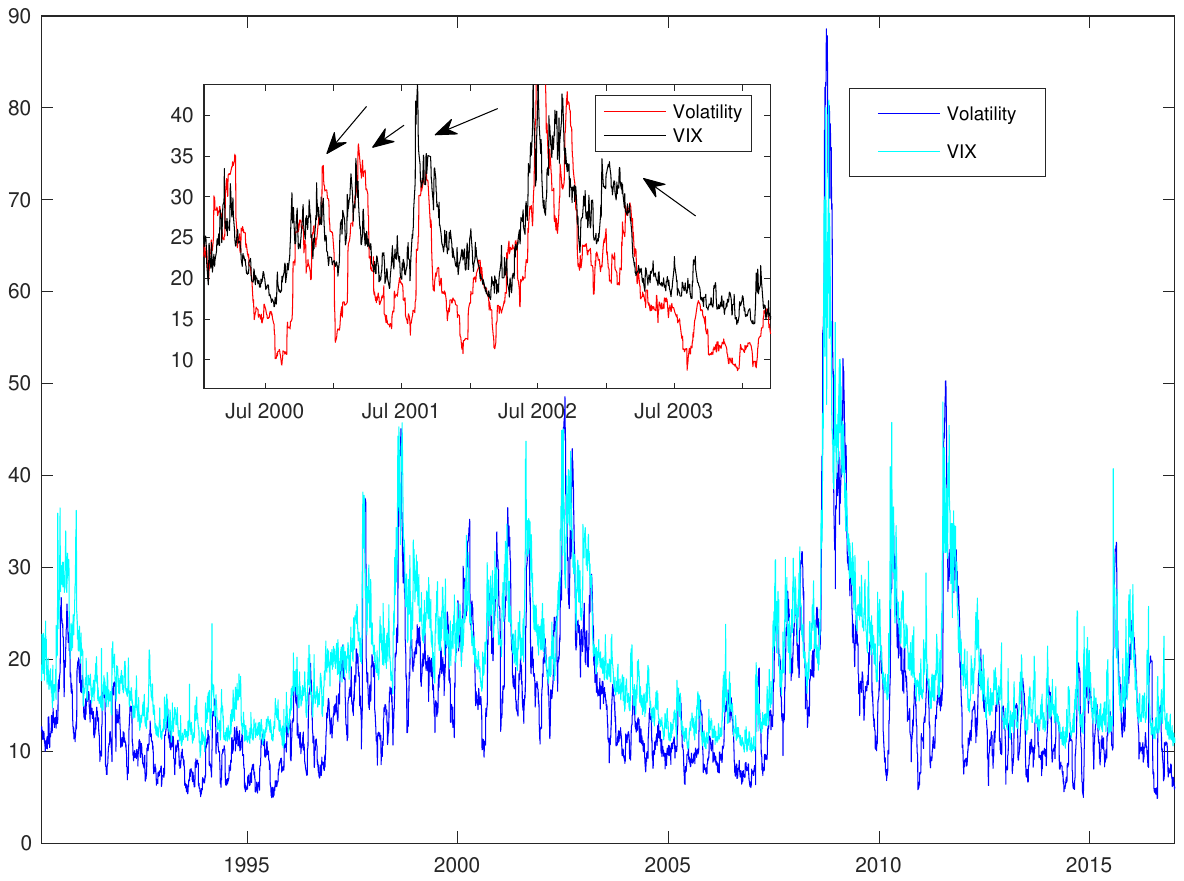}
\vspace*{-3cm}
\par
\end{centering}
\caption{The plot shows the comparison of VIX and the volatility of S$\&$P 500. It illustrates that VIX can be a good factor for predicting the volatilty of S$\&$P 500 index, by analyzing short time scales.\\}
\end{figure}

\subsubsection{The power of self-fulfilling prophecy in VIX}

 The index demonstrates investors' attitude toward a
 financial market. It can be interpreted as a market psychology
  that reveals through the dynamics of the price change 
  and the reaction of the traders to the price movement. 
 Since the index can have an influence on the dynamics
of the market and can reflect the investors' expectation,
it can act as a signal of the performance of the self-fulfilling prophecy.  
For example, Figure 9 highlights that there are periods, 
where the VIX index tends to increase, whereas the 
market goes down. The figure illustrates that despite
 the prices are low, the higher fear index leads to the increment 
 of the S$\&$P 500 index itself, which in turn, shows the power of 
 the self-fulfilling prophecy in VIX. Another example could be found in Figure 10, where it
 shows that on June 2016 the VIX raised by its close value of 20.97.
  This local peak was due to a global sell-off of U.S. equities.
   As a consequence, the investors around the world by feeling
    uncertainties in the market, made some decisions, to
     take gains or experienced losses, which in turn, increased
      the market volatility. The phenomena is the same what 
      R. Merton describes in his ``run in a bank'' example, 
      which describes the notion of the self-fulfilling prophecy.

    The values of VIX largely depend
 on investors' emotions and reactions and the relationship 
 between  S$\&$P 500 and VIX is a bull and bear cycle.

      Another remarkable and still open problem that attracted
 us, is the following: We used so far the available data of VIX
  covering the time since 1988 until present. However, there
   is no available data for VIX before 1988 on financial websites.
    So it has intrigued us to calculate the VIX index before this
     date, by using the corresponding parameters and formulas
      from CBOE website, for which there is a need of respective
       database. The idea is to understand how much influence
        VIX has over the stock market. Otherwise said, how would
         the stock prices fluctuate if there had not been defined
          a VIX or fear index. This idea is based again on the 
          notion of self-fulfilling prophecy, which will tell us 
           ``does the VIX index actually lead to the fear of traders
            and consequently influence the market?''

       \begin{figure}[!]
 \vspace*{-2cm}
\begin{centering}
\includegraphics[scale=0.5]{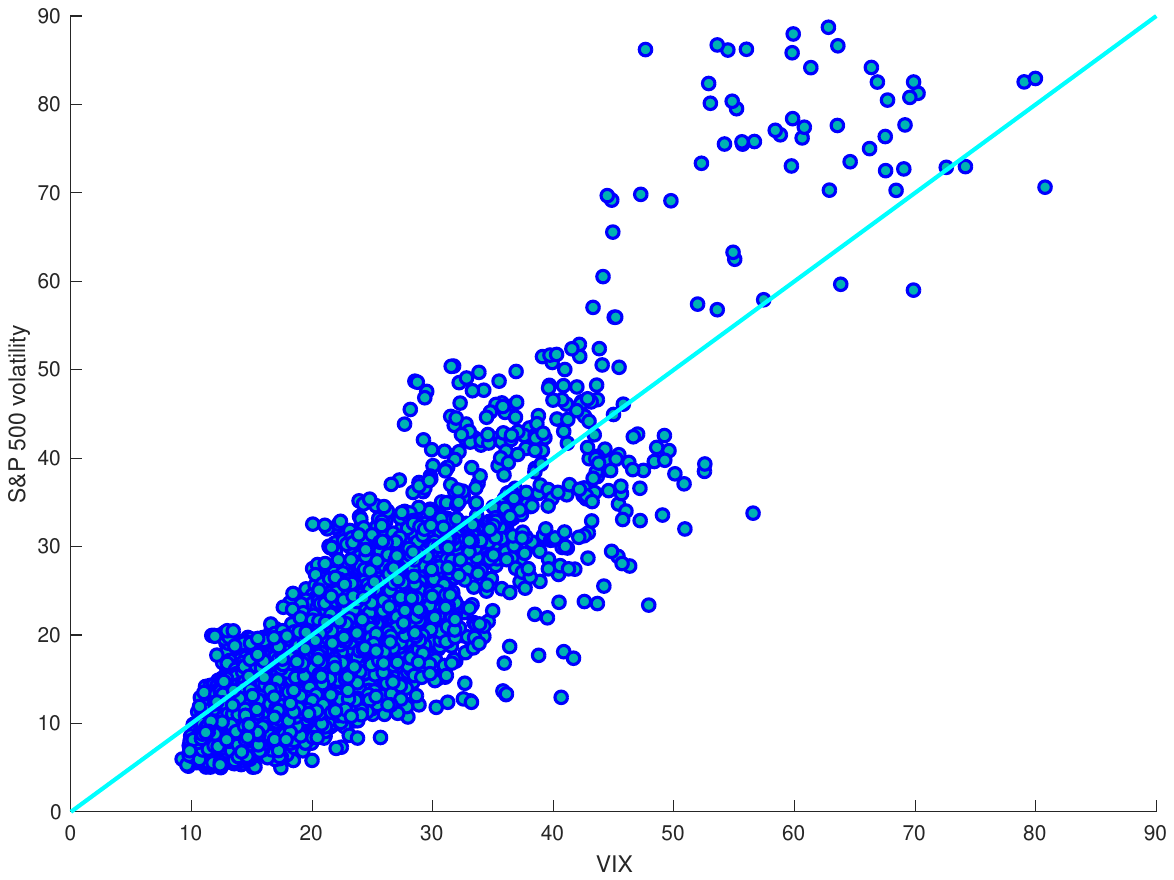}
\vspace*{-3cm}
\par
\end{centering}
\caption{Scatter plot of volatility vs. VIX.}
\end{figure}

\newpage

\subsubsection{Has the VIX predictive power or not?}
 
 Can the VIX predict the market volatility or not? This is the question
that intrigues many researchers and investors. In the literature
   one can find a contradictive debate. Some sources are
   criticizing its predictive value, e.g.~\cite{Der}, other scientific reports like
   \cite{Fleming, Bek} and recently Bloomberg \cite{Bloom}
    evaluate the performance of VIX as a forecast of stock market volatility, especially considering the calm periods of the market.
     While there are different theories
     about the predictive power, in this part of the thesis we give some comments,
     being optimistic on its predictive power. 
     
      \begin{figure}[!]
\vspace*{-2.5cm}
\begin{centering}
\includegraphics[scale=0.5]{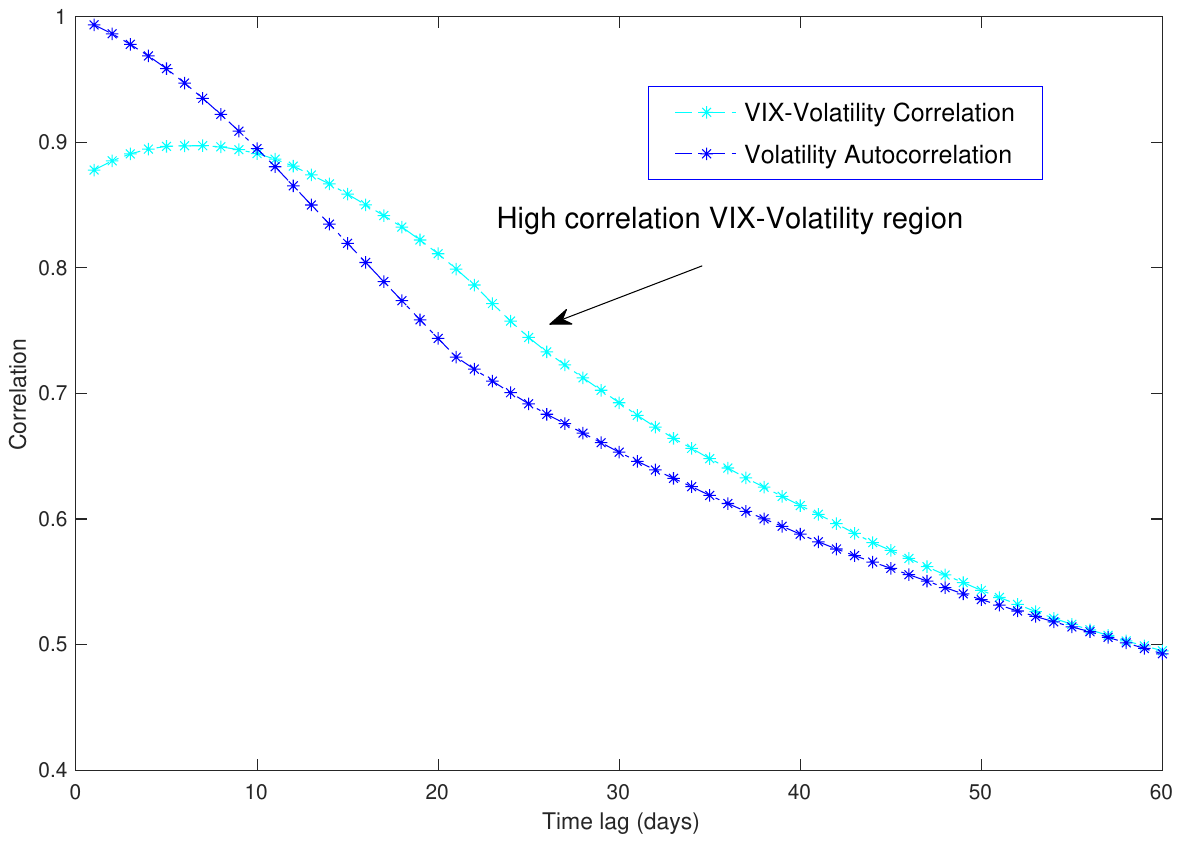}
\vspace*{-3cm}
\par
\end{centering}
\caption{The plot shows comparison of volatility autocorrelation
 and the VIX-volatility correlation. As is visible there are periods,
  where VIX-volatility correlation is higher than the volatility autocorrelation.\\}
\end{figure}

 Figure 10 illustrates a historical price chart of VIX and S$\&$P 500. One can clearly see that the VIX
     index exhibits a 
      strong relationship with the future volatility by illustrating that VIX can 
 be a good factor for predicting the volatility of the S$\&$P 500 index, 
 by considering high frequency data and small time scale. Second by 
      tracking the time series illustrated in the same figure 
       one can see
       that S$\&$P 500 volatility lags behind VIX, 
       which is already the first evidence that VIX can give an additional information for forecasting of the volatility. Finally because it embeds marke expectations, VIX could be a useful gauge for prediction purposes \cite{Fleming}.

The inset
  of the figure shows data for a smaller period enlarged, where we see more clearly that the volatility lags behind the VIX index. We like to mention that some findings here are not novel, such as the chart of VIX-volatility, but it is interesting to generate it and analyze the phenomenon qualitatively, by giving own comments.

When speaking about the predictability of the VIX it is important 
to know the time period that the investor is looking for,
 since VIX could give useful information for short
  term investments, but not for long term investments.

Bloomberg \cite{Bloom} argues that VIX can be thought of as the market's prediction of volatility, especially by considering low volatility periods. Let us illustrate that by looking at the S$\&$P 500 volatility vs.~VIX scatter plot.
 The cyan colored line in Figure 11 indicates the level of volatility
  predicted by VIX. 
    The most of the points lie below the cyan line,
which indicates that VIX overestimates volatility. One does not
experience a shock at low levels of the VIX. The relation
between points shows that volatility is not above 20
percent in cases where VIX is under 12 percent. So, in the range of low volatilities dramatic movement is unlikely (factoring out sudden events). The observation of high volatilities first starts when VIX exceeds 20 percent.

 In conclusion, VIX is the implied volatility and demonstrates the expectation value of annualized
 30-day S$\&$P 500 volatility. So it can react as the market
  prediction indicator. VIX plays as a warning sign and if we eliminate the 
hypothesis of sudden events, news, bankruptcies, political 
turmoil, etc., then dramatic moves and an immediate switch to a period with 
high volatility, in general
 are not common and unlikely.

 Another interesting finding is shown in Figure 12, which illustrates
   the comparison of volatility autocorrelation and VIX-volatility
    correlation. We consider the Pearson's correlation coefficient between the current value of VIX and the expected value of the volatility in the near future as follows:
    
\begin{equation}
C(\tau)=\frac{\langle \hbox{VIX}(t)\cdot\hbox{vol}(t+\tau)\rangle- \langle \hbox{VIX}\rangle \langle \hbox{vol} \rangle}{\sigma_{VIX}\sigma_{vol}}\nonumber,
\end{equation}

where $\tau$ indicates $21$ trading days in a following month, and $\sigma_{VIX}$ and $\sigma_{vol}$ are the standard deviations of the VIX and volatility time series. The current VIX is positively correlated with volatility in the future and the correlation function is not symmetric in the time lag. 

As shown in the figure, there are periods, where the
     VIX-volatility correlation is higher than volatility autocorrelation, 
     which is a striking result, showing that the VIX can be a
      predictor for the volatility of S$\&$P 500 index. In fact, in a region shown
       in the figure, we observe that the relationship and dependency between 
       VIX and volatility are remarkable than the volatility with itself. Given this,
        we conclude that the additional information originated from VIX can lead 
        to the forecasting of the future volatility of the S$\&$P 500 index.

\newpage
 
 \topskip0pt
\vspace*{\fill}
\thispagestyle{empty}
\begin{center}
\section*{\textbf{\LARGE{}{Part IV}}}
\section*{\textbf{\huge{}{Model-based analysis of volatility of cumulative production}}}
\end{center}
\vspace*{\fill}
\vspace*{10cm}
\newpage

\section{Model-based analysis of production processes}

In this chapter we study the stochastic properties
 of the cumulative production, based on Wright's or
  experience curves law, which describes the cost-experience-quantity
   relationship. Behind this fundamental hypothesis stands a large amount 
   of empirical evidence which motivated us to establish a model-based 
   theory for describing the relationship between volatility of production
    process, which in turn is of major importance to predict the volatility of the cumulative production by passing time.
   
   In the following sections, various quantities
    in production process and capital stock modeling based on the
     probability distribution functions of the processes are
      derived.
      
More precisely, in the first part, the focus is on normally
 distributed noise in the production process. By using the
  steepest descent method we demonstrate a model-based
   relationship 
  of volatility of the cumulative production and volatility of
 the production process itself. The result is stunning and 
 potentially powerful and it is tested for various products 
 and technologies. Furthermore, we generalize the concept
  and present a complete model for the relationship 
  between these quantities, now based on an arbitrary
   distribution function of noise in the process. In this chapter
    a recursion relation of integral type is derived that replaces 
    simulations by highly accurate numerical integration. A new
     solution yields a systematic control over the efforts done 
     before. Higher order moments, such as kurtosis and skewness
      will be addressed as well.  Moreover using the same procedure 
      the relation between capital stock and the investment is 
      explored. Let us begin with the technical explanations of the above statements.

\subsection{Experience curve}

First of all let us recast
 what the experience curves present and why they are useful. There exists 
a large literature starting with Wright's paper \cite{Wright}, 
 in which he argued that when the total aircraft 
   production doubled, the required labor time decreased by 
   10 to 15 percent. More general,  
the learning curve effect \footnote{Different names used in the
     literature might be confusing, therefore we want 
     to mention that all these names (experience curve,
      learning curve or simply Wright's Law) express the same effect.} proposes that for 
    each doubling of the total quantity of a particular product, costs decrease by a fixed proportion rate, (costs include marketing,
  manufacturing, distribution, etc).

   The Boston Consulting Group in the 1960 period observed the 
   experience curve effects for various industries and figured out that the production of any good and commodity gives
 evidence of the experience curve effect \cite{Handerson}. 
Based on empirical research, for the experience curves the following mathematical framework, which is a power law function, has been validated: 

\begin{equation}
P_{n}=P_{1}n^{{-b}},\nonumber
\end{equation} 

where $P_{1}$ is the cost of the first unit of production and
 $P_{n}$ is the cost of the $n$-th unit of production.  
 The cumulative volume of production is denoted by $n$ and $b$ is called an experience coefficient. This equation states that the costs decrease
 by a constant factor for every doubling of cumulative
  production, which creates a linear relationship between
   the log of the cost and the log of the cumulative production.
   
   Furthermore, there must be a characteristic
 pattern that causes the experience curve
  effect, for instance a development of better tools,
   automatization, training programs \cite{Terwiesch}-\cite{Arrow}, 
  prior experience and the work complexity task
\cite{Nembhard,Pananiswami}. Let us discuss some of them: The employee in the company regardless of position, needs time which is directly associated with the experience gained for learning the task, i.e. by passing time, labors become more adept and agile to the
   underlying task.  The experience of iteration helps them to spend
    less time and to make less mistakes for performing the same task. This is valid for 
     all employees in the particular firm and not just those who
      are directly connected with the production process. (Given this, it is clear that an expert and experienced employee should decrease the companies costs over time). 
      
Another
       reason to mention here is the network effects, which
        states that the more a specific product will be ubiquitous and widely-used
        then the more efficient the network and the demand
          are and as a consequence costs will be reduced per utility. 
          
The notion of experience
 curve could also describe the effect between business
  competitors, e.g. who is faster in reducing the costs,
   which in turn, could describe the complex interactions
    and network between them. Given this, the hypothesis may not only project the own firms costs, but costs relative to competitors are of major importance, which in turn will give some information about the market strategy. However
            the production process is a complex system 
            and many factors have contributions on its development,
             such as input prices, innovation, etc.
      
The analysis of experience curve also led to the idea (developed by the Boston Consulting group (BCG))
 that in the
 long run, preserving a high price for an item could create  
self-defeat, i.e. the opposite effect for the strategy can be observed, although in the short run it can be profitable. The 
 reason for this effect stems from the fact that it simply motivates competitors, 
 by triggering to produce the same item with lower price and the
 quantity of the product will raise in the market.
  But according to the experience curve effect if the prices would 
  decline as unit costs are reduced, then the item would not attract
   other competitors and their entrance to the market could be
    eliminated. This is one of the strategies that support to maintain
     share holders in the market safely. For supporting this phenomenon, the underpinned
      idea of BCG's growth share matrix  was developed \cite {matrix}.

We like to note, that from the power of experience the firm profits,
 yet in general producing products should be accompanied by innovation and
  invention, because the latter items can dramatically change
   the history of the firm products. For example all the experience
    of making only black and white television would be 
    worthless for a firm, if innovation provides in this case the color one.  
    
 It is also clear if the production of a product is not growing then the rate of cost declines gradually to zero. On the other hand, if competitors have the same relative market shares and almost the same experience for the specific product then their costs tend to remain the same in the market.

The curves of different products have a different slope 
and downward gradient, which means different source of cost reduction.
The reasons for this discrepancy stems from the fact
 that producing different products demand different time 
 and level of experience, e.g. production of a nuclear reactor
   and semiconductors demand different time,
   performance and experience.

The slope of the curve can also characterize the efficiency of learning, for example, a steep learning curve represents a quickly-learned task. Contrarily, difficult subjects require a longer duration, which leads to a shallower curve. 

In conclusion, the experience curve hypothesis states that costs follow a definite pattern due to the accumulated production experience. Knowing that constant increasing rate is for each year the same, one can figure out what would be the effect on cost in the upcoming year. For example, in \cite{Lafond} it has been found that experience curves can be used to estimate
future technology costs, considering the shape of the forecast error
distribution. The hypothesis can provide a significant understanding of the market strategy, for instance export potentials due to the knowledge of experience levels, the
prediction of future prices, given some information 
about the market costs
decrease by some consistent rate of decline, 
the applicability in risks
management, etc.\\

\subsection{Volatility of production for narrow distributions}

It was first discovered by Sahal \cite{Sahal} that the exponentially
increasing cumulative production and exponentially decreasing
costs gives an experience curve law, which indicates a linear
relationship between the log of the cost and the log of the cumulative production.  

 For not causing confusion let us mention
 that Wright's law describes cost-experience relationship,
  which is a power law. But experience itself grows exponentially,
   because we assumed that production grows exponentially,
    as a geometric random walk with drift. 
The exponential growth of production and thus experience gained from this
 comes from the observation which was found first in \cite{Sahal}. 
  Further investigations followed by, e.g. \cite{Nagy, Alberth}. 
 (Regarding the exponential growth of the production and experience
  the curious reader is
  deferred to figures 2 and 3 in \cite{Lafond}. Note that
    the constant growth rate leads to a perfectly exponential
     growth (deterministic as in \cite{Sahal}).

As discussed above, in our model we consider that empirically cumulative production 
growth follows a smooth exponential behavior in the presence
 of noise, by assuming that production is a geometric random
  walk with drift $g$ and variance $\sigma_a^2$. Within this model,
   cumulative production $Z_{t}$ is given by: 
\begin{equation}
Z{_t}=\sum_{j=0}^{t}{ {\rm e}^{gj}{{\rm e}^{a_{1}}...{\rm e}^{a_{j}}}},
\label{2}
\end{equation}
where $a_{1}, a_{2}$, ... are stochastic i.i.d.~variables, which describe the
 noise in the production process.

Let us first consider the special case, where $a_{1}, a_{2}$, ...~are normally
distributed i.i.d.~variables, with mean zero and variance
$\sigma_a^2$. As mentioned before, for the calculation of cumulative production
 and its volatility  for the narrow distribution the saddle point method is used.
The main idea of the saddle point is to approximate an integral by
taking into account only the range of the integration where the integrand
takes its maximum. A priory, this can only be correct for small variance
$\sigma_a^2$.

Next we give here short summary of the results and the calculation are deferred to the next section.

First we calculate the expectation value of cumulative production and its
variance, which lead to the multiple integral
over $a_i$
\begin{align}
\label{1}
E(\log Z)&=\int_{-\infty}^{\infty} \log Z\prod_{i=1}^t
\frac{da_i}{\sqrt{2\pi \sigma_a^2}}
\exp\Big[-\frac{a_i^2}{2\sigma_a^2}\Big]\nonumber\\
&=
\int_{-\infty}^{\infty}\prod_{i=1}^t
\frac{da_i}{\sqrt{2\pi \sigma_a^2}}{\rm e}^{S(\{a_i\})},
\end{align}
\label{DefProbl}
with $S(\{a_i\})=\log (\log Z)-\sum_{i=1}^t \frac{a_i^2}{2\sigma_a^2}$.

We would like to note that $\log{Z}$ is a function of random i.i.d. variables which have Gaussian distributions (with variance $\sigma_{a}^2$). Our independent variables are  $a_{i}$, $i=1....n$ and $Z(a_{1},...a_{n})$ is a function of them. Therefore one should integrate over independent $a_{i}$'s. Eq. (58) expresses this statement, namely, we have a product over identical Gaussian distributions for each variable $a_{i}$.

The applicability of our method essentially depends whether $\sigma_{a}$ is small enough. The final justification comes from comparison with empirical data, which agree with our result with reasonable accuracy.

 The essence of the saddle point method is to approximate the integral by taking into account only that portion of the range of the integration where the integrand assumes large values. More specifically, in our calculation, we find the maxima of the integrands and approximate fluctuations around these points keeping quadratic and neglecting higher order terms.

For $\sigma_a^2\ll 1$ the saddle point method yields explicit results, for
instance the variance of $\log Z(t)$
\begin{align}
\mbox{Var}(\log Z(t))=E(\log^2 Z)-E(\log Z)^2\nonumber\\
=\sigma_a^2 \left(\frac{2 {\rm e}^{g }+1}{1-{\rm e}^{2 g }}+t\right)
+\mathcal{O}(\sigma_a^4).
\end{align}
Finally, the main result of this method is {\em volatility}, i.e.~the variance of {\em volatility
variable} $\Delta \log Z:= \log Z_{t}-\log Z_{t-1}$ for large time $t\rightarrow
\infty $ and is given by the following expression (valid for $g>0$ and small $\sigma_{a}^2$):
\begin{align}
\mbox{Var}(\Delta \log Z)\equiv{\sigma_{x}}^2=\sigma_a^2 \tanh
\left(\frac{g}{2}\right)+\mathcal{O}(\sigma_a^4).
\label{saddle}
\end{align}

\begin{figure}[!]
\vspace*{-0.3cm}
\begin{centering}
\includegraphics[scale=0.5]{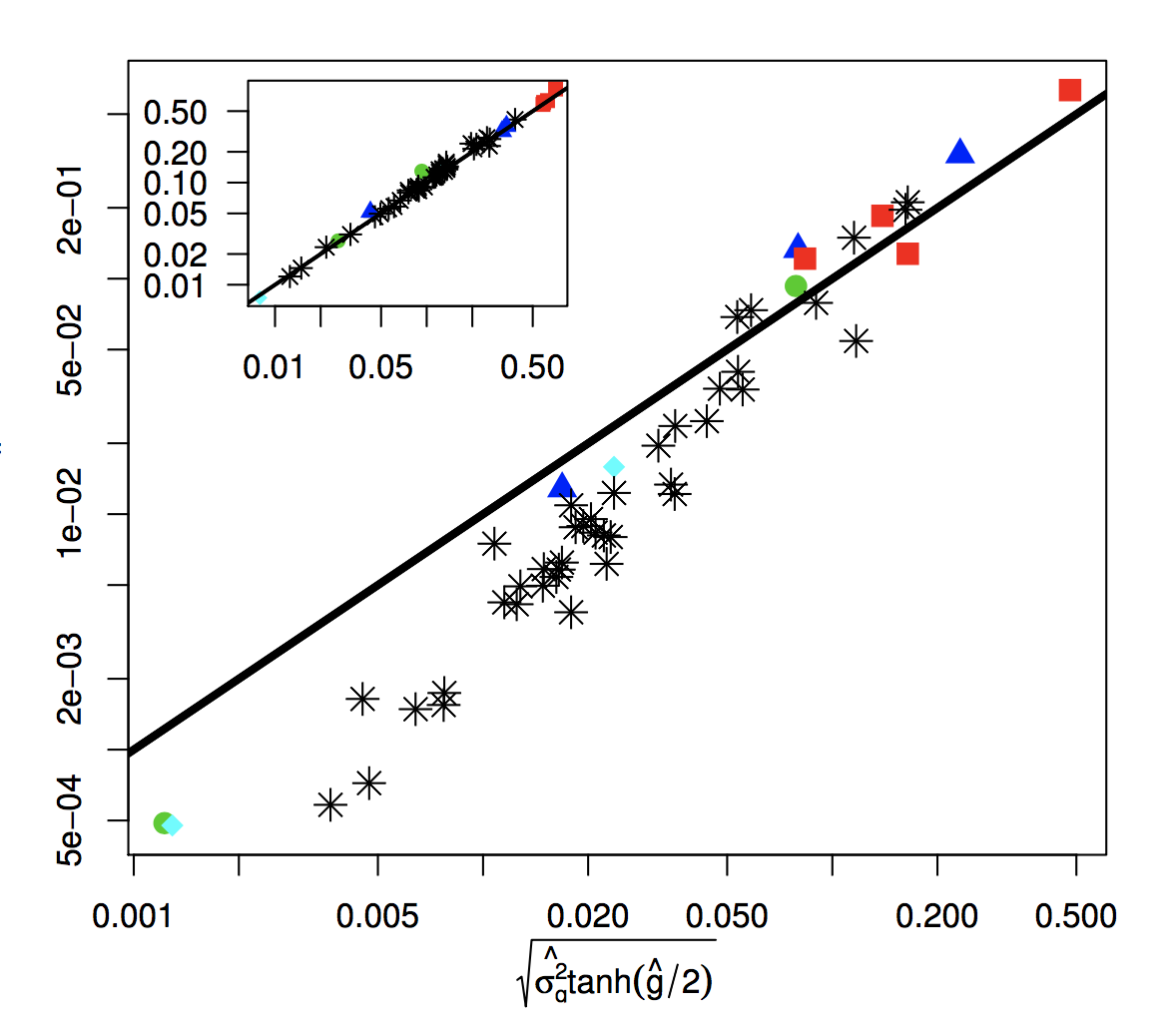} 
\par
\end{centering}
\protect\caption{The plot shows the 
comparison of the volatility of cumulative production $\sigma_{x}$ and the volatility of
the production $\sigma_{a}$ by empirically and analytically obtained results.
The plot is taken from our paper \cite{Lafond}. Asterisks
 demonstrate chemicals, red squares hardware, green dots consumer
  goods, triangles energy and light blue dots demonstrate food.}
\end{figure}

The result for the limit $g\to\infty$ is intuitively clear: $\hbox{Var}(\Delta\log{Z})$ has to be identical to $\hbox{Var}(g+a_{t})$, which is equal to $\sigma_{a}^2$. In the other extreme, namely $g=0$, the variable $Z_{t}$ does not grow exponentially with $t$. As $Z_{t}$ fluctuates weakly around the value $t$, we may intuitively understand in this case $\hbox{Var}(\Delta{\log{Z}})=0$. Furthermore, since 
$\tanh \left(\frac{g }{2}\right)<1$ we have always $\mbox{Var}(\Delta
\log Z)<\sigma_a^2$ an inequality, which means the volatility
 of cumulative production is lower than the volatility of production.
 This is not surprising, because the integration and 
    accumulation act as a low pass filter, which in our 
    case makes cumulative production smoother 
    than the production itself. This finding is also
     visible in empirical data of volatility of production compared
      to the volatility of the cumulative production. For the details
       we refer to Table $1$ in \cite{Lafond}. Note that in realistic 
       production processes, there are correlations induced, e.g. by seasonal effects.

We have tested this remarkably simple, but potentially
 powerful relationship using empirically available data.
  Figure 13 shows the test of this finding, which illustrates
   that the volatility generated by empirical data and the
    theoretical analysis based on Eq. (\ref{saddle}) coincides pretty well. 
  This is the first time that a relationship between these two has been derived.

For testing the formula one needs the values
$\sigma_{x}$, $\sigma_{a}$ and $g$ for each single dataset.
The data used here was created at the Santa Fe Institute and it
 can be accessed at pcdb.santafe.edu. The data is created from
  experience curves existing in the literature and for the below 
  listed technologies it is collected in Table 1 in our paper \cite{Lafond}:

 The emphasis was on 51 technologies including different 
 sectors (chemical, energy, hardware, food, etc.), where 
 chemicals demonstrate a dominant part of the database. 
 More specifically, the data studied here consists of the 
 following products and technologies: Automotive, Milk, 
 Neoprene Rubber, Phthalic Anhydride, Sodium, Pentaerythritol, 
 Methanol, Hard Disk Drive, Geothermal Electricity, Phenol,
  Transistor, Formaldehyde, Ethanolamine, Caprolactam, Ammonia,
   Acrylic Fiber, Ethylene Glycol, DRAM, Benzene, Aniline, VinylAcetate,
Vinyl Chloride, Polyethylene LD, Acrylonitrile, Styrene, Maleic 
Anhydride, Ethylene, Urea, Polyester Fiber, Bisphenol A, Paraxylene,
 Polyvinylchloride, Low Density Polyethylene, Sodium Chlorate, 
 TitaniumSponge, Photovoltaics, Monochrome Television, Cyclohexane,
  Polyethylene HD, Laser Diode, Aluminum, Beer, Primary Aluminum,
   Polystyrene, Primary Magnesium and Wind Turbine.\\

\subsubsection{Analytical treatment by the saddle point method}

First let us derive Eq. (\ref{saddle}) in detail.
 In order to obtain this,  we propose here a useful 
 and powerful method, which is based on the saddle
  point approximation, described in the second chapter. 

First of all we compute the expectation value of $z=\log Z$, where
\[
Z_{t}=1+\sum_{i=1}^te^{y_i},
\]
and 
\[
y_i=g\cdot i+a_1+\cdots +a_i.\]

For simplicity, in the further calculations we omit to write the index
 $t$ for $Z,$ since the derivation part is focused on the $a^*$ saddle point for a fixed $t$.

Assuming that $a_1,\ldots ,a_t$ are i.i.d. with distributions
\[
P(a_i=a)=\frac{1}{\sqrt{2\pi \sigma_{a}^2}} \exp{\left(-\frac{a^2}{2\sigma_{a}^2}\right)}.
\]
This leads to the (multiple) integral (the range of 
integration variables $a_i$ are $(-\infty,\infty)$)
\[
E(\log Z)=\int \log Z\prod_{i=1}^t
\frac{da_i}{\sqrt{2\pi \sigma_{a}^2}}
\exp\left(-\frac{a_i^2}{2\sigma_{a}^2}\right).
\]
It is unlikely that this integral can be computed exactly. We
will apply saddle point method which assumes that $\sigma_{a}^2\ll 1$.

The saddle point is defined from the system of
equations ($n=1,2,\ldots,t$ and $\partial_n
\equiv \partial/\partial_{a_n}$)
\[
\partial_n \left(\log (\log Z)-\frac{a_n^2}{2\sigma_{a}^2}\right)
=0,
\]
or
\[
\frac{\partial_nZ}{Z\log Z}-\frac{a_n}{\sigma_{a}^2}=0,
\]
which we rewrite as
\[
a_n=\sigma_{a}^2\,\frac{\partial_nZ}{Z\log Z}.\]

Since we are going to make calculations up to order $\sigma_{a}^2$,
it is legitimate to substitute $a_i=0$.
Thus for the saddle point $(a_1^*,\ldots ,a_t^*)$ we get
\[
a_n^*=\sigma_{a}^2 \left.\frac{\partial_nZ}{Z\log Z}\right\vert_{a_i=0}+\mathcal{O}(\sigma_{a}^4).
\]
Expanding up to order $\sigma_{a}^2$ for the $\log Z\vert_{a_i=a_i^*}$ we get
\begin{equation}
\log Z\vert_{a_i=a_i^*}=\left(\log Z+\sigma_{a}^2
\left.\frac{\sum_{i=1}^t\left(\partial_iZ\right)^2}{Z^2 \log Z}\right)\right\vert_{a_i=0}+\mathcal{O}(\sigma_{a}^4).
\end{equation}
The contribution of the Gaussian term is
\begin{equation}
\left.\exp\left(-\sum_{i=1}^t\frac{a_i^2}{2\sigma_{a}^2}\right)\right\vert_{a_i=a_i^*}
=1-\left.\frac{\sigma_{a}^2}{2}\frac{\sum_{i=1}^t(\partial_iZ)^2}{Z^2\log^2Z}
\right\vert_{a_i=0}+\mathcal{O}(\sigma_{a}^4).
\end{equation}

Below we should calculate the matrix of quadratic fluctuations.
For second derivatives we get
\[
\partial_n\partial_m\log\log Z=\frac{\partial_n\partial_mZ}{Z\log Z}
-\frac{\partial_nZ\partial_mZ(1+\log Z)}{Z^2\log^2Z}.
\]
Incorporating this (with factor $1/2$ coming from the Taylor's
formula for the second order coefficient) with the initial
Gaussian term, for the matrix of quadratic form we get
\[
G_{n,m}=\frac{1}{2\sigma_{a}^2}\left(
\delta_{n,m}+\sigma_{a}^2\left.\left(-
\frac{\partial_n\partial_mZ}{Z\log Z}
+\frac{(1+\log Z)\partial_nZ\partial_mZ}{Z^2\log^2Z}
\right)\right\vert_{a_i=0}+\mathcal{O}(\sigma_{a}^4)
\right).
\]
Recall now the Gauss formula for the integral
\[
\int \exp \left( -\sum_{n,m=1}^t\xi_n G_{n,m}\xi_m\right)
\prod_{i=1}^td\xi_i=
(\det G)^{-1/2}\pi^{t/2}.
\]
Using
\[
\det G=\exp (tr \log G).
\]
Up to desired order we get
\begin{eqnarray}
&&(2 \sigma_{a}^2)^{-t/2}(\det G)^{-1/2}=\\
&&=1+\left.\frac{\sigma_{a}^2}{2}
\left(\frac{\sum_{i=1}^t\partial_i^2Z}{Z\log Z}
-\frac{\sum_{i=1}^t(\partial_iZ)^2(1+\log Z)}{Z^2\log^2 Z}\right)
\right\vert_{a_i=0}+\mathcal{O}(\sigma_{a}^4).\nonumber
\end{eqnarray}
Multiplying all factors given by Eqs. (61), (62), (63) we finally get
\[
E(\log Z)=\left.\log Z+\frac{\sigma_{a}^2}{2}
\sum_{i=1}^t\left(
\frac{\partial_i^2 Z}{Z}
-\frac{(\partial_i Z)^2}{Z^2}\right)
\right\vert_{a_i=0}+\mathcal{O}(\sigma_{a}^4).
\]
Notice that
\begin{eqnarray}
&&Z\vert_{a_i=0}:= Z_0(t)=\sum_{i=0}^te^{ig}
=\frac{1-e^{g (t+1)}}{1-e^g}\\
&&\partial_i Z\vert_{a_i=0}=\partial_i^2
Z\vert_{a_i=0}=Z_0(t)-Z_0(i-1),
\end{eqnarray}
so that we can rewrite above expression as
\begin{eqnarray}
E(\log Z(t))=\log Z_0(t)+\frac{\sigma_{a}^2}{2}
\sum_{i=1}^t\left(
\frac{Z_0(t)-Z_0(i-1)}{Z_0(t)}
-\frac{(Z_0(t)-Z_0(i-1))^2}{Z_0(t)^2}\right)\nonumber\\
+\mathcal{O}(\sigma_{a}^4).
\end{eqnarray}
If $g>0$ in the large $t$ limit one gets
\begin{eqnarray}
E(\log Z(t))|_{t\rightarrow \infty}=
g  (t+1)-\log \left(e^{g }-1\right)+
\frac{\sigma_{a}^2}{4 \sinh (g )}+\mathcal{O}(\sigma_{a}^4).
\end{eqnarray}

Now let us calculate the expectation value $E(\log^2Z)$.

The integral we have to compute is
\[
\int \log^2 Z\prod_{i=1}^t
\frac{da_i}{\sqrt{2\pi \sigma_{a}^2}}\,
\exp-\frac{a_i^2}{2\sigma_{a}^2}.
\]

The saddle point equations
\[
\partial_n \left(2 \log (\log Z)-\frac{a_n^2}{2\sigma_{a}^2}\right)
=0,
\]
or
\[
2\frac{\partial_nZ}{Z\log Z}-\frac{a_n}{\sigma_{a}^2}=0,
\]
which gives
\[
a_n=2 \sigma_{a}^2\,\frac{\partial_nZ}{Z\log Z}.
\]
As in the previous case on the right hand side we can substitute $a_i=0$
and for the saddle point $(a_1^*,\ldots ,a_t^*)$ we get
\[
a_n^*=2 \sigma_{a}^2 \left.\frac{\partial_nZ}{Z\log Z}
\right\vert_{a_n=0}+\mathcal{O}(\sigma_{a}^4).
\]
Up to order $\sigma_{a}^2$ expansion for $\log^2 Z$ we get
\begin{equation}
\log^2 Z\vert_{a_i=a_i^*}=\log^2 Z+4 \sigma_{a}^2
\left.\frac{\sum_{i=1}^t\left(\partial_iZ\right)^2}
{Z^2}\right\vert_{a_i=0}+\mathcal{O}(\sigma_{a}^4).
\end{equation}
The contribution of the Gaussian term is
\begin{equation}
\left.\exp\left(-\frac{a_i^2}{2\sigma_{a}^2}\right)\right\vert_{a_i=a_i^*}
=1-\left.2\sigma_{a}^2\,\frac{\sum_{i=1}^t(\partial_iZ)^2}{Z^2\log^2Z}
\right\vert_{a_i=0}+\mathcal{O}(\sigma_{a}^4).
\end{equation}

The matrix of quadratic fluctuations is
\[
G_{n,m}=\frac{1}{2\sigma_{a}^2}\left(
\delta_{n,m}+2 \sigma_{a}^2\left.\left(
-\frac{\partial_n\partial_mZ}{Z\log Z}
+\frac{(1+\log Z)\partial_nZ\partial_mZ}{Z^2\log^2Z}
\right)\right\vert_{a_i=0}+\mathcal{O}(\sigma_{a}^4)
\right).
\]

Up to desired order we get
\begin{eqnarray}
&&(2 \sigma_{a}^2)^{-t/2}(\det G)^{-1/2}=\\
&&=1+\left.\sigma_{a}^2
\left(\frac{\sum_{i=1}^t\partial_i^2Z}{Z\log Z}
-\frac{\sum_{i=1}^t(\partial_iZ)^2(1+\log Z)}{Z^2\log^2 Z}\right)
\right\vert_{a_i=0}+\mathcal{O}(\sigma_{a}^4).\nonumber
\end{eqnarray}
Multiplying all factors we finally get
\begin{eqnarray}
&&E(\log^2 Z)=\log^2 Z+\nonumber\\
&&+\left.\sigma_{a}^2\sum_{i=1}^t
\left(\frac{(\partial_iZ)^2}{Z^2}+
\left(\frac{\partial_i^2Z}{Z}-
\frac{(\partial_iZ)^2}{Z^2}
\right)\log Z
\right)\right\vert_{a_i=0}+\mathcal{O}(\sigma_{a}^4).\qquad\qquad
\end{eqnarray}
For the variance we get
\begin{eqnarray}
E(\log^2 Z)-(E(\log Z))^2=
\left.\sigma_{a}^2\sum_{i=1}^t
\frac{(\partial_iZ)^2}{Z^2}
\right\vert_{a_i=0}+\mathcal{O}(\sigma_{a}^4).
\end{eqnarray}
or
\begin{eqnarray}
\mbox{Var}(\log^2 Z(t))=\sigma_{a}^2\sum_{i=1}^t
\frac{(Z_0(t)-Z_0(i-1))^2}{Z_0(t)^2}+\mathcal{O}(\sigma_{a}^4).
\end{eqnarray}
If $g>0$ and $t\rightarrow \infty $ we get
\begin{eqnarray}
\mbox{Var}(\log^2 Z(t))=\sigma_{a}^2
\left(\frac{2 e^{g }+1}{1-e^{2 g }}+t\right)
+\mathcal{O}(\sigma_{a}^4).
\end{eqnarray}
In a similar way we compute the
quantity
\begin{eqnarray}
&&E(\log Z(t) \log Z(t+1))-E(\log Z(t))E( \log Z(t+1))\nonumber\\
&&\left.=\sigma_{a}^2\,\sum_{i=1}^{t+1}\frac{\partial_iZ(t)\partial_iZ(t+1)}{ Z(t)Z(t+1)}\right\vert_{a_i=0}+\mathcal{O}(\sigma_{a}^4)=\nonumber\\
&&=\sigma_{a}^2\sum_{i=1}^t
\frac{(Z_0(t)-Z_0(i-1)(Z_0(t+1)-Z_0(i-1))}{Z_0(t)Z_0(t+1)}+\mathcal{O}(\sigma_{a}^4).
\end{eqnarray}
For $g >0$ and large $t$ the result is
\begin{eqnarray}
E(\log Z(t) \log Z(t+1))-E(\log Z(t))E( \log Z(t+1))\nonumber\\
=\sigma_{a}^2\left(\frac{e^{g }+2}{1-e^{2 g }}+t\right)+\mathcal{O}(\sigma_{a}^4).
\end{eqnarray}
Incorporating (74), (76) for
the variance of $\Delta \log Z$ we get:
\begin{align}
\mbox{Var}(\Delta \log Z)=E((\log Z(t+1)-\log Z(t))^2)
-(E(\log Z(t+1)-\log Z(t)))^2\nonumber\\
=\mbox{Var}(\log Z(t+1))+\mbox{Var}
(\log Z(t))\hspace{6cm}\nonumber\\
-2 (E(\log Z(t)\log Z(t+1))-E(\log Z(t))E(\log Z(t+1)))\nonumber\\
\qquad =\frac{\sigma_{a}^2}{2}\sum _{i=1}^{t+1} \left(\frac{Z_0(t)-Z_0(i-1)}{Z_0(t)}-\frac{Z_0(t+1)
-Z_0(i-1)}{Z_0(t+1)}\right)^2+\mathcal{O}(\sigma_{a}^4)=\nonumber\\
\frac{\sigma_{a}^2 \left(e^{g  (t+1)}-e^{g  (t+2)}\right)^2 \left(2 e^{g }-2 e^{g  (t+2)}+e^{2 g  (t+2)}-2 e^{g  (t+3)}+e^{2 g } (t+2)-t-1\right)}{\left(e^{2 g }-1\right) \left(e^{g  (t+1)}-1\right)^2 \left(e^{g  (t+2)}-1\right)^2}.\nonumber\\
\end{align}
The final result for $g>0$ in large $t$ limit is
\begin{eqnarray}
\mbox{Var}(\Delta \log Z)\equiv{\sigma_{x}}^2=
\sigma_{a}^2 \tanh \left(\frac{g }{2}\right)+\mathcal{O}(\sigma_{a}^4).
\end{eqnarray}
Since
\[
\tanh \left(\frac{g }{2}\right)<1,
\]
always
\[
\mbox{Var}(\Delta \log Z)<\sigma_{a}^2.
\]

In this model-based result we observe the dependence of the volatility on the drift of the process. The finding will allow us to predict the volatility of the cumulative production, by knowing the drift and volatility of the experience or production itself. 
The method and the solution can shed light also on other important areas, for instance later we will use the finding for the derivation of capital from the net capital and the investment.

Note the implication only holds in the limit $\sigma_{a}^2\to 0$, otherwise there are correction terms $\mathcal{O}(\sigma_{a}^4)$. Therefore in the next section we will consider the general case which is valid for any $\sigma_{a}$ and arbitrary distribution.

\newpage

\subsection{Volatility of the production for an arbitrary distribution}

The core result of this section is the investigation of the volatility of
cumulative production for more general distribution functions of $a'_{i}s$ than
considered above. Let us assume that in Eq.~(\ref{2}), $a_1$, $a_2$,
... are stochastic i.i.d.~variables which are distributed according to some
distribution function $\rho_a$ of any shape and width.

We are interested in the distribution function of the log of cumulative production
$z_t:=\log Z_t$ which we call $\rho_{z_t}$. From this distribution function we
can calculate all important characteristic quantities of the
system. Surprisingly, the distribution function can be shown
to satisfy a useful recursion relation for successive times:
\begin{equation}
\rho_{z_{t+1}}(x)=\frac{1}{1-\exp(-x)}\cdot{\rho_{a}*\rho_{z_t}}(\log(\exp(x)-1)-g),
\label{wichtig}
\end{equation}
where $\ast$ denotes convolution $f\ast g(x)=\int dy
  f(x-y)*g(y)$ and the argument of the convolution in (\ref{wichtig}) is a
  non-linear function of $x$ with derivative appearing as prefactor. For
   the comprehensive derivation of Eq. (79) we refer to the next subsection.

Generally Eq.~(\ref{wichtig}) has to be solved numerically by
 recursions. Numerical analyses can be done to high accuracy
  and completely replace simulations, which are time consuming 
  and sometimes inaccurate.

It is possible to obtain analytic results at least for two cases.
 In the case of a distribution of $\rho_{a}$ with main weight around some $x_0$ and a
value of $g$ such that $g+x_0>0$ we find for large $t$ an asymptotic
solution. In this case only large values of $x$ matter and (\ref{wichtig}) linearizes to
$\rho_{z_{t+1}}(x)=\rho_{a}*\rho_{z_t}(x-g)$. The second case 
is a narrow distribution of $\rho_{a}$, which we already discussed in the last section.

Eq.~(\ref{wichtig}) is highly useful in numerical calculations,
 especially because the
convolution integral can be carried out efficiently and the
convergence for increasing time is fast. Of course it would be desirable to
treat the time evolution of the probability distribution function for
arbitrary $t$ fully analytically such 
as in \cite{Zadourian}, where
 the exact probability distribution functions for the Parrondo's
games are derived (for the detailed explanation we refer
 to the chapter five). The
analytical solution is the subject of current investigation.\\

\subsubsection{Probability distribution functions of production processes}

 Let us now derive Eq.~(\ref{wichtig}) in detail.
In the first step, the preliminary analysis was done, for 
getting closer to the main goal and results. To this end,
 we present here the expectation values of products of $Z$'s,
  which lead to the further investigations of the variance of
   the cumulative production $Z_{t}$.

Let us recall
\begin{equation}
Z_t:=\sum_{j=0}^t Q_j,
\end{equation}
where
\begin{equation}
Q_j:={\rm e}^{gj}{\rm e}^{a_1}...{\rm e}^{a_j},\qquad (Q_0=1)
\end{equation}
where the drift $g$ is a constant and $a_1$, $a_2$, ... are stochastic
i.i.d.~variables.

We further define the expectation values
\begin{equation}
\langle {\rm e}^{a_j} \rangle=:x,\qquad \langle {\rm e}^{2a_j} \rangle=:y.
\end{equation}
Note that these objects are independent of the index $j$. Also note that the
expectation values of the exponentials of $a_j$ are not the exponentials of the
expectation values of $a_j$.

Now we can easily calculate the expectation value of $Q_j$ and $Z_t$. We find
\begin{equation}
\langle Q_j \rangle={\rm e}^{g j}\langle {\rm e}^{a_1}...{\rm e}^{a_j}
\rangle={\rm e}^{g j}\langle {\rm e}^{a_1}\rangle ...\langle {\rm e}^{a_j} \rangle,
\end{equation}
because the exponentials of the different $a_i$ are independent. Hence
\begin{equation}
\langle Q_j \rangle={\rm e}^{g j} x^j=\left({\rm e}^{g} x\right)^j,\qquad
\langle Z_t \rangle=\sum_{j=0}^t \left({\rm e}^{g} x\right)^j=
\frac{\left({\rm e}^{g} x\right)^{t+1}-1}{\left({\rm e}^{g} x\right)-1}.
\end{equation}

The calculation of expectation values of products of $Z$'s like $\langle Z_t
Z_s\rangle$ is more complicated:
\begin{equation}
\langle Z_t Z_s\rangle=\sum_{j=0}^t \sum_{i=0}^s \langle Q_jQ_i\rangle.
\end{equation}
In the product of $Q_jQ_i$ we have factors like ${\rm e}^{a_n}$ and ${\rm
  e}^{2a_m}$. For $j< i$ we find
\begin{equation}
Q_jQ_i={\rm e}^{g (j+i)} {\rm e}^{a_1}...{\rm e}^{a_j}\cdot {\rm
  e}^{a_1}...{\rm e}^{a_i}
={\rm e}^{g (j+i)} {\rm e}^{2a_1}...{\rm e}^{2a_j}\cdot {\rm
  e}^{a_{j+1}}...{\rm e}^{a_i}.
\end{equation}
In this case the expectation value is
\begin{equation}
\langle Q_jQ_i\rangle={\rm e}^{g (j+i)} \langle{\rm e}^{2a_1}...{\rm
  e}^{2a_j}\cdot {\rm e}^{a_{j+1}}...{\rm e}^{a_i}\rangle=
{\rm e}^{g (j+i)}y^j x^{i-j}.
\end{equation}
Here we calculate $\langle Z_t Z_s\rangle$ for the case $s=t$ where we regroup the
terms in the double sum in diagonal and off-diagonal terms
\begin{equation}
Z_t Z_t=\sum_{j=0}^t \sum_{i=0}^t Q_jQ_i=
\sum_{j=0}^tQ_jQ_j+2\sum_{j=0}^t \sum_{i=j+1}^t Q_jQ_i.
\end{equation}
We find
\begin{equation}
\langle Z_t Z_t\rangle=
\sum_{j=0}^t\langle Q_jQ_j\rangle+2\sum_{j=0}^t \sum_{i=j+1}^t \langle
Q_jQ_i\rangle
=\sum_{j=0}^t{\rm e}^{g (2j)}y^j + 
2\sum_{j=0}^t \sum_{i=j+1}^t {\rm e}^{g (j+i)}y^j x^{i-j}.
\end{equation}
Doing the sums we find
\begin{equation}
\langle Z_t Z_t\rangle=
2\frac{\left({\rm e}^{g}x\right)^{t+1}}{{\rm e}^{g}x-1}
\frac{\left({\rm e}^{g}y/x\right)^{t+1}-1}{{\rm e}^{g}y/x-1}
-\frac{{\rm e}^{g}x+1}{{\rm e}^{g}x-1}
\frac{\left({\rm e}^{2g}y\right)^{t+1}-1}{{\rm e}^{2g}y-1}.
\end{equation}

Now the variance can be calculated $\langle Z_t Z_t\rangle-\langle Z_t\rangle^2$. This is one of the approach for calculation the varianace of the $\log{Z}$, but below we will propose a better mechanism for that.

\bigskip

Let us now propose another mechanism for deriving the recursion 
expression (\ref{wichtig}).  Note that the modified
object
\begin{equation}
\tilde Z_t:=\sum_{j=0}^t {\rm e}^{g j}{\rm e}^{a_2}...{\rm
e}^{a_{j+1}}
\end{equation}
has the same distribution function as $Z_t$, because we have used different,
but independent and identically distributed $a_i$'s. So $\tilde z_t:=\log
\tilde Z_t$ is distributed according to $\rho_{\tilde z_t}=\rho_{z_t}$.  Note
that
\begin{equation}
1+{\rm e}^{g+a_1}\tilde Z_t=
1+\sum_{j=0}^{t} {\rm e}^{g (j+1)}{\rm
e}^{a_1}{\rm e}^{a_2}...{\rm e}^{a_{j+1}}\\
=1+\sum_{i=1}^{t+1} {\rm e}^{g i}{\rm
e}^{a_1}{\rm e}^{a_2}...{\rm e}^{a_{i}}
=Z_{t+1}.
\end{equation}
Therefore we have
\begin{equation}
z_{t+1}=\log(1+\exp(g+a_1+\tilde z_t)=f(g+a_1+\tilde z_t),
\end{equation}
where we have used the definition of the function $f$:
\begin{equation}
f(x):=\log(1+\exp(x)).
\end{equation}
The stochastic variable $a_1+\tilde z_t$ is distributed according to the
convolution of $\rho_a$ with $\rho_{z_t}$. The distribution of $g+a_1+\tilde
z_t$ is the convolution with a subsequent shift of the argument:
\begin{eqnarray}
&\rho_{a_1+\tilde z_t }=\rho_{a_1}*\rho_{\tilde z_t}=\rho_{a}*\rho_{z_t},\nonumber\\
&\rho_{g+a_1+\tilde z_t }(x)=\rho_{a}*\rho_{z_t}(x-g).
\end{eqnarray}
With (93) and (95) we can calculate the distribution function $\rho_{z_{t+1}}$
of $z_{t+1}$. If we use the arguments $x$ for $\rho_{g+a_1+\tilde z_t }$ and
$y=f(x)$ for $\rho_{z_{t+1}}$ we find
\begin{equation}
\rho_{z_{t+1}}(y)dy = 
\rho_{g+a_1+\tilde z_t }(x)dx,
\end{equation}
and from this
\begin{equation}
\rho_{z_{t+1}}(y)=[f'(x)]^{-1}\rho_{g+a_1+\tilde z_t }(x)\\
=[f'(x)]^{-1} \rho_{a}*\rho_{z_t}(x-g).
\end{equation}
Now we use $f'(x)=\exp(x)/(1+\exp(x))$ and $x=f^{-1}(y)=\log(\exp(y)-1)$ and
reach one of our main findings:
\begin{equation}
\rho_{z_{t+1}}(x)=\frac 1{1-\exp(-x)}.\\
\rho_{a}*\rho_{z_t}(\log(\exp(x)-1)-g).
\end{equation}

\begin{figure}[h!]
\begin{centering}
\vspace*{-4cm}  
\centerline{\includegraphics[scale=0.6]{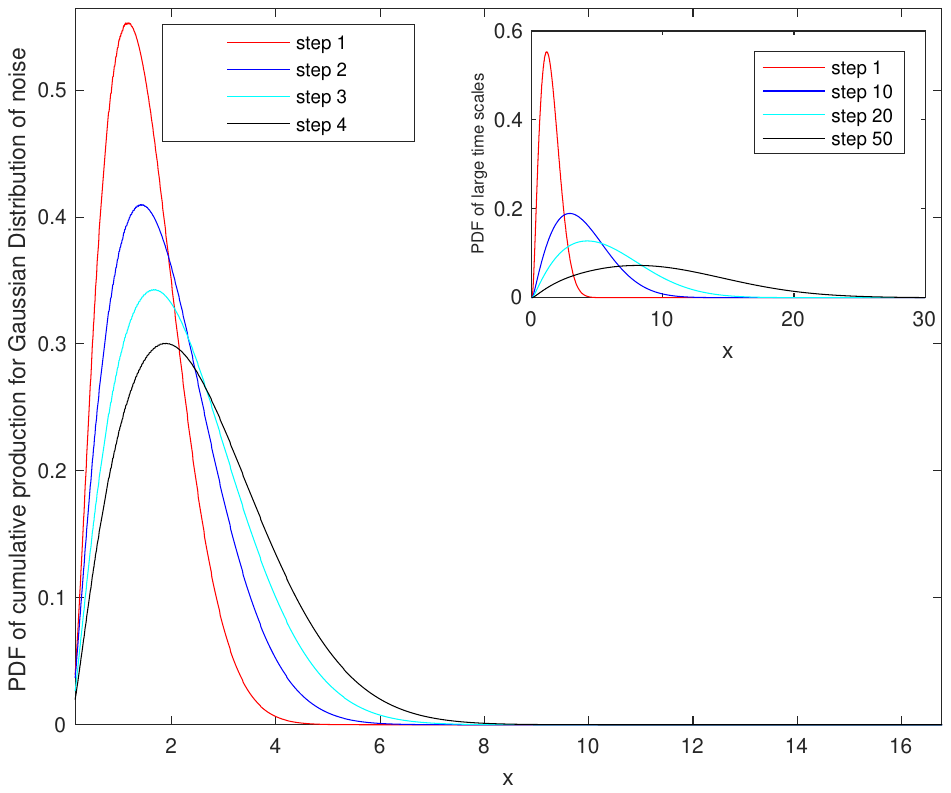}}
\end{centering}
\vspace*{-4cm}
\caption{Plot of probability distribution functions of 
$\log Z_t$ of the Gaussian distributed $\rho_a$ for 
different time steps, where $g=0.2$ and $\sigma_{a}=1$ 
are chosen. The inset shows curves on a large time scale.}
\label{fig1.fig}
\end{figure}

Figures $14$, $15$ and $16$ show the distributions of
 cumulative production for different types of noise and several time steps.  Note that for
  Gaussian $\rho_{a}$ the distribution functions for $\log Z_t$ have positive
  skewness and the asymptotic growth of the mean is linear in $t$ (see
  Figs.~$14$, $15$). For Lorentzian $\rho_{a}$ the distribution for $\log Z_t$
  has a peak at zero (see Figure~$16$). This means that there is a sizable
  probability for no production for a long time which stems from the fat tail
  for negative increments of the production which causes resting. In contrast
  to the Gaussian case the location of the maximum of the distribution
  function for $\log Z_t$ looks stationary.

\begin{figure}[H]
\begin{centering}
%\centerline{\includegraphics[scale=0.6]{bar}}
\vspace*{-4cm}  
\centerline{\includegraphics[scale=0.6]{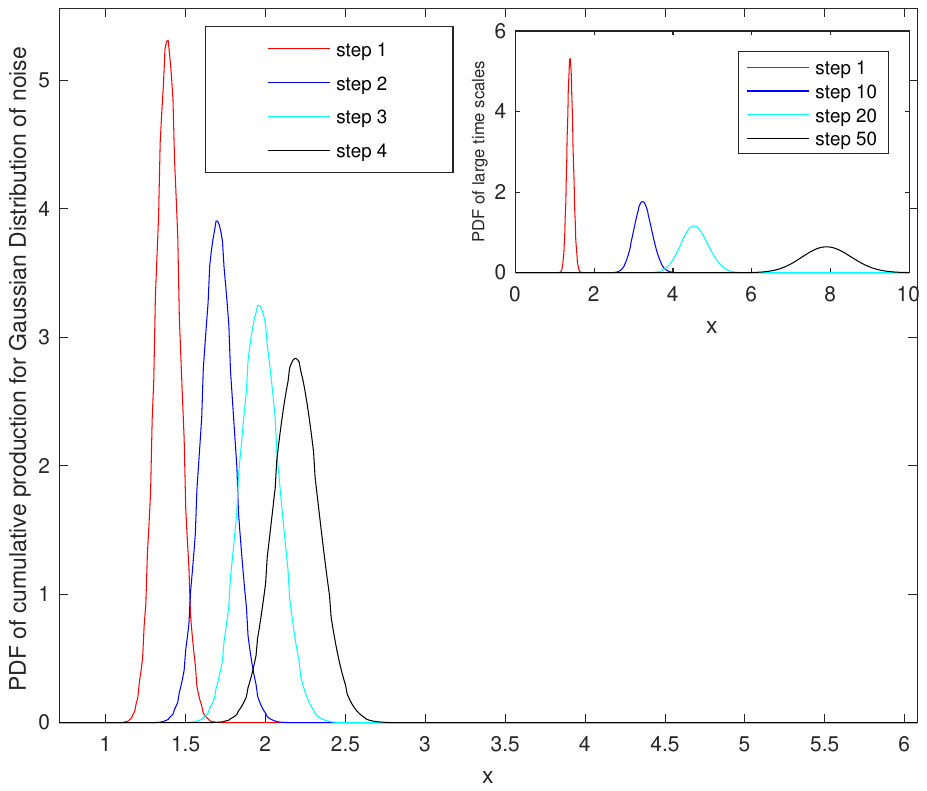}}
\end{centering}
\vspace*{-4cm}  
\caption{Plot of probability distribution functions of $\log Z_t$
 of Gaussian distributed noise for different time steps, where 
 $g=0.2$ and $\sigma_{a}=0.1$ are chosen. The inset 
 shows curves on a large time scale.}
\label{fig2.fig}
\end{figure}

\begin{figure}[H]
\begin{centering}
%\centerline{\includegraphics[scale=0.6]{bar}}
\vspace*{-4cm}  
\centerline{\includegraphics[scale=0.6]{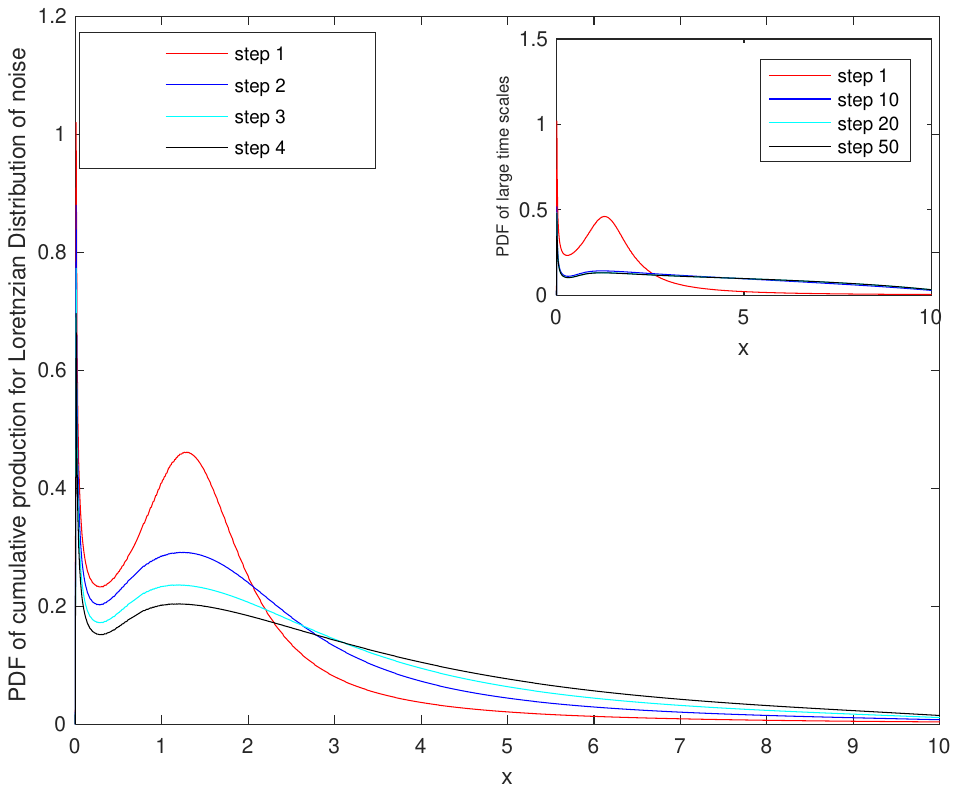}}
\end{centering}
\vspace*{-4cm}  
\caption{Plot of probability distribution functions of $\log Z_t$ of
 Lorentzian distributed noise for different time steps, where $g=0.2$
  and width equal to 1 are chosen. The inset shows curves on a large time scale.}
\label{fig3.fig}
\end{figure}

\subsection{The distribution of volatility} 
The quantity of our interest is the volatility, which is 
by definition the variance of the distribution function $z_{t}-z_{t-1}$.  We observe
\begin{eqnarray}
\Delta{z_{t}}&:=&z_t-z_{t-1}=\log Z_t-\log Z_{t-1}\nonumber\\
&=&-\log\left(1-\frac{ Z_t-Z_{t-1}}{Z_t}\right).
\end{eqnarray}
We define:
\begin{equation}
Y_{t}:=\frac{Z_t}{Z_t-Z_{t-1}}.
\end{equation}

In the next subsection we show $Y_{t}$ is distributed as $Z_{t}$ in Eq.~(\ref{2}) with
$g\to -g$ and $a_{i}\to -a_{i} $, see Eq.~\ref{eqnine}.

From the above expression we get the following result for the distribution
function of the variable $y_{t}=\log{Y_t}$:

%\rho_{\Delta{z_{t}}}(x)=\frac{1}{x}\rho_{z_{t}^{v}}(-\log{x}),
\begin{equation}
\rho_{\Delta{z_t}}(x)=\frac{1}{\exp(x)-1}\rho_{y_{t}}(-\log(1-\exp(-x))),
\label{eqnine}
\end{equation}

where $\rho_{y_{t}}$ satisfies the same recursion as $\rho_{z_{t}}$, after
changing the signs of $g$ and $a_{i}$, as mentioned above.\\

\subsubsection{Derivation of volatility}

Let us now derive Eq. (\ref{eqnine}) in detail:

According to the definition of $Y_t$ we have $Y_t=\frac{Z_{t}}{Q_{t}}$, with
\begin{equation}
Z_t:=\sum_{j=0}^t Q_j,\qquad Q_j:={\rm e}^{gj}{\rm e}^{a_1}...{\rm e}^{a_j},
\end{equation}
where $g$ is a constant, and $a_1$, $a_2$, ... are stochastic i.i.d.~variables
which are distributed according to some distribution function $\rho_a$.  Now
-- luckily -- $Y_{t}$ has the same structure as $Z_t$ if we replace the constant
$g$ and the stochastic variables $a_i$ by $g$ and $-a_i$:
\begin{equation}
Y_{t}:=\frac{Z_t}{Q_t}=\sum_{j=0}^t \frac{Q_j}{Q_t},\qquad 
\frac{Q_j}{Q_t}={\rm e}^{g(j-t)}{\rm e}^{-a_{j+1}}...{\rm e}^{-a_t}.
\end{equation}
Next we define
\begin{equation}
\tilde{g}:=-g,\qquad \tilde a_1:=-a_t,\quad \tilde a_2:=-a_{t-1},\quad ...,\\
\tilde a_t:=-a_{1}
\end{equation}
And indeed
\begin{equation}
\frac{Q_j}{Q_t}={\rm e}^{g (j-t)}{\rm e}^{-a_{j+1}}...{\rm e}^{-a_t}
={\rm e}^{\tilde{g} (t-j)}{\rm e}^{\tilde a_1}...{\rm e}^{\tilde a_{t-j}}.
\end{equation}
Hence
\begin{equation}
Y_t=\sum_{j=0}^t{\rm e}^{\tilde{g} (t-j)}{\rm e}^{\tilde
a_1}...{\rm e}^{\tilde a_{t-j}}=\sum_{j=0}^t{\rm e}^{\tilde{g} j}{\rm
e}^{\tilde a_1}...{\rm e}^{\tilde a_{j}}.
\end{equation}
As mentioned above we find that the distribution function of $y_t:=\log{Y_t}$
corresponds to that of $z_t:=\log{Z_t}$, by taking into account sign changes
of $g$ and $a_i$. Hence it satisfies the recursion relation derived in the subsection 4.2.1.
\begin{equation}
\rho_{y_{t+1}}(x)=\frac 1{1-\exp(-x)}.\\
\rho_{-a}*\rho_{y_t}(\log(\exp(x)-1)+g).
\end{equation}
\bigskip
\begin{figure}[!]
\begin{centering}
%\centerline{\includegraphics[scale=0.6]{bar}}
\vspace*{-4cm}  
\centerline{\includegraphics[scale=0.6]{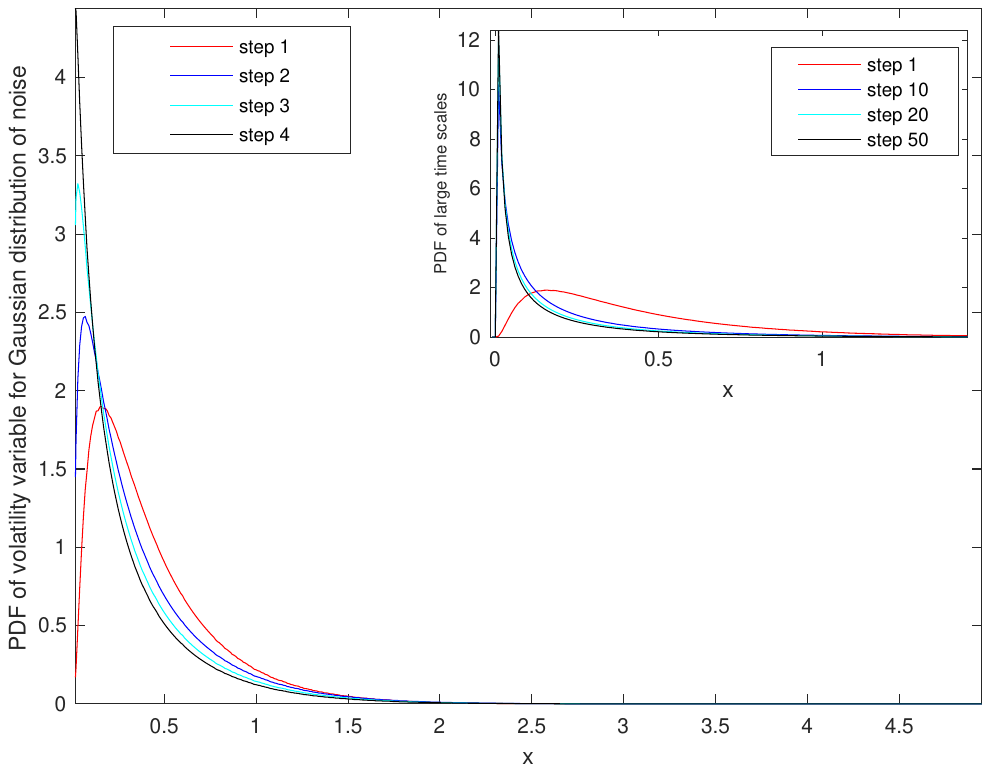}}
\end{centering}
\vspace*{-4cm}  
\caption{Plot of probability distribution functions of the volatility variable
  for the Gaussian case for $\rho_{a}$ and different time steps, where $g=0.2$
  and $\sigma_{a}=1$. The inset is considered for a large time scale.}
\label{fig4.fig}
\end{figure}

Figures $17$ and $18$ illustrate volatility distributions
  $\rho_{\Delta{z_t}}$,  considering for
$\rho_{a}$ normal and Lorentzian distributions. The
 figures show the different behaviors of $\rho_{z_t}(x)$,
with singular (but integrable) characteristics at $x=0$ for
i.e.~$g\geq0$.  The first few time steps show sizable
changes whereas only small changes happen at larger times.

\begin{figure}[h!]
\begin{centering}
%\centerline{\includegraphics[scale=0.6]{bar}}
\vspace*{-4cm}  
\centerline{\includegraphics[scale=0.6]{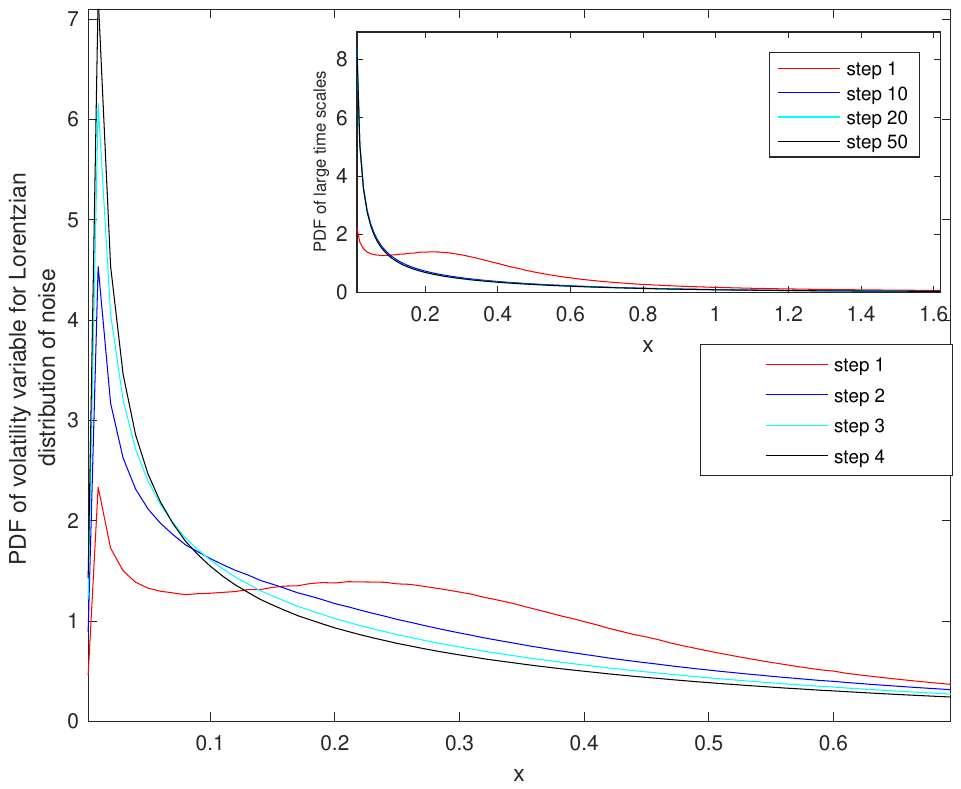}}
\end{centering}
\vspace*{-4cm}  
\caption{Plot of probability distribution functions of the volatility variable
  for the Lorentzian case for $\rho_{a}$ and different time steps, where $g=0.2$
  and width is equal to 1. The inset shows curves on a large time scale.\\}
\end{figure}

\subsubsection{Comparison to the saddle point result}

In order to demonstrate the usefulness of Eq.~(\ref{wichtig}) and (\ref{eqnine}) let us derive
from it  Eq.~(\ref{saddle}), which is the result of the saddle point approximation.

$\rho_a$ is given by a narrow
distribution around $0$ and variance $\sigma_{a}^2$ and correspondingly
$\rho_{y_t}$ is defined by the mean ${\bar y_t}=\log(\sum_{j=0}^t{\rm
  e}^{-jg})$ and ${\sigma_t}^2$.

Due to the additivity feature of the variance under convolution, the narrow
$\rho_{a}*\rho_{y_t}$ has the
variance equal to ${\sigma_t}^2+{\sigma_a}^2$.

Let us use variable transformation in Eq. (\ref{eqnine}).
 Then we get  ($\sigma_t:=\sigma_{y_t}$)

\begin{equation}
\frac{d(\log({\rm e}^x-1)+g)}{dx}{\sigma_{t+1}}=\sqrt{{\sigma_t}^2+{\sigma_a}^2},
\end{equation}
where $x=\bar y_t$. For large times $t \to \infty$
we get 
\begin{equation}
{\sigma_\infty}^2=\frac{1}{{\rm e}^{2g}-1}{\sigma_a}^2.
\end{equation}
Let us now calculate the main result, namely $\sigma_{\Delta{z_{t}}}$, the
volatility for narrow distributions. To this end we need to transform
variables in Eq.~(\ref{wichtig})
\begin{equation}
\sigma_{\Delta{z_{t}}}=({\rm e}^x-1){\sigma_t}.
\end{equation}

For $x$ equal to its maximum we have
\begin{equation}
-\log(1-{\rm e}^{-x})=\bar y_{t}
\end{equation}
and for $t \to \infty $ this amounts to $x=g$.

Finally, we obtain the result for the special, narrow distributed function,
which above was obtained by the saddle point
approximation
\begin{equation}
\sigma_{\Delta{z_{t}}}=\sqrt{\tanh{(\frac{g}{2}})}{\sigma_a}.
\end{equation}

We also compared our numerical results for general values of $\sigma_a^2$ with
the saddle point result obtained in Eq. ($\ref{saddle}$). These analyses show that the
general treatment of the probability distributions and the saddle point
approximation coincide for small variance $\sigma_a^2$.

\begin{figure}
\begin{centering}
%\centerline{\includegraphics[scale=0.6]{bar}}
\vspace*{-4cm}  
\centerline{\includegraphics[scale=0.6]{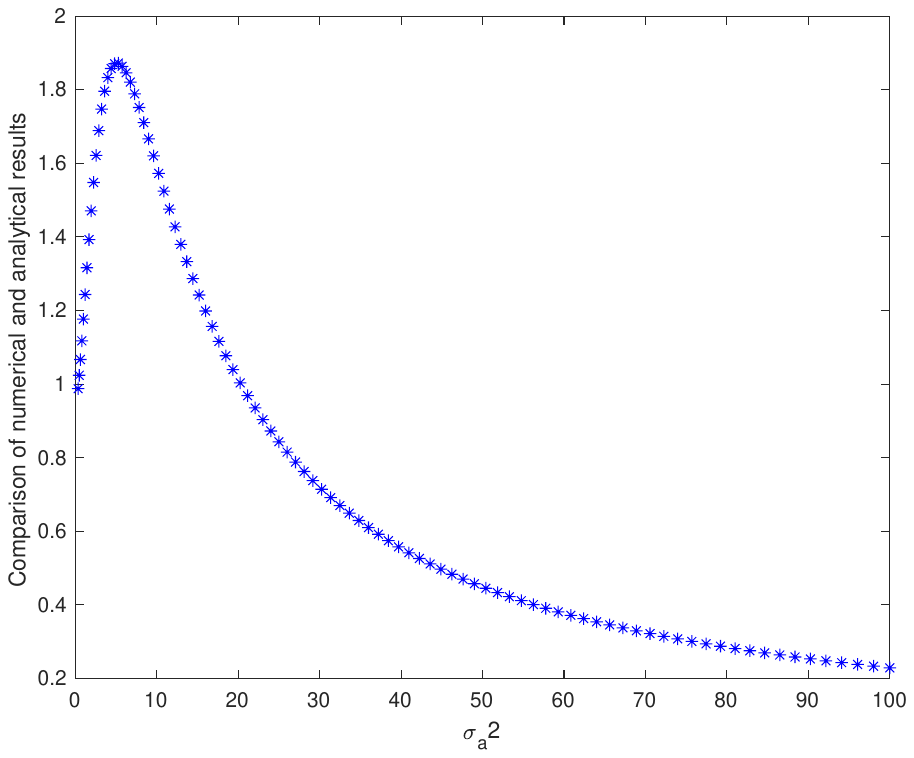}}
\end{centering}
\vspace*{-4cm}  
\caption{Plot of the ratio of the numerically exact value
 of volatility and the saddle point approximation versus 
 various values of the variance ${\sigma_{a}}^2$ of noise,
  and fixed drift, $g=0.1$.}

\end{figure}

Figure 19 shows that for small values of $\sigma_a$ the analytical approximation
and the numerical results coincide whereas for larger $\sigma_a$ we see
sizable deviations. The volatility for intermediate (large) values of
$\sigma_a$ takes larger (smaller) values than the result of the saddle point
approximation.

\newpage

\subsection{Volatility of the process without memory in the noise} 

Here we discuss about a model, which does not possess a memory. 
In general such stochastic models are known as a Markov process,
 in which the probability of an event depends only on the current state. 
Such processes are used in finance and economy to interpret a 
variety of different phenomena such as dynamics of macroeconomics
 and Markov Switching Multifractal models, see e.g.  \cite{Cal}.
 Another goal of the analysis of Markov processes is to make prediction, 
 based on asset prices, taking into account the present value of the asset, regardless of its history.

Let us go back and consider the Gaussian distribution of
 the noise in the cumulative production. Another interesting
  problem is to modify the model and consider a chain without
   memory, i.e. the sum of the past noise in the cumulative
    production is removed. That means we replace

\begin{equation}
y_{i}=gi+a_{1}+ a_{2}... a_{i} \qquad  {\rm by} \qquad  y_{i}=gi+a_{i}.
\end{equation}

In the new model, the first component of $y$ is the 
deterministic or trend component and the second part
 is the stochastic component.
                                                                             
The first model described in the subsection 4.1.1 was 
difference stationary: its first difference has a constant 
mean and variance. It's called integrated of order $1$ or 
$I(1)$. We could show that $\log\,Z$ is $I(1)$ if $\log\,Q$ is $I(1)$,
 which was non-trivial. In statistics and probability theory this
  approach is called unit root test, which tests the stationarity 
  or trend stationarity in the sample. More generally, the integrand
   of the order $d$ or $I(d)$ simply implies that after $d$ times
    differentiation the series will be stationary.

Now let us suppose that $Q$ follows the new model
 without considering the past noise. This new model is called
  trend stationary because if we remove the trend 
  then it is stationary. We do not need to differentiate to
   obtain stationarity, because noise does not accumulate. 
   So it is an integrated of order $0$ or $I(0)$. It is a challenging 
   question in economy whether the integrand is of order
    $1$ or $0$, which leads to the huge literature of the 
     unit root tests. Let us now calculate the volatility for 
     the cumulative production by using the steepest descent
      method for this new model. 

We follow again the same procedure as in subsection 4.1.1,
 with the difference that in this case the process is trend
  stationary as described above.

We compute the expectation value of $z=\log Z$, where
\[
Z_{t}=1+\sum_{i=1}^te^{y_i},
\]
assuming that the distribution of $y_t$ is
\[
P(y_t=Y)=\frac{1}{\sqrt{2\pi \sigma_{a}^2 t}} \exp{\left(-\frac{(Y-g t)^2}{2\sigma_{a}^2t}\right)}.
\]
This leads to the (multiple) integral (the range of
 integration variables $y_i$ are $(-\infty,\infty)$)
\[
\int \log Z\prod_{i=1}^t
\frac{dy_i}{\sqrt{2\pi \sigma_{a}^2 i}}
\exp\left(-\frac{(y_i-g i)^2}{2\sigma_{a}^2i}\right).
\]

In analogy,  we apply saddle point method which assumes that $\sigma_{a}^2\ll 1$.

It is convenient to make change of variables $y_i\rightarrow
\sqrt i y_i$ after which our integral becomes
\[
\int \log Z\prod_{i=1}^t
\frac{dy_i}{\sqrt{2\pi \sigma_{a}^2}}\,
\exp{\left(-\frac{(y_i-g \sqrt i)^2}{2\sigma_{a}^2}\right)},
\]
but $Z$ now is given by
\[
Z(y_1,\ldots,y_t)=1+\sum_{i=1}^te^{\sqrt i\,y_i}.
\]
The saddle point is defined from the system of
equations ($n=1,2,\ldots,t$ and $\partial_n
\equiv \partial/\partial_{y_n}$)
\[
\partial_n \left(\log (\log Z)-\frac{(y_n-g \sqrt n)^2}{2\sigma_{a}^2}\right)
=0,
\]
or
\[
\frac{\partial_nZ}{Z\log Z}-\frac{y_n-g \sqrt n}{\sigma_{a}^2}=0,
\]
which we rewrite as
\[
y_n=g \sqrt n+\sigma_{a}^2\,\frac{\partial_nZ}{Z\log Z}.
\]
Since we are going to make calculations up to order $\sigma_{a}^2$,
it is legitimate to substitute in second term $y_i=g \sqrt i$.
Thus for the saddle point $(y_1^*,\ldots ,y_t^*)$ we get
\[
y_n^*=g \sqrt n+\sigma_{a}^2 \left.\frac{\partial_nZ}{Z\log Z}\right\vert_{y_n=g \sqrt n}+\mathcal{O}(\sigma_{a}^4).
\]
Expanding up to order $\sigma_{a}^2$ for the $\log Z(y_i^*)$ we get
\begin{equation}
\log Z(y_i^*)=\log Z+\sigma_{a}^2
\left.\frac{\sum_{i=1}^t\left(\partial_iZ\right)^2}{Z^2 \log^2Z}\right\vert_{y_i=g \sqrt i}+\mathcal{O}(\sigma_{a}^4).
\end{equation}
The contribution of the Gaussian term:
\begin{equation}
\left.\exp\left(-\frac{(y_i-g \sqrt i)^2}{2\sigma_{a}^2}\right)\right\vert_{y_i=y_i^*}
=1-\left.\frac{\sigma_{a}^2}{2}\frac{\sum_{i=1}^t(\partial_iZ)^2}{Z^2\log^2Z}
\right\vert_{y_i=g \sqrt i}+\mathcal{O}(\sigma_{a}^4).
\end{equation}
Below we should calculate the matrix of quadratic fluctuations.
For second derivatives we get
\[
\partial_n\partial_m\log\log Z=\frac{\partial_n\partial_mZ}{Z\log Z}
-\frac{\partial_nZ\partial_mZ(1+\log Z)}{Z^2\log^2Z}.
\]
Incorporating this (with factor $1/2$ coming from the Taylor's
formula for the second order coefficient) with the initial
Gaussian term, for the matrix of quadratic form we get
\[
G_{n,m}=\frac{1}{2\sigma_{a}^2}\left(
\delta_{n,m}+\sigma_{a}^2\left.\left(-
\frac{\partial_n\partial_mZ}{Z\log Z}
+\frac{(1+\log Z)\partial_nZ\partial_mZ}{Z^2\log^2Z}
\right)\right\vert_{y_i=g\sqrt i}+\mathcal{O}(\sigma_{a}^4)
\right).
\]
Recall now the Gauss formula for the integral
\[
\int \exp \left( -\sum_{n,m=1}^t\xi_n G_{n,m}\xi_m\right)
\prod_{i=1}^td\xi_i=
(\det G)^{-1/2}\pi^{t/2}.
\]
Using
\[
\det G=\exp (tr \log G),
\]
we up to desired order we get
\begin{eqnarray}
&&(2 \sigma_{a}^2)^{-t/2}(\det G)^{-1/2}=\\
&&=1+\left.\frac{\sigma_{a}^2}{2}
\left(\frac{\sum_{i=1}^t\partial_i^2Z}{Z\log Z}
-\frac{\sum_{i=1}^t(\partial_iZ)^2(1+\log Z)}{Z^2\log^2 Z}\right)
\right\vert_{y_i=g \sqrt i}+\mathcal{O}(\sigma_{a}^4).\nonumber
\end{eqnarray}
Multiplying all factors given by Eqs. (114), (115), (116) we finally get
\[
E(\log Z)=\log Z_0+\frac{\sigma_{a}^2}{2}
\left(
\frac{\sum_{i=0}^tie^{i g}}{Z_0}
-\frac{\sum_{i=1}^tie^{2ig}}{Z_0^2}
\right)+\mathcal{O}(\sigma_{a}^4),
\]
where
\[
Z_0\equiv\sum_{i=0}^te^{ig}=\frac{1-e^{g (t+1)}}{1-e^g}.
\]
We can rewrite above expression as
\begin{eqnarray}
E(\log Z)=\log Z_0+\frac{\sigma_{a}^2}{4 Z_0^2}\partial_g(Z_0^2-Z_2)
+\mathcal{O}(\sigma_{a}^4),
\end{eqnarray}
where we have introduced notation
\begin{eqnarray}
Z_2=\sum_{i=0}^te^{2ig}=\frac{1-e^{2g(t+1)}}{1-e^{2g}}.
\label{Z02_def}
\end{eqnarray}
If $g<0$, in large $t\rightarrow \infty$ limit
\[
Z_0\rightarrow \frac{1}{1-e^g}\,\,,\qquad
Z_2\rightarrow \frac{1}{1-e^{2g}},
\]
hence
\[
E(\log Z)|_{t\rightarrow \infty}=
\log \frac{1}{1-e^g}+\frac{\sigma_{a}^2}{2}
\left(
\frac{e^g}{1-e^g}-\frac{e^{2g}}{(1+e^g)^2}
\right)+\mathcal{O}(\sigma_{a}^4).
\]
In the opposite case $g>0$ we'll have
\[
Z_0\rightarrow \frac{e^{g (t+1)}}{e^g -1}\,\, ,\qquad
Z_2\rightarrow \frac{e^{2g (t+1)}}{e^{2g }-1},
\]
which leads to
\begin{eqnarray}
E(\log Z)|_{t\rightarrow \infty}=g-\log \left(e^{g }-1\right)&+&
\frac{ \sigma_{a}^2}{2} \left(\frac{1}{\left(e^{g }+1\right)^2}
+\frac{1}{1-e^{g }}\right)\nonumber\\
&+&t \left(g+\frac{\sigma_{a}^2}{e^{g }+1}
\right)+\mathcal{O}(\sigma_{a}^4).
\end{eqnarray}
We see that drift gets correction and becomes
\begin{eqnarray}
g+\frac{\sigma_{a}^2}{e^{g }+1}.
\end{eqnarray}

Now let us calculate the expectation value $E(\log^2Z)$.

The integral we have to compute is
\[
\int \log^2 Z\prod_{i=1}^t
\frac{dy_i}{\sqrt{2\pi \sigma_{a}^2}}\,
\exp-\frac{(y_i-g \sqrt i)^2}{2\sigma_{a}^2}.
\]

The saddle point equations
\[
\partial_n \left(2 \log (\log Z)-\frac{(y_n-g \sqrt n)^2}{2\sigma_{a}^2}\right)
=0,
\]
or
\[
2\frac{\partial_nZ}{Z\log Z}-\frac{y_n-g \sqrt n}{\sigma_{a}^2}=0,
\]
which gives
\[
y_n=g \sqrt n+2 \sigma_{a}^2\,\frac{\partial_nZ}{Z\log Z}.
\]
As in previous case in second term we can substitute $y_i=g \sqrt i$
and for the saddle point $(y_1^*,\ldots ,y_t^*)$ we get
\[
y_n^*=g \sqrt n+2 \sigma_{a}^2 \left.\frac{\partial_nZ}{Z\log Z}
\right\vert_{y_n=g \sqrt n}+\mathcal{O}(\sigma_{a}^4).
\]
Up to order $\sigma_{a}^2$ expansion for $\log^2 Z(y_i^*)$ :
\begin{equation}
\log Z(y_i^*)=\log^2 Z+4 \sigma_{a}^2
\left.\frac{\sum_{i=1}^t\left(\partial_iZ\right)^2}{Z^2}\right\vert_{y_i=g \sqrt i}+\mathcal{O}(\sigma_{a}^4).
\end{equation}
The contribution of the Gaussian term is
\begin{equation}
\left.\exp\left(-\frac{(y_i-g \sqrt i)^2}{2\sigma_{a}^2}\right)\right\vert_{y_i=y_i^*}
=1-\left.2\sigma_{a}^2\,\frac{\sum_{i=1}^t(\partial_iZ)^2}{Z^2\log^2Z}
\right\vert_{y_i=g \sqrt i}+\mathcal{O}(\sigma_{a}^4).
\end{equation}

The matrix of quadratic fluctuations is
\[
G_{n,m}=\frac{1}{2\sigma_{a}^2}\left(
\delta_{n,m}+2 \sigma_{a}^2\left.\left(
-\frac{\partial_n\partial_mZ}{Z\log Z}
+\frac{(1+\log Z)\partial_nZ\partial_mZ}{Z^2\log^2Z}
\right)\right\vert_{y_i=g\sqrt i}+\mathcal{O}(\sigma_{a}^4)
\right).
\]

Up to desired order we get
\begin{eqnarray}
&&(2 \sigma_{a}^2)^{-t/2}(\det G)^{-1/2}=\\
&&=1+\left.\sigma_{a}^2
\left(\frac{\sum_{i=1}^t\partial_i^2Z}{Z\log Z}
-\frac{\sum_{i=1}^t(\partial_iZ)^2(1+\log Z)}{Z^2\log^2 Z}\right)
\right\vert_{y_i=g \sqrt i}+\mathcal{O}(\sigma_{a}^4).\nonumber
\end{eqnarray}
Multiplying all factors we finally get
\[
E(\log^2 Z)=\log^2 Z_0+\sigma_{a}^2
\left(\frac{(\log Z_0-1)\sum_{i=1}^tie^{2ig}}{Z_0^2}+
\frac{\log Z_0\sum_{i=0}^tie^{i g}}{Z_0}
\right)+\mathcal{O}(\sigma_{a}^4),
\]
or
\begin{eqnarray}
E(\log^2 Z)=\log^2 Z_0+\frac{\sigma_{a}^2}{2}
\left(\frac{(\log Z_0-1)\partial_g Z_2}{Z_0^2}+
\partial_g \log^2 Z_0
\right)+\mathcal{O}(\sigma_{a}^4).
\end{eqnarray}
It remains to calculate the variance
\begin{eqnarray}
E(\log^2 Z)-(E(\log Z))^2=
\frac{\sigma_{a}^2}{2} \,\frac{\partial_g Z_2}{Z_0^2}+\mathcal{O}(\sigma_{a}^4),
\end{eqnarray}
Explicitly we get
\begin{eqnarray}
&&E(\log^2 Z)-(E(\log Z))^2\nonumber\\
&&=\frac{\sigma_{a}^2}{2} \,\frac{e^{2 g }-e^{2 g  (t+1)}+t \left(e^{2 g  (t+2)}
-e^{2 g  (t+1)}\right)}{\left(e^{g }+1\right)^2 \left(e^{g  (t+1)}-1\right)^2}+\mathcal{O}(\sigma_{a}^4).
\end{eqnarray}
If $g>0$ and $t\rightarrow \infty $ we get
\begin{eqnarray}
E(\log^2 Z)-(E(\log Z))^2=
\frac{\sigma_{a}^2}{2} \,\frac{\left(e^{2 g }-1\right) t-1}{\left(e^{g }+1\right)^2}
+\mathcal{O}(\sigma_{a}^4).
\end{eqnarray}
In a similar way we compute the
quantity
\begin{eqnarray}
E(\log Z(t) \log Z(t+1))-E(\log Z(t))E( \log Z(t+1))\nonumber\\
=\sigma_{a}^2\,\frac{\partial_g Z_2(t)}{2 Z_0(t)Z_0(t+1)}+\mathcal{O}(\sigma_{a}^4).
\end{eqnarray}
Remind that (see (\ref{Z02_def}))
\[
Z_0(t)=\frac{1-e^{g(t+1)}}{1-e^{g}}
\,; \qquad Z_2(t)=\frac{1-e^{2g(t+1)}}{1-e^{2g}}.
\]
Incorporating this result with previous ones, for
the variance of $\Delta \log Z$ we get:

\begin{align}
&\mbox{Var}(\Delta \log Z)=\nonumber\\
=&E((\log Z(t+1)-\log Z(t))^2)-(E(\log Z(t+1)-\log Z(t)))^2=\nonumber\\
=&\mbox{Var}(\log Z(t+1))+\mbox{Var}(\log Z(t))\nonumber-2 (E(\log Z(t)\log Z(t+1))\\
&-E(\log Z(t))E(\log Z(t+1)))\nonumber\\
=&\frac{\sigma_{a}^2}{2}\left(\frac{\partial_g Z_2(t)}{Z_0(t)^2}
+\frac{\partial_g Z_2(t+1)}{Z_0(t+1)^2}
-\frac{2\partial_g Z_2(t)}{Z_0(t)Z_0(t+1)}
\right).
\end{align}
As already mentioned for $g>0$ in large $t$ limit
\[
Z_0(t)\approx\frac{e^{g(t+1)}}{e^{g}-1}
\,; \qquad Z_2(t)\approx\frac{e^{2g(t+1)}}{e^{2g}-1},
\]
which leads to
\begin{eqnarray}
\mbox{Var}(\Delta \log Z)=
\frac{\sigma_{a}^2  \left(e^{g }
-1\right)^2 \left(2+e^{g }+2 \left(e^{g }+1\right) t
\right)}{e^{g }\left(e^{g }+1\right)^2}.
\end{eqnarray}

Here as it is visible we see the dependency on $t$
 which implies that the process is not stationary. It is possible
  with the same procedure to check higher order derivatives and
   to analyze the root test of the process. The finding tells us
    that if we consider the production process as a chain without
     memory in the noise, then the  variance is not stationary. In reality the variance should decay and has a finite limit for $t \to \infty$, but in this model the variance goes to infinity for large $t$. The behavior of this model is qualitatively different from that studied in the section 4.2, and also different from empirical findings. Therefore we discard the further investigation of this model. Note that in contrast, the first difference of the model obtained in the section 4.2
       has a constant mean and variance, which is in turn non-trivial
       and important result for economy. \\

 \subsection{Skewness and kurtosis of the production processes}

It is also interesting to calculate the third and fourth 
cumulants (polynomial combinations of the moments)
 of the distribution functions, as the additional knowledge
  characterizes more closely the distribution function and
   quantifies derivations from for instance normal distributions.
    The important characteristics of cumulants is their 
    additive feature, for sums of independent random
     variables. It is notable, that for the Gaussian distribution all cumulants
      of order larger than two are identically zero.

In probability theory and statistics the third cumulant 
or skewness describes asymmetry of the probability
 distribution about its mean. If the skew is negative,
  then the tail on the left side of the probability density
   function is heavier than that on the right side. 
   Conversely, positive skew shows that the tail on the 
   right hand side is heavier than on the left hand side \cite{wik}. 

As mentioned above and according to the definition of the
 skewness, it is obvious
 that the skewness of the normal distribution is zero. 
 But there are often cases where the distributions are
  not symmetric and skewness can describe this asymmetry
   and is given by the following expression:

\begin{equation}
c_{3}=\frac{\langle(x-m)^3\rangle}{\sigma^3},
\end{equation}

 where $m$ denotes the mean and $\sigma^2$ is the variance of the process.
  Skewness plays a crucial role in finance and investing
 strategies. Stock prices and asset returns have mostly
  either positive or negative skew. Knowing which way 
  data skews, an investor can better assess whether it 
  is time to invest or not \cite{invest}. 

In analogy to the concept of skewness, kurtosis
 or the fourth moment also describes the shape of 
 the probability distribution. The concept is related to the tail of the distribution.
 Kurtosis also quantifies the deviation of a distribution 
from Gaussian, i.e it can be understood as a measure
 of the distance between a given distribution and a normal 
 distribution and is given by the following formula:

\begin{equation}
c_{4}=\frac{\langle(x-m)^4\rangle}{\sigma^4}-3.
\end{equation}

We derive here the third and fourth moments of the
 distribution function of the production process. As
  discussed above the distribution function of $\rho_{a}$
 has the mean $c_{1}=0$, variance $c_{2}=\sigma_{a}^2$ 
 and skewness $c_{3}=0$. Similarly we define for the
  distribution function of $\rho_{y_{t}}$ the mean, variance
   and skewness as follows: $c_{1}=b_{t}$, $c_{2}=\sigma_{t}^2$, $c_{3}=s_{t}$.

For the mean, variance, and skewness of the convolution
 $\rho_{a}*\rho_{y_{t}}$ of the distributions, we have 
 consequently the following expressions: $c_{1}=b_{t}$, $c_{2}=\sigma_{a}^2+\sigma_{t}^2$ and  $c_{3}=s_{t}$ and similarly for the kurtosis we have
   $c_{4}(\rho_{a}*\rho_{y_{t}})=c_{4}(\rho_{a})+c_{4}(\rho_{y_{t}})$.
    Here we use the nice additivity feature of cumulants under convolution.

The main problem to solve is to understand how skewness
 and variance behave under the variable transformation. For
  the latter one we get the following expression

\begin{equation}
\sqrt{\bar{m_{2}}}=|f'(k)|\sqrt{m_{2}},
\end{equation}

where, $m_{2}$ is the second moment, $k=-\log(1-e^{-g})$ and $f(x)=\log(exp(x)-1)+g$, $f(-\log(1-e^{-g}))=-\log(1-e^{-g})$.

Furthermore, using the variable transformation for
 the kurtosis of $\rho_{y_\infty}$ and $m=-\log(1-e^{-g})$ we obtain the following

\begin{equation}
c_{4}(\rho_{y_\infty})=\frac{e^{-4g}}{1-e^{-4g}}c_{4}(\rho_{a}),
\end{equation}

where $c_{4}(\rho_{a})\simeq-3\sigma_{a}^4 $ 

Hence we get

\begin{equation}
c_{4}(\rho_{\Delta{z}})=\frac{g^3}{4}c_{4}(\rho_{a})=-3\sigma_{a}^2\frac{g^3}{4}.
\end{equation}

In analogy for the skewness of the distribution we obtain

\begin{align}
c_{3}(\rho_{\Delta{z}})={(e^g-1)^3}3{\sigma_{a}^4}\left[\frac{1}{(1-e^{-2g})^2}-\frac{(1-e^{-g})^4}{(1-e^{-4g})^2}\right]=\nonumber\\
3\sigma_{a}^4e^{4g}\frac{e^g-1}{(e^g+1)^2}\left[1-\frac{(e^g-1)^4}{(e^2g+1)^2} \right]=\frac{3}{4}g\sigma_{a}^4.
\end{align}

The analytical result shows that the skewness
 for our case is positive, which coincides with results
  shown in plots, because the tails of the plots (see, e.g. figures 14 and 15),  are 
  heavier on the right hand side.
  
The saddle point approximation is an appropriate method when we deal
 with a distribution function like the Gaussian. But as discussed above,
  in general distribution functions have non-vanishing skewness and 
  kurtisis. In this section the goal was to calculate such corrections to 
  the Gaussian and the saddle point approximation.\\

\subsection{Aggregation of capital stocks by depreciation} 

Most countries estimate capital stocks from investment, 
which is drawn upon the perpetual inventory method. In
 business and industrial activities it is important to know
  how the price of a particular commodity declines over 
  time passing by. The amount by which
     the capital stock decreases
       in the period of time is called depreciation rate. Otherwise said, 
        as years pass by, the value of stock price
   usually falls and the rate of depreciation is the descriptor
    for this change \cite{Dep}. 

This phenomenon occurs because most capital assets have
 finite lifetime, due to some reasons such as repair costs 
 are rising, so keeping them is not economical; there is an 
 immense technological progress in the market, which leads 
 to an innovation of the market, and finally some assets are 
 simply destroyed in the period of time. 

In calculating depreciation, researchers developed  a number
 of methods, aimed to estimate how asset values decline 
 during their lifetimes. In the vast majority of alternatives, 
 mostly the following three depreciation methods are used: 
 straight-line (if asset prices are falling by a constant amount each year), 
 geometric (if asset prices are falling by a constant rate each year)
  and sum-of-the-years-digits. These methods presume, 
  that the systematic decline in the value of an asset over
   a period of time is based only on the initial value of the 
    asset and its expected lifetime \cite{Berle, Gold}. 

The geometric depreciation method is applicable for 
 the cases,  where the ability for producing capital services
  for the given assets reduces by the largest amount in
   the first year. Thus the geometric depreciation is not a
    useful method for the cases, where assets will have an 
    increasing amount of preservation as they get older.

The sum-of-the-years-digits depreciation method is also
 based on the concept that the largest fall in efficiency 
 is at the beginning of the asset's service life. The main 
 difference between these two methods is that there is 
 a huge change on the initial and final asset price in the 
 latter method. In contrast to these two methods, 
 the straight-line depreciation method shows a linear
  decline in an age of efficiency, i.e. the asset price falls
   by the same absolute amount for each period.

We are going to use some of the techniques we introduced
 for the study of cumulative production, aimed 
 to construct the relationship between the capital and the investment.  We denote the
  capital of the stock by $Z_{t}$ at time $t$ and gross 
  investment at any time by $Q_{t}$. By assuming
   geometric depreciation at a constant rate $d$ we can 
   rewrite the capital stock at time $Z_{t+1}$ as a function
    of its previous value and the investment as follows

\begin{equation}
Z_{t+1}=(1-d) Z_{t}+ Q_{t+1},\ 0<d<1.
\end{equation}

The strategy is to reduce the new problem to that one studied in section 4.2.

In previous model we had:

\begin{equation}
Z_t:=\sum_{j=0}^t Q_j,\qquad Q_{t}:={\rm e}^{g t}{\rm e}^{a_1}...{\rm e}^{a_t}.
\end{equation}

Otherwise said, in the old model we have:

\begin{equation}
Z_{t+1}= Z_{t}+ Q_{t+1}, 
\end{equation}

where $d=0$ and the solution for volatility is also known: 

\begin{equation}
\mbox{Var}(\Delta \log Z)=
\sigma_a^2 \tanh \left(\frac{g}{2}\right)+\mathcal{O}(\sigma_{a}^4).
\end{equation}

Now it is desirable to obtain the solution for (137). 

For this purpose we replace $Z_{t}:=(1-d)^t\hat{Z}_{t}$. Let us
 substitute this in Eq. (137), so we get 
\begin{equation}
(1-d)^{(t+1)}\hat{Z}_{t+1}=(1-d)^{(t+1)}\hat{Z}_{t}+Q_{t+1}.
\end{equation}

By dividing the above equation by $(1-d)^{(t+1)}$, we obtain

\begin{equation}
\hat{Z}_{t+1}=\hat{Z}_{t}+(1-d)^{(-t-1)}Q_{t+1}.
\end{equation}

Let us denote

\begin{equation}
\hat{Q}_{t+1}:= (1-d)^{-t-1}Q_{t+1}.
\end{equation}

Thus we have 

\begin{equation}
\hat{Z}_{t+1}:= \hat{Z}_{t}+\hat{Q}_{t+1},
\end{equation}

our old formula, namely Eq. (139).

From Eq. (142) we can easily read that 

\begin{equation}
\hat{Q}_{t}= (1-d)^{-t}Q_{t}
\end{equation}

Let us substitute Eq. (138) in (143), so we get

\begin{equation}
\hat{Q}_{t}=(1-d)^{-t}{\rm e}^{g t}{\rm e}^{a_1}...{\rm e}^{a_t}=\\e^{(g-\ln(1-d)t)}{\rm e}^{a_1}...{\rm e}^{a_t}={\rm e}^{\tilde{g}t}{{\rm e}^{a_{1}}...{\rm e}^{a_{t}}},
\end{equation}

where $\tilde{g}:=g-\ln(1-d)$.  We see that $\hat{Z}_{t}$ 
satisfies the old equation with $Q \to \hat{Q}_{t}$.

So, our main analytical result, for the volatility of the cumulative
 capital, for $g>0$ in the large $t$ limit is
\begin{equation}
\mbox{Var}(\Delta \log Z)=
\sigma_a^2 \tanh \left(\frac{\tilde{g}}{2}\right)+\mathcal{O}(\sigma_{a}^4),
\end{equation}
where $\tilde{g}:=g-\ln(1-d)$.

If we consider the case $|d|\ll1$, we get $\tilde{g}:=g+d$. Hence, for this case we have 
\begin{equation}
\mbox{Var}(\Delta \log Z)=
\sigma_a^2 \tanh \left(\frac{g+d}{2}\right)+\mathcal{O}(\sigma_{a}^4).
\end{equation}

We find counterintuitively, that $Z_{t}$ grows less with $t$ for $0<d<1$ than in section 4.2, but the variance of $\Delta\log{Z}$ is larger.

The depreciation stems from the companies investing activities and strategies and has an influence on the asset value. Furthermore, in order to be able to calculate the current capital stock, one needs the dataset of the investment data, information about the initial capital stock and the rate of depreciation of the current capital stock.  

In fact here we showed how to reduce the more general 
problem to the special case we treated before in section 4.2, by deriving the
 full solution and performing the calculations for a completely
  new area and problem. Eq. (148) shows how the volatility of 
  aggregated capital stock is related to the volatility of the investment 
  and depreciation, by opening the research of the predictive power
   of capital through an investment.

\newpage
 
 \topskip0pt
\vspace*{\fill}
\thispagestyle{empty}
\begin{center}
\section*{\textbf{\LARGE{}{Part V}}}
\section*{\textbf{\huge{}{Exact probability distribution functions of Parrondo's games}}}
\end{center}
\vspace*{\fill}
\vspace*{10cm}
\newpage

\section{Exact probability distribution of Parrondo's games}

 The Parrondo's games are related to Brownian ratchets \cite{pr92}-\cite{gn13} with applications in physics, biology, engineering and financial risk.
 
 Phrased in simple words, the phenomenon describes games, each with higher
probability of losing than winning. The Parrondo's paradox
 suggests that it is possible to obtain the winning strategy
  after playing the games alternately. Otherwise said, 
 an agent tosses biased coins using one of two strategies
(games), and both strategies are losing. In some cases a random combination of
the losing games is a winning game.

  The phenomenon is
fundamentally related to portfolio optimization \cite{so15}, and
corresponds to the ``volatility pumping'' strategy in portfolio
optimization. For a two-asset portfolio one half of the capital is kept in the
first asset, the other half in the second asset with high volatility
\cite{vo}. Applications have also been linked to quantum models \cite{quan1, quan2}, where the classical coin toss is replaced by the measurement of a qubit. One of the studies is by Meyer et al. \cite{meyer} in which they show that Parrondo's paradox can be modelled by probabilistic lattice gas automata. {Their work introduces a quantum analogue of the ratcheting mechanism seen in the original Parrondo's game, with possible applications in quantum computing. Further investigations on quantum Parrondo's games \cite{q1}-\cite{q5}  yielded other variants, in the process shedding light on the roles of entanglement and coherence on game outcomes.} Almeida et al. have considered how two chaotic systems can give rise to order, in the form of quadratic maps. This is related to Parrondo's paradox, having a lose + lose = win situation \cite{almeida}. {Other phenomena like the presence of stable states within chaos \cite{danca, chau} have been understood under the framework of Parrondo's games.} The paradox has also been considered in reliability theory. Starting with subunits of a system being less reliable than the subunits of another system, Crescenzo has shown that by randomly choosing units from these two systems can allow one to obtain a system that is more reliable than the initial two \cite{crescenzo}. 

In evolutionary biology, Wolf et al. have found that if different environmental states are selected for different cell states, and if cells are unlikely to sense environmental transitions or are subject to long signal transduction delays relative to the time-scale of environmental change, then a time-varying environment can select for random phase variation (RPV) \cite{wolf}. They noted that the success of RPV can be understood as a variant of Parrondo's paradox \cite{wolf}, in which random alternations between losing strategies (in this case, cells that choose the wrong phase variation or sequence of variations) produce a winning strategy. Separately, the evolution of less accurate sensors has been modelled and explained in terms of the paradox \cite{cheong1}. In ecology, the periodic alternation of certain organisms between nomadic and colonial behaviors has also been suggested as a manifestation of the paradox \cite{cheong2}. Recently, there is an intriguing finding of Parrondo-like phenomena in a Bayesian approach to the modelling of the work by a jury \cite{gu16}: the unanimous decision of its members has a low confidence. The developments of Parrondo's paradox have cut across many disciplines, with a wide range of possible applications.  

Another interesting variant of Parrondo's paradox is the Allison mixture \cite{ab14}-\cite{ad09a}, where random mixing of two random sequences creates autocorrelation \cite{ad09a}. The Allison mixture has some resemblance to the thermodynamic picture of the Feynman-Smoluchowski ratchet: at equilibrium there is detailed balance, time-reversibility, and no net displacement; whereas out of equilibrium detailed balance is broken, with time-irreversibily leading to net displacement \cite{ad09a, ad10a}. {It is worth noting that the Allison mixture process mixes two random sequences in a time-irreversible way \textit{only} when symmetry in the governing transition probabilities is broken. Here we refer to \cite{ab10, ad10a, Rub} for more information on the Allison mixture.}  There have been discussions to link the Allison mixture to applications in encryption and optimization of file compression \cite{ab14}.

After this introduction part let us begin with the explanation of the process by focusing on the probability of capital and variances of distributions: In the case of Brownian ratchets, a particle
moves in a potential, which randomly changes between two options. For each
there is a detailed balance condition. However, for random switches between
the two potentials, there is on average a directed motion.

%Parrondo's games have been considered either on the one dimensional axis with
%some periodic potential, or by looking at the time dependent version of the
%game parameters.

%For the first case the
The state of the system is characterized by the current value of the capital
$X$, and the choice of the strategy. $X$ is defined on a
  one-dimensional axis with discrete points, a ``chain''. For the study
   of Parrondo's paradox we consider the probability of a position corresponding to the probability of capital. In analogy to
  ratchets, there may be a periodicity $M$ in the rules, how capital $X$ can
  increase or decrease. This version corresponds to a
  particle moving on a ladder geometry with several rungs \cite{sa12}.
%% The rules of the game depend on
%% Mod$(X,M)$, where $M$ is the period of the ``potential''.
Originally $M=3$ games were considered, then $M=2$ versions of Parrondo's
games were constructed \cite{ab02}, \cite{as05}. For the history dependent
versions of the game the current rules of the game depend on the past, whether
there was
%%% a
growth of capital in the previous rounds or not.

{As the variances of distributions (volatilities) are important in economics,
  we calculated the variance for the history independent case with specific
  parameters in \cite{sa16}. And vice versa, for obtaining the
  unknown parameters of a model that describes empirical data, the complete
  distributions have to be available. This is the problem we address here by
  applying a Fourier transform technique to solve exactly the probability
  distribution function. This approach allows for an efficient calculation of
  the long time asymptotics from saddle point contributions. We discover that
  under certain conditions
%AK
sub-leading saddle points become degenerate in absolute value with the leading
saddle point. Still, the degenerate saddle points differ in phases which leads
to strong fluctuations.

%We apply this method to the random walks on a {\color{red} ladder, a strip of
%  1-dimensional discrete sequences of states (chains) } \cite{sa12}, which is
We apply this method to random walks on chains and ladders
  corresponding to capital dependent Parrondo's models. We calculate the
  entire probability distribution for the capital, then solve the same problem
  for history dependent games. 
  
 In passing we revisit the grounds of Parrondo's paradox. We calculate from our efficient formulas for the capital growth examples of two losing strategies that jointly but randomly applied yield a winning strategy.\\ 

\subsection{A biased discrete space and time random walk}

As an illustration let us consider the discrete time random walk on
%AK
a chain, where the probability of right and left jumps are $p$ and $q$,
respectively. We can write the master equation for the probability $P(n,t)$
at position $n$ after $t$ steps:
\begin{eqnarray}
&&P(n,t+1)=\barp P(n-1,t)+q P(n+1,t)+(1-\barp-q)P(n,t).
\end{eqnarray}
\label{e1}
%%% We have
The initial distribution is $P(n,0)=\delta_{n,0}$.

For the motion on the infinite axis we can always write a Fourier
transform like
\begin{eqnarray}
\label{e2} P(n,t)&=&\int_{-\pi}^{\pi}dk \e^{\i kn}\bar
P(k,t),\nonumber\\
\bar P(k,t)&=&\frac{1}{2 \pi}\sum_n P(n,t)\e^{-\i kn}.
\end{eqnarray}
%%% Let the particle start at $n=0$, thus
For the initial distribution the Fourier transform is $\bar P(k,0)={1}/{2\pi}$.

By considering the recursion relation for $\bar P(k,t)$, 
for the link hand side of the Eq. (178) we have}:
 
 \begin{equation}
P(n,t+1)=\int_{-\pi}^{\pi}dke^{ikn}\bar P(k,t+1)\nonumber,
 \end{equation}
 
and the right hand side is equal to: 
 
\begin{align}
 pP(n-1,t)+qP(n+1,t)+(1-p-q)P(n,t)=\nonumber\\
 =\int_{-\pi}^{\pi}dk(pe^{-ik}+qe^{ik}+1-p-q)e^{ikn}\bar P(k,t)\nonumber.
\end{align}
 
Here we used: 
 
 \begin{equation}
 P(n-1,t)=\int_{-\pi}^{\pi}dk\underbrace{e^{ik(n-1)}}_{e^{-ik}.e^{ikn}}\bar P(k,t)\nonumber.
 \end{equation}

Eq.~(149) transforms into
\begin{eqnarray}
\label{e3} \bar P(k,t+1)= \left[p\e^{-\i k}+q\e^{\i k}+(1-p-q)\right]\bar P(k,t).
\end{eqnarray}

Hence we get the following solution:

\begin{equation}
\tilde P(k,t)=(pe^{-ik}+qe^{ik}+1-p-q)^t\bar P(k,0)\nonumber.
\end{equation}

We obtain:
\begin{eqnarray}
\label{e4} \bar P(k,t)=\lambda^t(k)\cdot\bar P(k,0),\nonumber\\
\lambda(k)= \left[p\e^{-\i k}+q\e^{\i k}+(1-p-q)\right].
\end{eqnarray}
As $\lambda$ is a linear polynomial in $\e^{\i k}$ and $\e^{-\i k}$,
$\lambda^t$ is a polynomial with
%%% powers from $\e^{-tik}$ to $\e^{tik}$.
monomials ranging from $\e^{-t\i k}$ to $\e^{t\i k}$.

%%% We can write the solution as
The Fourier transform from momentum to spatial representation yields
\begin{eqnarray}
\label{e5} P(n,t)=
%\frac{1}{2 \pi}
\int_{-\pi}^{\pi} dk \e^{tV(\i k)+\i kn} \bar P(k,0),\nonumber\\
 V(\kappa):=\ln \left[p \e^{-\kappa}+q \e^{\kappa}+(1-p-q)\right],
\end{eqnarray}
where we defined the function $V(\kappa)$. Note that $V$ is not a potential and for the large time we only need the expansion to the second order. Using that $\lambda^t$ has a
finite expansion in powers of $\e^{\i k}$ we obtain with $\bar
P(k,0)={1}/{2\pi}$
\begin{eqnarray}
\label{e6} P(n,t)=\frac{1}{2 t}\sum_{m=1}^{2t}
\e^{tV(\i m\pi/t)+\i mn\pi/t}.
\end{eqnarray}

{Also for the study of more involved models, we will use both representations
  (\ref{e5}), (\ref{e6}).}

Note that Eqs.~(\ref{e5}) and (\ref{e6}) are exact for any $t$ and $n$. By use
of the saddle point approximation we derive the large $t$ and $n=xt$
asymptotics.  We are allowed to move the integration contour because of the
analytic dependence of the integrand on $k$ resulting in ($\kappa=\i k$)
\begin{eqnarray}
\label{e7} P(n,t)=\frac{\exp[tu(x)]}{\sqrt{2 \pi t V''(\kappa)}},
\end{eqnarray}
and
\begin{eqnarray}
x=-V'(\kappa),\quad u(x)=V(\kappa)+\kappa x\nonumber.
\end{eqnarray}
As $u(x)$ is the Legendre transform of $-V(\kappa)$, we also have
$V''(k)=-1/u''(x)$ and hence
\begin{equation*}
P(n,t)=\frac{1}{\sqrt{2 \pi t}}\exp\left[tu(x)+1/2\log|u''(x)|\right].
\end{equation*}
Let us assume an expansion for $V(\kappa)$
\begin{eqnarray}
\label{e8}V(\kappa)=-r\kappa+K \kappa^2/2+\mathcal{O}(\kappa^3).
\end{eqnarray}

In general the expression  is not Gaussian, but for the large time limit it approaches a Gaussian.

 In fact the expression (153) yields 

\begin{equation}
r=p-q\nonumber
\end{equation}

and 

\begin{equation}
K=q+p-(p-q)^2.\nonumber
\end{equation}

From (156) we obtain the Legendre transform

\begin{equation}
u(x)=-\frac{(x-r)^2}{2K}+\mathcal{O}((x-r)^3),\nonumber
\end{equation}

where the high order terms vanish in case of vanishing high order terms in (156). In the long time limit, these higher order terms can be neglected and a Gaussian distribution evolves. Higher order terms in $V(\kappa)$, namely $\mathcal{O}(\kappa^3)$ leads to higher order terms in $u(x)$.

For the case $p+q=1$ we get a nice expression: 

\begin{equation}
u(x)=-\frac{1+x}{2}\ln\frac{1+x}{2p}-\frac{1-x}{2}\ln\frac{1-x}{2q}.\nonumber
\end{equation}

It then follows that
\begin{eqnarray}
\label{e9} \langle n\rangle=rt,\nonumber\\
\langle(n-\langle n\rangle)^2\rangle=Kt.
\end{eqnarray}

In this scalar case we do not observe any Parrondo effect.
 Note that the rates are given by $r=p_i-q_i$. If $p_1$ and $p_2$ are losing,
  $p_1-q_1<0$ and $p_2-q_2<0$, then also $\frac{p_1+p_2}{2}- \frac{q_1+q_2}{2}<0$, 
  which is the rate for the combined strategy with $p=\frac{p_1+p_2}{2}$ and $q=\frac{q_1+q_2}{2}$.

\subsection{Random walks with periodicity}

%\subsection{The general case}

Consider the case of a random walk on an axis, using rules
with period $M$. We divide the entire $x$ axis in intervals of length $M$,
$[(n-1)M,nM[$, and label points by $(n,l)$ where $l=\rm{Mod} (X,M)$. We
    represent the sets of $p_X$ (the discrete probability distribution of the
    capital value) by $P_l(n,t), 0\le l< M$. The integer $t$ represents
    time.
%AK
This bookkeeping allows us to consider the geometry with periodicity $M$ as a
multi-rung ladder or as a chain of unit cells containing $M$ points.

We study the following master equation \cite{mo00}
\begin{eqnarray}
\label{e10}
P_l(n,t+1)&=&p_{l_-}P_{l_-}(\hat n,t)+q_{l_+}P_{l_+}(\bar
n,t)+(1-p_{l}-q_{l})P_{l}(n,t),
\end{eqnarray}
where $l_-={\rm Mod}(l-1,M),$ ${l_+={\rm Mod}(l+1,M)}$, $\hat n=n$ for $l-1\ge
0$ and $\hat n=n-1$ for $l-1< 0$; $\bar n=n$ for $l+1<M$ and $\bar n=n+1$ for
$l+1> M$.  Thus the model is characterized by the parameters $p_l,q_l$, where
$p_l$ and $q_l$ are the probabilities to win and lose for capital $X$ with
$l={\rm Mod} (X,M)$. The model comes from the literature and the goal is 
to analyze the distribution function of this process as we did in section 5.1.

Again we consider the Fourier transform
\begin{eqnarray}
\label{e11}
P_l(n,t)=
%v_l(k) P(n,t)=
\int_{-\pi}^{\pi}dk \e^{\i kn}\bar P_l(k,t),
 \end{eqnarray}
and obtain
 \begin{align}
 \label{e12}
\bar P_l(k,t+1)=p_{l_{-}}\e^{\i k(\hat n-n)}\bar P_{l_{-}}(k,t)
+q_{l_+}\e^{\i k(\bar n-n)} P_{l_{+}}(k,t)+
(1-(p_{l}+q_{l}))\bar P_l(k,t)\nonumber\\
\equiv \sum_{m=0}^{M-1} \hat Q_{lm}(\i k)\bar P_m(k,t).
 \end{align}
Using the eigenvalues and eigenvectors $\lambda_m(\kappa),v_{ml}(\kappa)$
($m=0,...,M-1$), of the matrix $\hat Q(\kappa)$,
% \begin{eqnarray}
% \label{e13}
% \end{eqnarray}
we find
\begin{eqnarray}
\label{e13} \bar P_l(n,t)=
\int_{-\pi}^{\pi}dk\e^{\i kn}\sum_mc_m\exp[tV_m]v_{ml},
\end{eqnarray}
where $V_m(\kappa):=\ln (\lambda_m(\kappa))$. The factors $c_m(\kappa)$ are
determined by the initial distribution.
{For using Eq.~(\ref{e7}), we choose as $V(\kappa)$ the eigenvalue function
  $V_m(\kappa)$ with the largest saddle point. For generic
%  parameters. Later we encounter the possibility for different sub-phases in
%  our model, related to the existence of different eigenvalues degenerate in
  parameters and close to the maximum of the distribution $u(x)$, the choice
  is unique. Saddle points with smaller value do not contribute to the large
  time asymptotics. However, we will encounter the possibility of several
  eigenvalues degenerate in absolute value.}
%AK I do not know the ``optimal coding theory''. That is why I commented out
%the next lines.
%
%The large deviation (decoding error) probability in optimal coding theory
%\cite{gal} is similar to our case and has several sub-phases.
To compare our results for the {gain/loss} rate
with the formulas in \cite{ab02} we have to multiply the rate $r$ in
Eq.~(\ref{e9}) for the case of the multi-rung ladder
%several chains
with a factor $M$, as one step
in $n$ in our approach equals $M$ ordinary steps.

The periodicity reflects the capital independent strategy of portfolio
 management. Ladder geometries with rungs that contain $p>1$ 
 points represent invested capital in one out of $p$ different stocks. 
 Motions along the ladder represent gain or loss, motions along rungs
  represent change of the composition of the portfolio.\\

\subsection{The eigenvalues $\pm 1$}
Next we investigate more closely the case of zero probability for holding the
capital at the current value, i.e.~$p_{l}+q_{l}=1$ for all $l$ in (\ref{e12}).

Consider first the case of odd $M$: $\hat Q(0)$ has one eigenvalue $+1$ with
left eigenstate $(1,1,1,1...)$, and $\hat Q(\pi \i)$ has one eigenvalue $-1$
with left eigenstate $(1,-1,1,-1...)$. Hence, in Fourier representation the
two ``momenta'' $\kappa=0$ and $\kappa=\pi \i$ contribute to the asymptotic
behaviour.

Let us now consider the matrices $\hat Q(\kappa)$ and $\hat Q(\kappa+\pi \i)$
for arbitrary $\kappa$. It is easy to see that the spectra are simply
related. Let $(x_0,x_1,x_2,...)^T$ be a right eigenvector of $\hat Q(\kappa)$
with eigenvalue $\lambda(\kappa)$, then $(x_0,-x_1,x_2,...)^T$ is a right
eigenvector of $\hat Q(\kappa+\pi \i)$ with eigenvalue $-\lambda(\kappa)$.

Let us assume an expansion like Eq.~(156) for the leading $V(\kappa)$ near
$\kappa=0$, then
\begin{eqnarray}
\label{e14}V(\pi +\kappa)=\pi \i+ r\kappa+K \kappa^2/2.
\end{eqnarray}
Let $v^+$ and $v^-$ be the right eigenstates of $\hat Q(0)$ and $\hat Q(\pi \i)$
with eigenvalues $+1$ and $-1$. There are constants $\alpha$ and $\beta$ such
that
%
%Assuming that there is a projection $\alpha |+>$ for the $\hat
%\hat Q(0)$ and $\beta |->$ for the $\hat \hat Q(\pi)$, we obtain
%
%
%\begin{eqnarray}
%\label{e15} P_l(n,t)=\frac{v_l^+\alpha}{\sqrt{2Kt
%\pi}}\e^{-\frac{(n-rt)^2}{2Kt}}+(-1)^n\frac{v_l^-\beta}{\sqrt{2\bar
%Kt \pi}}\e^{-\frac{(n-\bar rt)^2}{2\bar Kt}}
%\end{eqnarray}
%We verified that there are oscillations for the odd M, which is
%possible for $\bar r=r,\bar K=K$.
%
\begin{eqnarray}
\label{e16} P_l(n,t)=\frac{\alpha v_l^++(-1)^{n+t}\beta v^-_l}{\sqrt{2\pi Kt
}}\e^{-\frac{(n-rt)^2}{2Kt}}.
\end{eqnarray}
We see oscillations caused by the rapid sign change of the second term. The
coefficients $\alpha$ and $\beta$ are determined by the initial probability
distribution. If this was peaked at $n=l=0$ then $\alpha=\beta$
(with $v^+$ and $v^-$ related as pointed out above). In this case $P_l(n,t)$
is non-zero (zero) for even (odd) $l+n+t$.

%and for the odd n
%\begin{eqnarray}
%\label{e17} P_l(n,t)=\frac{v_l^+\alpha-v^+_l\beta}{\sqrt{2Kt
%\pi}}\e^{-\frac{(n-rt)^2}{2Kt}}
%\end{eqnarray}

Consider now the case of even $M$. Here, $\hat Q(0)$ has one eigenvalue 1 with
left eigenstate $(1,1,1,1...)$ and one eigenvalue $-1$ with left eigenstate
$(1,-1,1,-1...)$. Hence, only the ``momentum'' $\kappa=0$ contributes in the
Fourier representation to the asymptotic behaviour, but with two eigenvalues.
Let $(x_0,x_1,x_2,...)^T$ be a right eigenvector of $\hat Q(\kappa)$ with
eigenvalue $\lambda(\kappa)$, then $(x_0,-x_1,x_2,...)^T$ is also a right
eigenvector of $\hat Q(\kappa)$, but with eigenvalue $-\lambda(\kappa)$. Now
we find similar to the case above
\begin{eqnarray}
\label{e16} P_l(n,t)=\frac{\alpha v_l^++(-1)^{t}\beta v^-_l}{\sqrt{2\pi Kt
}}\e^{-\frac{(n-rt)^2}{2Kt}}.
\end{eqnarray}
Note that the oscillating factor does not depend on $n$. For a probability
initially peaked at $n=l=0$ we find $P_l(n,t)$ is non-zero (zero) for
even (odd) $l+t$.

The findings in both cases, odd $M$ and even $M$, can however be summarized:
$P_l(n,t)$ is non-zero (zero) for even (odd) $l+n M+t$.

It is quite interesting to consider the quantity
\begin{eqnarray}
\label{e18} \hat P(n,t)=\sum_{l=0}^{M-1}P_l(n,t).
\end{eqnarray}
It shows non-zero oscillations in dependence on $n$ and $t$ for odd $M$. Such
oscillations do not exist for even $M$. The reason for this is easily
understood: the term entering $\hat P(n,t)$ with a $(-1)^t$ factor is
$\sum_lv_l^-$. This is the scalar product of $(1,1,1,1...)$ with
$(x_0,-x_1,x_2,-x_3,...)^T$ which are the left and right eigenvectors of $\hat
Q(0)$ with different eigenvalues $+1$ and $-1$, and hence this product must be
zero.\\

\subsection{Expressions for the capital growth rates}

Consider the capital depending Parrondo's game with $p_1,...,p_M$
for the winning probabilities and periodicity $M$.
%%% for the corresponding Mod$(X,M)$.
We find the corresponding $\hat Q$ matrix
\begin{equation}
\hat Q(\kappa)=\begin{pmatrix}
0&{q_2}&...&p_{M}\e^{-\kappa}\\
{p_1}&{0}&.&.\\
.&{p_2}&.&{q_M}\\
q_1\e^\kappa&{0}&...&0
\end{pmatrix}.
\end{equation}
Applying the method of \cite{ab02} gives
 \begin{eqnarray}
 \label{c1}
r=\frac{\sum_i(p_i-q_i)x_i}{\sum_ix_i}.
 \end{eqnarray}

The goal of the game is moving the capital into positive direction.
We prove that Eqs.~(\ref{e8},\ref{e9}) give the same result.  Let us denote by
$\lambda(\kappa)$ the largest eigenvalue of $\hat Q(\kappa)$ with left and
right eigenstates $\langle y(\kappa)|$ and $|x(\kappa)\rangle$. For $\kappa=0$
we have $\lambda(0)=1$ and $\langle y(0)|=(1,1,...,1)$. The growth rate $r$ is
the first derivative of $\log\lambda(\kappa)$ at $\kappa=0$. As
$\lambda(0)=1$ we have $r=\lambda'(0)$, and hence
 \begin{eqnarray}
 \label{c2}
r=\frac\partial{\partial\kappa}\frac{\langle y(\kappa)|
  \hat Q(\kappa)|x(\kappa)\rangle}{\langle y(\kappa)|x(\kappa)\rangle}
=\frac{\langle y(0)|\hat Q'(0)|x(0)\rangle}{\langle y(0)|x(0)\rangle},
 \end{eqnarray}
where the last equality follows from the Hellmann-Feynman theorem. Using the
explicit form of the matrix $\hat Q(\kappa)$, $\langle y(0)|=(1,1,...,1)$, and
$|x(0)\rangle=(x_1,x_2,...,x_M)^T$ we find
\begin{equation}
\label{c3}
r=\frac{p_Mx_M-q_1x_1}{\sum_ix_i}.
\end{equation}

Now we prove the equivalence of Eq.~(\ref{c1}) and Eq.~(\ref{c3}). The
eigenvalue equation for the right eigenstate $(x_1,x_2,...,x_M)^T$ of $\hat
Q(0)$ for eigenvalue 1 is
\begin{equation}
 \label{c4}
p_{i-1}x_{i-1}+q_{i+1}x_{i+1}=x_i,
\end{equation}
for all $i$. From this we derive
\begin{equation*}
 \label{c5}
p_{i-1}x_{i-1}-q_ix_i = x_i-q_{i+1}x_{i+1}-q_ix_i=
p_ix_i-q_{i+1}x_{i+1},
\end{equation*}
where the first equality is simply (\ref{c4}) and the second equality is due to
$q_{i+1}=1-p_{i+1}$. Hence $p_{i-1}x_{i-1}-q_ix_i$ is independent of $i$
and the sum over this term for all $i$ is simply $M$ times the first
term for $i=1$. The sum over all terms can be written as
\begin{equation}
\sum_i(p_i-q_i)x_i=M(p_0x_0-q_1x_1) = M(p_Mx_M-q_1x_1),
\end{equation}
where we used the cyclic ``boundary condition'' $x_0=x_M$.
This completes the proof.\\

\subsection{M=3 Parrondo's games}

 Let us apply the theory of the previous
subsection to the concrete case of the $M=3$ Parrondo's game.  We have two
{elementary} games. The first game is a random walk on the 1-d
axis with probability $h$ for the right jumps and probability $(1-h)$ for the
left jumps. For the second game the jump parameters depend on the capital
value.  The probability for the right jumps is $\r$ for mod$(X,3)\ne 0$ and
$\s$ for the case mod$(X,3)= 0$. We randomly choose the game every round.
{For
this we have an effective $M=3$ Parrondo's game with probability for right
jumps $(h+\r)/2$ for mod$(X,3)\ne 0$ and $(h+\s)/2$ for the case mod$(X,3)=
0$.}

We solve the master equation (158) for calculating the probability distribution
after $t$ rounds. The results of iterative numerics are given in figures 20 and
21.
We see that after $t=50$ there is an oscillation near the maximum,
then as time passes the number of oscillations grows.
\begin{figure}[!]
\centerline{\includegraphics[width=.6\textwidth]{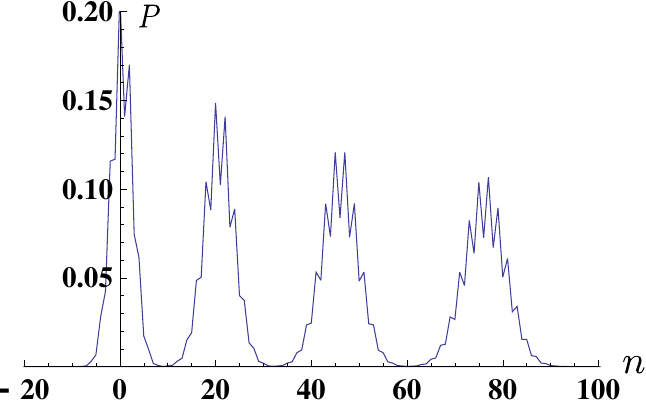}}
\caption{
%% explain l, numerics/analytics
The probability distribution for the capital growth $\hat P(n,t)$, see
Eq.~(165), for $t=50,100, 150, 200$ for the $M=3$ Parrondo's model with
$p=0.5-\ep, p_1=0.75-\ep, p_2=0.1-\ep, \ep=0.005$, [cf.~Eq.~(172)]. {For a proper
  illustration of the distributions we moved the graphs horizontally. Without
  this shift, the maximum of the distribution after $t$ rounds is located at
  the point $n=0.0052\, t$.} Our analytical results by Eq.~(154) are identical
to the results of the numerics.} \label{fig2}
\end{figure}

For the analytic solution with Eqs.~(160), (161) we obtain the matrix
$\hat Q(\kappa)$
 \begin{eqnarray}
  \label{e19}
 \left (\begin{array}{ccc}
0 & (1-p_1) & p_2\e^{-\kappa}\\
p_1 & 0 &(1-p_2) \\
(1-p_1)\e^\kappa & p_1 & 0 \end{array}\right ),
\end{eqnarray}
where $p_1=(h+\s)/2$ and $p_2=(h+\r)/2$ are parameters that we choose for the strategies.

Let us check it numerically: we have the following three strategies:

\begin{itemize}
\item $p_{1}=p_{2}=h$,  $\qquad$  for mod$(X,3)\ne0$. 
\item $p_{1}=h_{0}$, $p_{2}=h_{1}$,   $\qquad$ for mod$(X,3)\ne0$,

and the joint strategies

\item $p_{1}=\frac{h+h_{1}}{2}, p_{2}=\frac{h+h_{0}}{2}$,  $\qquad$ for mod$(X,3)\ne0$.
\end{itemize}

First let us calculate the leading eigenvalue of $\hat{Q}(\kappa)$, namely  of Eq.~(172), for 
all three strategies. From this and (168) we determine the
 (growth) rate $r$. We find the joint strategy has the opposite sign,
  which is exactly the parrondo effect.  For numerical calculation
   we take: $\s=0.745$, $h=0.495$ and $\r=0.095$. 
   For the fixed
     values of $\s, h$ and $\r$ we obtain the following 
     expressions for the rate $r$ and the variance $K$ for the corresponding strategies mentioned above:

\begin{itemize}
\item $ 
  r=-0.002898428905,  \qquad  K=0.05281583411$,         

\item $ 
   r=-0.0033333333,  \qquad   K=0.1111111110 $,

\item $ r=0.005234741795,   \qquad   K=0.09707276331$.    
   
\end{itemize}

The demonstration of this example shows that two losing
 strategies may result in a combined strategy, that is
  winning and exactly this effect is called Parrondo effect.
   We like to note that in our approach this effect is built
    on the fact that eigenvalues of a matrix are not linearly
     dependent on the entries with the trivial exception of 
     the scalar case, see section 5.1. That is why the Parrondo is known as a nonlinear effect.

The $M=3$ game with zero probability for holding the capital at the current
value, has peculiar properties: the probability distribution is non-zero for
odd differences in the capital after an odd number of time steps, and for even
differences after even time steps.  We checked that there are smooth limiting
distributions for even and odd $n$'s, see Figure~21.

\begin{figure}[!]
\centerline{\includegraphics[width=.6\textwidth]{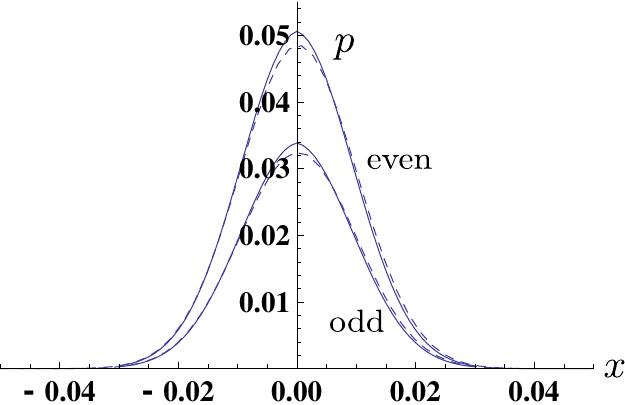}}
\caption{Illustration of $p(x)=\hat P(n,t), x=n/t-r$ with $t=1000$ for even
  $n$ (upper two lines) and for the odd $n$ (lower two lines) of the $M=3$
  Parrondo's game with the same parameters as for Figure~20. The smooth lines are
  derived according to our asymptotic formulas and Eq.~(163), the dashed lines
  correspond to the numerics.} \label{fig6}
\end{figure}

{We have seen above that strong oscillations exist in the case of zero
  probability for holding the capital at the current value, $p_l+q_l=1$. It is
  interesting and important to understand if this, namely the existence of
  degenerate saddle points
% resp.~sub-phases,
may also appear under other conditions.}

%\subsection{M=2 Parrondo's game}

Consider the $M=2$ case. Again, the first game is a random walk on the one dimensional
axis with probability $p$ for right jumps and probability $q$ for left
jumps.  For the second game we have the right jump probabilities
$\barp_1,\barp_2$ and left jump probabilities $\barq_1,\barq_2$.  {For the
  random combination of the games we have the matrix} $\hat Q(\kappa)$
%% 2nd game, combined?
\begin{eqnarray}
  \label{e18}
 \left (\begin{array}{cc}
 1-\frac{p_1+q_1+p+q}{2} &\frac{q+q_2}{2}+ \frac{p+p_2}{2}\e^{-\kappa}\\
\frac{p+p_1}{2}+\frac{q+q_1}{2}\e^{\kappa} &  1-\frac{p_2+q_2+p+q}{2}
\end{array}\right ).
\end{eqnarray}
%%% We give the characteristic equation to define the function $V(k)$
%%% in the appendix.

%\begin{figure}
%\centerline{\includegraphics[width=.4\textwidth]{fig6.eps}}
%\caption{The comparison of the asymptotic expression and
%numerical result for the distribution $p(x),x=l/(2t)-r$ (smooth
%line) of the $M=2$ Parrondo's model with the parameters
%$p_1=(h+s)/2,p_2=(h+r)/2, q_1=1-p_1,q_2=1-p_2-ep$, where
%$h=0.5-ep,s=0.1,r=0.75,ep=0.005$ with the probability distribution
%calculated from the solution of the master equation (dashed line)
%after $t=1000$ iterations.} \label{fig2}
%\end{figure}

\subsection{Games with state dependence on history}

Consider the case of random walks with memory. We define the current state by
$(X,\alpha_1,\alpha_2)$ where $X$ is the current value of the capital,
$\alpha_1$ is $+$/$-$ if the last change of the capital was gain/loss, and
$\alpha_2$ is $+$/$-$ if the second last change of the capital was gain/loss.
The parameters of the motion are allowed to depend on
$\alpha_1,\alpha_2$ and we get
 \begin{eqnarray}
P(X,+,\alpha,t+1)&=&{\sum_\beta} P(X-1,\alpha,\beta,t)p_{\alpha,\beta}, \label{e19}\\
P(X,-,\alpha,t+1)&=&{\sum_\beta}
P(X+1,\alpha,\beta,t)(1-p_{\alpha,\beta}).\nonumber
 \end{eqnarray}
Let us introduce $w(X,t),y(X,t),z(X,t),h(X,t)$ for the cases
$(-,-),(-,+),(+,-),(+,+)$, with corresponding probabilities
of the right jumps $p_1,p_2,p_3,p_4$.
Then we have the master equations
 \begin{eqnarray}
 \label{e20}
w(X,t+1)&=&w(X+1,t)(1-p_1)+z(X+1,t)(1-p_3),\nonumber\\
y(X,t+1)&=&w(X-1,t)p_1+z(X-1,t)p_3,\nonumber\\
z(X,t+1)&=&y(X+1,t)(1-p_2)+h(X+1,t)(1-p_4),\nonumber\\
h(X,t+1)&=&y(X-1,t)p_2+h(X-1,t)p_4.
 \end{eqnarray}
Performing the Fourier transform and subsequent analysis as above we get
\begin{eqnarray}
\label{e21}
w(X,t)=v_{1}\exp[tu(X/t)],\, y(X,t)=v_{2}\exp[tu(X/t)],\nonumber\\
z(X,t)=v_{3}\exp[tu(X/t)],\, h(X,t)=v_{4}\exp[tu(X/t)],\
\end{eqnarray}
where $u(x)$ is obtained from the largest eigenvalue $\lambda$ of the system
of equations
 \begin{eqnarray}
 \label{e22}
\lambda v_1&=&(1-p_1)\e^{\kappa}v_1+(1-p_3)\e^{\kappa}v_3,\nonumber\\
\lambda v_2&=&p_1\e^{-\kappa}v_1+p_3\e^{-\kappa}v_3,\nonumber\\
\lambda v_3&=&(1-p_2)\e^{\kappa}v_2+(1-p_4)\e^{\kappa}v_4,\nonumber\\
\lambda v_4&=&p_2\e^{-\kappa}v_2+p_4\e^{-\kappa}v_4.
 \end{eqnarray}

In conclusion, we considered general versions of Parrondo's games.
% which, discovered two decades ago, describe a counter-intuitive phenomenon
% related to the games.
For applications it is most important to find the capital growth rate and the
variance of the distribution. We analyzed not only these characteristics of
the models, but also found an exact distribution function. Furthermore, we
calculated analytically the asymptotics of the distribution $u(x)$ for large
$t$. The function $u(x)$ satisfies a highly non-linear differential equation,
but has an explicit expression as the Legendre transform of a computable
function, where for $M>1$ we have to carry out an eigenvalue analysis.
%AK Should we cite somebody?
Before our work, the simple matrix $\hat Q(0)$ has been used to analyze
Parrondo's games \cite{pa01}. The average growth of the capital is determined by the
eigenstate with the maximum eigenvalue 1. Here, we found that
the matrix $\hat Q(\pi)$ and its eigenvalue $-1$ lead to fundamental changes
of the characteristics of the distribution function. The existence of this
eigenvalue creates oscillations in the probability distribution of the capital
%after several rounds of playing the games}
and results into the existence of two limiting distributions. This is a typical
situation with real data of stock fluctuations in financial markets, and it is
{interesting that our simple model} describes this phenomenon. 
We gave general  formulas how to derive the capital growth and variance.
The capital growth formula is already known in the literature, the 
expressions for the variance are rather cumbersome, we just used them numerically.
%AK Here I am not sure about the sub-phases. I think that sub-leading saddle
%points are irrelevant as the sub-leading terms are exponentially smaller than
%the leading term. That is why I commented out the next lines.
%
%We also see that different sub-phases (the regimes with the tails of
%distributions are described by different analytical functions in the exponent)
%are possible when looking at large deviations from the mean values, i.e.~the
%case in optimal coding \cite{gal}.
%

\subsection{Exact probability distribution of the two-envelope problem}

The two-envelope problem has attracted researchers
and numerious work is devoted to it \cite{Christiansen}-\cite{Brams}.
 It is a choice between two cases. Such problems related to uncertainty
  and statistics play an important role in a number of fields including 
  physics, economics, game theory and finally probability theory.

The orgin of the problem dates back to \cite{Kraitchik}, where
 first the ``necktie paradox'' was presented. Later the important 
 properties of this problem were presented in
  \cite{Kraitchik1, Gardner} as the ``wallet game''. In 1988 
  the problem was recast in its present form as the so-called ``two-envelope problem'' \cite{Zabell}.
There are some similarities between the two-envelope
 game and the ``Monty Hall problem'' \cite{Savant,Flitney},
  ``Newcomb’s paradox'' \cite{Wolpert} and the
   ``St.~Petersburg paradox'' \cite{Bernoulli}, but of course it
    contains some distinguishable features.

In the two-envelope problem, the symmetry is also preserved 
if the envelopes are closed; broken when the envelopes are opened 
and an observation is being made \cite{ad10a}. {The two-envelope
 problem has been explored showing interesting similarities with
  volatility pumping on the stock market}, modeling statistical
   distribution of words in a human language, language of information 
   theory and quantum game setting \cite{ad10a}, for the 
   Schroedinger version of the game see \cite{sr}. The
    two-envelope problem possesses counterintuitive 
    dynamics as a result from symmetry breaking in discrete time ratchet phenomenon.

Let us describe it in detail. There are two-envelopes. There is money $x$ in 
one envelope and $2x$ in the other envelope, where $x$ is obtained from
 some distribution, $\rho(x)$. In each round, the player randomly chooses 
 an envelope and observes the amount inside. The question
is, should the player keep that envelope or now swap it for the other one, in order
to maximize payoff? Using the analogy in \cite{ad10a}, if the observed
 amount is $y\in{(x, 2x)}$, then the other envelope must contain
  either $2y$ or $y/2$ dollars. One may be inclined to swap because
   this will mean either he gains by a net $y$ dollars or drops back
    by $y/2$ dollars. The same argument holds if the other envelope
     was selected first, hence we have an apparent paradox \cite{ad10a}. 
     Any apparent paradox is generally due to treating what is actually 
     a conditional probability as an unconditional probability \cite{explain}.
      In this part, we focus on the two-envelope problem attributed by
       the information theorist Thomas Cover \cite{co}. The amount $x$ is
        unbounded and there is a large ensemble of independent games.

The average rate of capital has been calculated in \cite{ad10a} 
and some mathematical results have been derived using functional
 equations as part of an optimization problem. Here, we will give an
  integral representation for the exact probability distribution of the
   model, then calculate both the mean capital growth rate and
    variance of the capital distribution after a large number of games. 

We first assume that the player can either take the money from
 the envelope that has been opened, or choose another envelope.
  The choice is described via some probability function, $P(x)$.
   We need to find the average rate and variance { for the
    player's capital after $t$ rounds, for $t$ large}. The
     rate calculated in \cite{ad10a} makes use of the Cover's 
     switching function \cite{co}. It makes a biased random
      choice where the bias is conditioned on the observed value of one of the states.
%The switching probability is a some function on of observed amount \cite{co}. }%

\begin{figure}[!]
%\vspace*{-4.5cm}  
\centerline{\includegraphics[width=.7\textwidth]{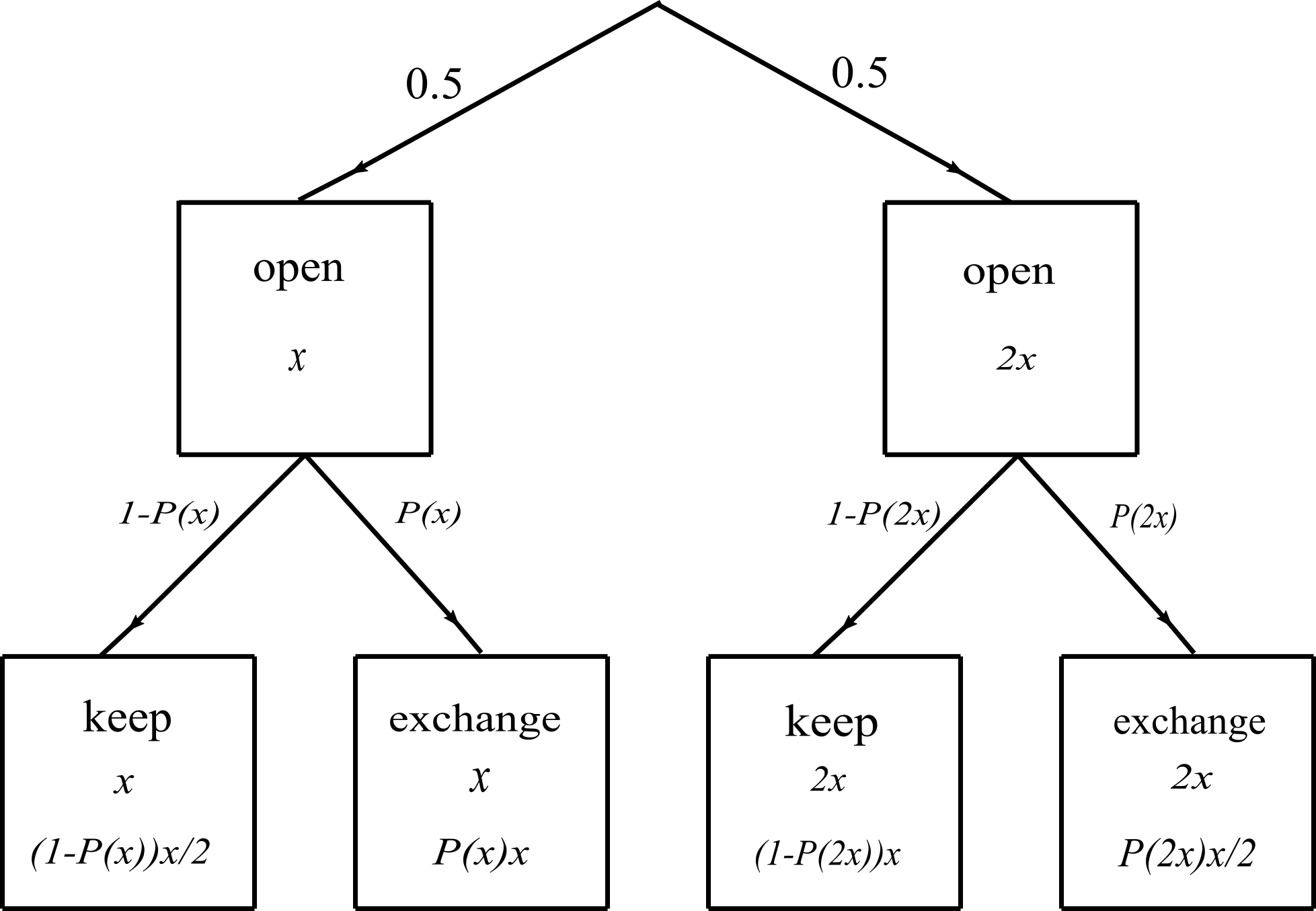}}
%\centerline{\includegraphics[scale=0.8]{Screen.pdf}}
\vspace*{0.8cm}  
\caption{The transition scheme for the two-envelope problem, related to Eq.(178). We show the processes and corresponding probabilities. In the lower boxes, we give the corresponding average change of capital according to Eq.(178).}
\end{figure}

Let us now consider an ensemble of distributions, with probability $p(t,x)$ of having capital $x$ after $t$ trials. When $\rho(x)=\delta(X-x)$,

\begin{multline}
 p(t+1,z)= p(t,z-X)\frac{1-P(X)}{2}+
p(t,z-2X)\frac{P(X)}{2}+\\+
p(t,z-2X)\frac{1-P(2X)}{2}+ p(t,z-X)\frac{P(2X)}{2}.
 \end{multline}
 \label{e13}
{ We use the probabilities $\frac{1-P(X)}{2}, \frac{P(X)}{2}$ and $\frac{1-P(2X)}{2}, \frac{P(2X)}{2}$ as the player chooses the envelope randomly.

Figure 22 illustrates the scheme for the two-envelope problem as described above, taking into account Eq.(178).}
%\begin{figure}[h!]
	%\center
	%\includegraphics[scale=0.3]{fig3.eps}
	%\caption{ The transition scheme for the two-envelope problem, related to Eq.(13). We show the processes and corresponding probabilities. In the lower boxes, we give the corresponding average change of capital according to Eq.(13). 	}
%\end{figure}

Similarly, we can obtain the general distribution in the same manner:
\begin{align}
 p(t+1,z)=
\int dx \rho[x][p(t,z-x)\frac{1-P(x)}{2}+\nonumber
p(t,z-2x)\frac{P(x)}{2}+\\
+p(t,z-2x)\frac{1-P(2x)}{2}+ p(t,z-x)\frac{P(2x)}{2}]=\nonumber
\int dx \rho[x][p(t,z-x)\frac{1-P(x)}{2}+\\+ p(t,z-x)\frac{P(2x)}{2}]
+\int dx \rho[x/2]p(t,z-x)\frac{P(x/2)}{4}+
p(t,z-x)\frac{1-P(x)}{4}.
 \end{align}
 \label{e14}

We follow the same procedure used in the last section, namely the Fourier transformation:  

\begin{eqnarray}
\label{e15} P(t,z)=\int_{-\pi}^{\pi}dk e^{iknz}\bar
P(k,t),\nonumber\\
\bar P(k,t)=\frac{1}{2 \pi}\int dz P(t,z)e^{-ikz}.
\end{eqnarray}
{ We will make use of the analogy between an abundance of money supply and particle to aid in our derivations here.} Assume that the particle starts at $n=0$ and $\bar
P(k,0)=\frac{1}{2\pi}$.

Eq.(179) can be easily transformed into
\begin{eqnarray}
\label{e16} \bar P(k,t+1)= e^{tK(-ik)}\bar P(k,t),
\end{eqnarray}
where we denote $p=-ik$,
\begin{eqnarray}
 \label{e17}
e^{K(p)}=
\int dx \rho[x]e^{-px}[\frac{1-P(x)}{2}+ \frac{P(2x)}{2}]\nonumber\\
\int dx \rho[x/2]e^{-px}[\frac{P(x/2)}{4}+\frac{1-P(x)}{4}].
 \end{eqnarray}

We can then write the solution as:
\begin{eqnarray}
\label{e18} P(t,z)=
%\frac{1}{2 \pi}
\int_{-\infty}^{\infty} dk e^{tK(ik)+ikz} \bar P(k,0).
\end{eqnarray}
{ Thus, we have obtained an integral representation for the exact probability distribution after $t$ rounds, akin to the solution to Parrondo's games \cite{sa16}. It will now be useful to calculate the average rate of capital growth and the variance of capital distribution after $t$ rounds to have a better understanding of the process. }

 \begin{figure}[!]
%\vspace*{-8cm}  
\centerline{\includegraphics[width=0.6\textwidth]{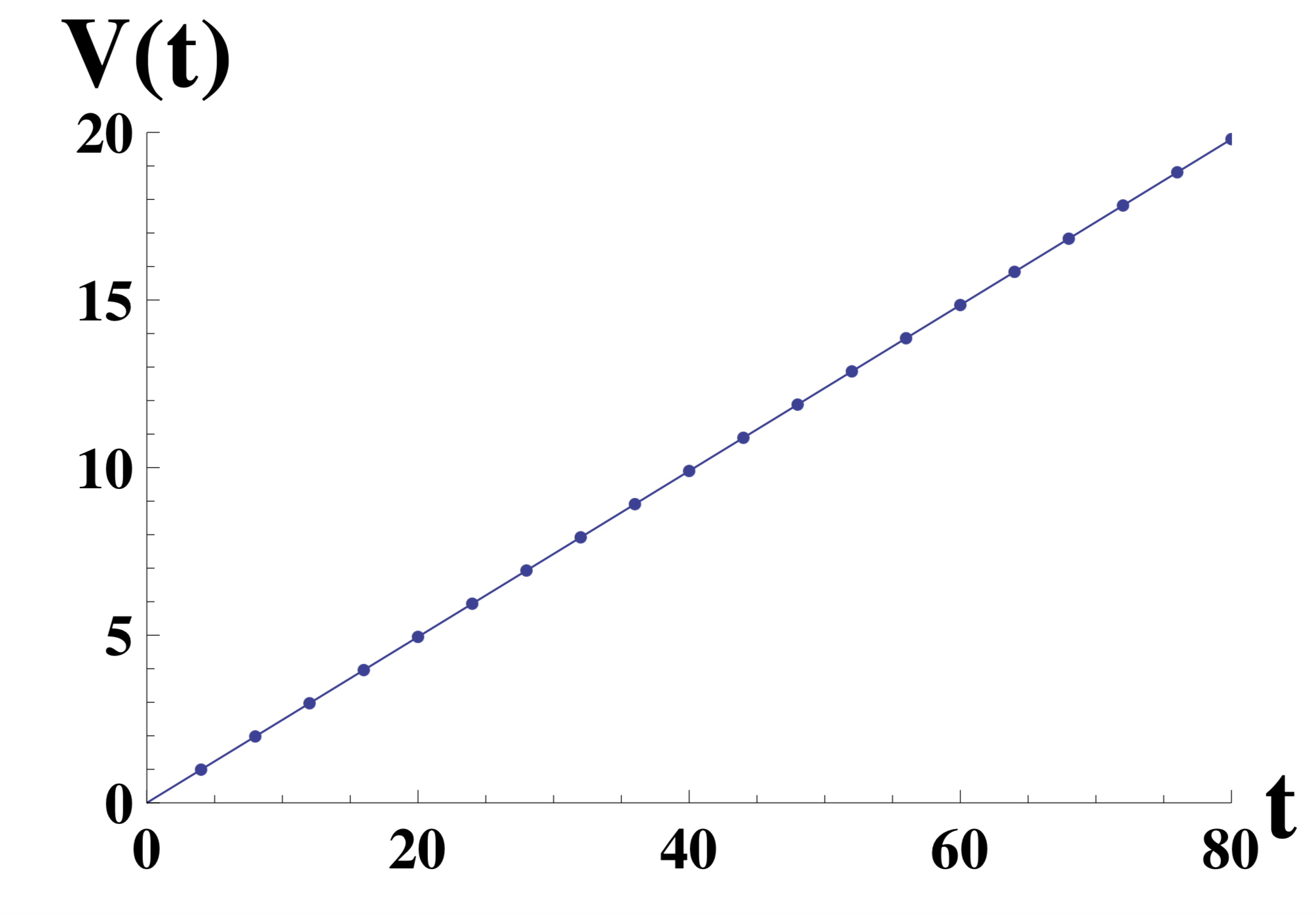}}
%\vspace*{-7cm}  
\caption{The mean capital variance  $V(t)\equiv <z^2>-<z>^2$ versus time $t$ for the model by Eq.(179) with $X=1$. The smooth line is given by our numerics using Eq.(179), whereas the solid dots are due to analytics Eq.(187) and a function $P(X)$ with $P(1)=0.2, P(2)=0.3$ was used.}
\end{figure}

Let us use an expansion
\begin{eqnarray}
\label{e19} K(p)=-rp+vp^2/2+\mathcal{O}(p^3).
\end{eqnarray}
Eqs. (183) and (184) give
 \begin{eqnarray}
 \label{e20}
r=
\int dx \rho[x]x[\frac{1-P(x)}{2}+ \frac{P(2x)}{2}]\times \nonumber\\
\times \int dx \rho[x/2]x[\frac{P(x/2)}{4}+\frac{1-P(x)}{4}].
 \end{eqnarray}

 Eq.(180) says that $r$ is the average growth rate ($r=\frac{d}{dt}<z>$) of capital in the large time limit: after $t$ rounds of the game, the player has a gain $rt$. 
In a similar way, we can also derive an expression for $v$:
 \begin{eqnarray}
 \label{e21}
(r^2+v)=
\int dx \rho[x]x^2[\frac{1-P(x)}{2}+ \frac{P(2x)}{2}]+\nonumber\\
+\int dx
\rho[\frac{x}{2}]x^2[\frac{P(\frac{x}{2})}{4}+\frac{1-P(x)}{4}].
\end{eqnarray}
{ where $V(t)\equiv vt$ is the variance of the capital distribution after $t$ rounds.} 

For the simple model by Eq.(178), Eqs. (185) and (186) give 
 \begin{eqnarray}
\label{e22}
r=(1.5-(p_2-p_1)/2)X\nonumber\\
v=(0.25-(p_2-p_1)^2/4)X^2,
\end{eqnarray}
where $p_1\equiv P(X),p_2=P(2X)$, where the saddle point $X$ turns out to be $1$.

{Figure 23 depicts the graph for the mean capital variance $V(t)\equiv <z^2>-<z>^2$ with respect to time $t$ for the model given by Eq.(179) at $X=1$, whereas the solid dots are given by Eq.(187), where a function $P(X)$ has been chosen with $P(1)=0.2,P(2)=0.3$. They are clearly in agreement with each other.}

It is not difficult to generalize the model. We can simply choose between two correlated random numbers. When we observe one of them, we naturally obtain information about the second one as well. In such a case, we can construct an algorithm to obtain a better choice among these random assets. We can then formulate the model for the general distribution $\rho(x_1,x_2)$ and minimize the variance.

\newpage
 \topskip0pt
\vspace*{\fill}
\thispagestyle{empty}
\begin{center}

\section*{\textbf{\huge{}{Conclusion and outlook}}}
\end{center}
\vspace*{\fill}
\vspace*{10cm}
\newpage

\section{Conclusion and outlook}

In this thesis different systems with many degrees 
of freedom are investigated. One of the main characteristics
 of such systems is the ``statistical'' behavior, which is
  based on the fact that the system includes uncertainties,
   randomness and is incompletely defined. Investigations
    of asymmetries, complexity, stochasticity and non-linearity
     of such systems were attempts developed in this thesis,
      which led to various theoretical and empirical analysis of
       important quantities in the financial and economic world. 

There are a number of statistical measurements and tools
 for random variables. For instance, the investigation of 
 correlation coefficients is a common way to describe the
  dynamics and dependency between random variables. 
  The moments of the distribution can also describe the
   phenomenon of interest, but the full investigation can
    only be achieved if the {\em probability distribution 
    functions} of the phenomenon are well known, whose analyses
     are intensively considered in this PhD work, for different systems, 
     by using various methods taken from statistical mechanics.

Since volatility is of major importance in financial sectors such as
 risk management, portfolio optimization and for the performance of prediction, 
 different studies of it were carried out. The first 
 analysis was based on empirical data without making
  any model assumptions, which was useful for better
   understanding the raw data themselves and to approach data-based predictions. As a result of 
    such analyses, a stunning time asymmetry was found
     between overnight and intraday volatilities. The result
      is striking and completely unexpected, as far as its
       significance and robustness are notable. 

Intraday volatilities can be considered with higher time
 resolution (even on the scale of seconds) than overnight 
 volatilities, which consists of just a simple jump. The 
 intraday price fluctuations seem to show a kind of step 
 response pattern. The fact that overnight and next day
  intraday volatilities are strongly coupled may indicate that
   intraday volatilities are rather driven by the overnight jumps
    than by any short term intraday perturbations.

It is known that there is a strong non-linear correlation
between price returns \cite{mil}. Therefore they can be used for prediction of bursts
in time series and associated risk assessment, by using non-linear
approaches. This is mostly relevant in cases of a localized breakdown of the
symmetry between gain and losses, see \cite{Mikhail, Bunde}. Accordingly the
observed strong correlations between overnight and intraday volatilities
could lead to an earlier predictability of gains/losses in financial
markets, and further improve the risk assessment, giving an earlier
Value-at-Risk estimate. It is also challenging to formulate a mechanistic
model that allows for an investigation of the distribution function and its volatility by
analytic techniques as in \cite{Zadourian, Lafond}.

Furthermore the analyses of sentiment and fear index 
were also addressed here. Such investigations are useful 
for forecasting time series (such as Dax and S$\&$P 500 index)
 which is one of the challenges in finance. The results are
  interesting and have shown that for example the fear (VIX) index
   can be a predictor for S$\&$P 500 index, by considering a 
   small time scale and by comparing volatility-VIX correlation 
   and S$\&$P 500 index volatility autocorrelation.
   
   \bigskip

Another investigation was devoted to production processes,
 as production in real life requires an immense understanding
  of supportive and inhibitive factors. To this end, a model
   analysis of the production process was carried through with
    positive drift and different distribution of noise. The model 
    is based on the experience curve hypothesis, which has been
     widely used in different domains, such
as industrial engineering and operations management services, aimed to
estimate the future costs, to reduce the production costs, to evaluate
workers’ learning profile, etc. It plays also a major role in some strategic
tasks such as estimation of capacity,  pricing and employment. 
The concept is also important for the analysis and forecasting of the price drop of products
with the number of total products produced, which can lead to
 the prediction of reduction in cost of manufacture, export potentials 
 of a product can be approximated and the stability of prices can be predicted.

 \bigskip

In this thesis potentially powerful analytical and numerical
 results are found for showing a link between cumulative
  production and production itself. In the first steps, by use
   of normally distributed noise in the process a stunning 
   relationship between the volatility of cumulative production
    and volatility of production itself was found. 
    The results suggest that cumulative production is almost
     equivalent to an exogenous time trend for predicting technological progress.  
    The evidence and finding is important and
      our empirical analysis shows how well this model-based finding sets to the dataset, which in turn
   will allow us to predict the volatility of the cumulative
    production time series quite well.

In the next step more general types of noise were considered
 as in industrial activities there exists a large number of cases
  where the distribution describing a complex phenomenon is
   not Gaussian. For this variety of applications we found a 
   comprehensive analytical approach, which is based on the 
   arbitrary probability distribution function of the model and 
   which can describe the marketing and movement of production
    process. We derived a recursion relation of integral type that
     replaces simulations by highly accurate numerical integration. 
     The results show how different types of noise affect the cumulative
      production within the model, based on the learning by doing 
      hypothesis. The distribution functions of the volatility are hardly
       characterized by the mean and variance and show rather
        interesting, sometimes singular behavior. 

Naturally, the cumulative production variables for 
successive production periods are highly correlated 
even if noise variables are independent. 
Another approach shows the occurrence of skewness
 and kurtosis by non-linear effects. This realizes a rather
  satisfactory study of the volatility of cumulative production
   within the model framework and also shows promising 
   results in the comparative analysis with empirical data. 

Our results are valid for arbitrary distribution functions
 of the production process and yield a systematic control
  over the validity of the saddle point approximation. 
  Knowing such an important quantity fosters a deeper understanding of the
industrial activities. It allows us to also understand the volatile and
complex feature of the system and accordingly to calculate the significant
quantities, by envisioning the opportunities of the model for future
investigations in risk management.

\bigskip

Finally, one of the important findings in this thesis is
 the exact distribution function for Parrondo's 
 games, which are related to Brownian ratchets and are
  striking and interesting phenomena at the intersection
  of multidisciplinary fields. For instance,
 Luenberger’s
 volatility pumping \cite{vo} is one of the
 simple and similar models to Parrondo's model,
which describes nicely the ideas of winning with
 poorer stocks in a clear way.

Parrondo's paradox is a well-known situation where 
combinations of losing strategies or harmful effects 
turn into a winning effect. Over the past decades 
numerous works reported on the importance and 
applicability of this phenomena. Parrondo's paradox 
mostly occurs where non-linearity of stochastic 
behavior exists in the process. Moreover, Parrondo's
 paradox awoke an interest in the area of finance \cite{Johnson} 
 because of the existing asymmetry in the model. 

For most applications it is critical to find the capital growth
rate and the variance of the distribution. Here are calculated not
only these characteristics for a variety of models, rather we 
found the exact probability distribution functions of the model and also 
 the
asymptotics of the distribution for large $t$. This function
satisfies a highly non-linear differential equation, but has an
explicit expression as the Legendre transform of a quite 
explicitly known function
of the momentum. 
 The tail of the distribution
is particularly interesting for applications. 

 An interesting finding is the existence of
oscillations in the probability distribution of the capital after
some rounds of gambling. Indications of such oscillations first appeared in the analysis of real financial data, but now in this thesis the phenomenon is found in model systems and a
theoretical understanding thereof. Parrondo's games
 assume a possibility to use a simple switch (degree of mixing
between several strategies), either strengthening the system or
attenuating it. The latter
situation \cite{sa05} with the possibility of anti-resonance, is typical for
the complex enough living systems.

In a similar vein, we have also considered the two-envelope problem 
and found its exact probability distributions. Probability distribution
 fitting related to the two-envelope problem can now be carried out
  for certain phenomena. Predictive analysis can then take place, for
   instance, to forecast the frequency of occurrence of the magnitude
    of the phenomenon in a certain interval of interest. It can also be
     used to obtain the unknown parameters of the model describing certain data.

\newpage

\section{Appendix}

In the Appendix one can find two of the numerical
Matlab programs that were used in this thesis. 
 The first program is based on the histogram method
 for the analysis of mutual information
  and the second program illustrates the analysis
   of the mutual information by using $K$-nearest neighbor 
 statistics. 
   
 \subsection{Numerical analysis based on histogram method for the estimation of MI}
   
   %\lstinputlisting{test21.pdf} 
%\includepdf{test21.pdf}
%\begin{lstlisting} 
%\lstinputlisting{test21.m} 
\begin{lstlisting}


% Estimation of mutual information by using histogram/binning method

N=1000;
Nbins=10;

stdwhite=0.5;
nn=100;

xmi=zeros(1,nn);
xa=zeros(1,nn);

for iii=1:100
   a=1/100*iii;
   xa(1,iii)=a;
   
% create two vector
x=zeros(1,N);
y=zeros(1,N);
% initialize
x(1,1)=rand();
y(1,1)=x(1,1);

for i=1:N-1
     x(1,i+1)=a*x(1,i)+stdwhite*randn(1);
     y(1,i+1)=a*x(1,i);

end
sxy=zeros(Nbins,Nbins);

hbinsx=(max(x)-min(x))/(Nbins-1);
hbinsy=(max(y)-min(y))/(Nbins-1);

for k=1:N
     for i=1:Nbins
         if (x(1,k)>=min(x)+hbinsx*(i-1)) && (x(1,k)<min(x)+hbinsx*(i))
              ki=i;
         end
     end
     for j=1:Nbins
         if (y(1,k)>=min(y)+hbinsy*(j-1)) && (y(1,k)<min(y)+hbinsy*(j))
              kj=j;
         end
     end
     sxy(kj,ki)=sxy(kj,ki)+1;
end
jprob=sxy/sum(sum(sxy));
  

  MI=0;
  for i=1:Nbins
       for j=1:Nbins
           if (jprob(i,j)>0)
               MI=MI+jprob(i,j)*log2((jprob(i,j)/
	           (sum(jprob(i,:))*sum(jprob(:,j)))));
           end
       end

  end
   xmi(1,iii)=MI;
end

  plot(xa,xmi)
  
  
\end{lstlisting}

% \includepdf[pages=1]{test20.pdf}
% \includepdf[pages=2]{test20.pdf}
% Insert the second page of the PDF on the same page
%\includepdfmerge[nup=1x2, pages={1,2}]{test21.pdf}

%\noindent\includepdf[pages=2, scale=0.8, pagecommand={}, fitpaper]{test21.pdf}

%\end{lstlisting}

\subsection{Numerical analysis based on KNN statistics for the estimation of MI}
%\includepdf[pages=1]{knn85.pdf}
%\includepdf[pages=2]{knn85.pdf}
%\includepdf[pages=3]{knn85.pdf}
%\lstinputlisting{knn85.m} 
\begin{lstlisting} 


% Estimation of mutual information using K nearest neighbor statistics 


% determine constants
N=1000;
K=5;
nn=100;
% B=zeros(1,N);

xmi=zeros(1,nn);
xa=zeros(1,nn);

for iii=1:100
   a=1/100*iii;
   xa(1,iii)=a;


% time series
x=zeros(1,N);
y=zeros(1,N);
% initialize
x(1,1)=rand();
y(1,1)=x(1,1);

for i=1:N-1
     x(1,i+1)=a*x(1,i)+0.5*randn();
     y(1,i+1)=a*x(1,i);

end
% calculate all distances from each point

for i=1:N 
    for j=1:N
        
        if i~=j
        
        ep_x_temp(1,i)=abs(x(1,i)-x(1,j)); 
        ep_y_temp(1,j)=abs(y(1,i)-y(1,j));
        ep_temp(i,j)=max(ep_x_temp(1,i),ep_y_temp(1,j));
        

        end
        

    end
end
        

    nx=zeros(1,N);
    ny=zeros(1,N);
    ep=zeros(1,N);

     for i=1:N
         Brow=ep_temp(i,:);          
           B=sort(Brow);
           B(:,1)=[];
          ep(1,i)=B(1,K-1);
       

        for T=1:N
            

           
       if ( abs(x(1,i)-x(1,T)))<ep(1,i)
            
            nx(1,i)=nx(1,i)+1;
            
            
       end
       if ( abs(y(1,i)-y(1,T)))<ep(1,i)
            
                ny(1,i)=ny(1,i)+1;
       end
               
        end
    end
       
   
       
%      using second estimator for MI       
        
    sum_n=0;
    for i=1:N
        
          if nx(1,i)~=0 && ny(1,i)~=0  
        
       sum_n= (psi(nx(1,i))+psi(ny(1,i)))+sum_n;
          end

        
    end

       MI=psi(K)-1/K-(1/N)*(sum_n)+psi(N);
        xmi(1,iii)=MI;
 end
  
 plot(xa,xmi)


 
\end{lstlisting}

\newpage

%\bibliography{Literature}

\newpage

%\includepdf{Ver.pdf}
\begin{lstlisting} 



\end{lstlisting}

\end{document}